%% file: main.tex
\documentclass[twocolumn]{aastex63}

\usepackage[utf8]{inputenc}
\DeclareUnicodeCharacter{2212}{-}
\usepackage{graphicx}
\graphicspath{{./}{./figures/}}
\usepackage{amsmath}
\usepackage{afterpage}
\usepackage{enumitem}
\usepackage{xcolor}
\usepackage{float}
\setenumerate{itemsep=0mm}
\usepackage{hyperref}

\usepackage{lineno}

\definecolor{mypink}{cmyk}{0, 0.7808, 0.4429, 0.1412}
\definecolor{newblue}{cmyk}{1,0.7,0,0}
\definecolor{cadmiumgreen}{rgb}{0.0, 0.42, 0.24}
\definecolor{orchid}{rgb}{0.85, 0.44, 0.84}

\input{commands}

\input{numbers}

\shorttitle{DELVE DR1}

\shortauthors{DELVE Collaboration}

\begin{document}

\reportnum{\footnotesize FERMILAB-PUB-21-075-AE-LDRD}

\title{The DECam Local Volume Exploration Survey: Overview and First Data Release}

\input{authors}

\correspondingauthor{Alex Drlica-Wagner}
\email{kadrlica@fnal.gov}

\begin{abstract}

The DECam Local Volume Exploration survey (DELVE) is a 126-night survey program on the 4\,m Blanco Telescope at the Cerro Tololo Inter-American Observatory in Chile. 
DELVE seeks to understand the characteristics of faint satellite galaxies and other resolved stellar substructures over a range of environments in the Local Volume.
DELVE will combine new DECam observations with archival DECam data to cover $\roughly \widearea \deg^2$ of high Galactic latitude ($|b| > 10\deg$) southern sky to a $5\sigma$ depth of $g,r,i,z \sim \widedepth$ mag.
In addition, DELVE will cover a region of $\roughly \mcarea \deg^2$ around the Magellanic Clouds to a depth of $g,r,i \sim \mcdepth$ mag and an area of $\roughly \deeparea \deg^2$ around four Magellanic analogs to a depth of $g,i \sim \deepdepth$ mag.
Here, we present an overview of the DELVE program and progress to date.
We also summarize the first DELVE public data release (DELVE DR1), which provides point-source and automatic aperture photometry for $\roughly 520$ million astronomical sources covering $\roughly 5000 \deg^2$ of the southern sky to a $5\sigma$ point-source depth of $g{=}\maglimpsfg$, $r{=}\maglimpsfr$, $i{=}\maglimpsfi$, and $z{=}\maglimpsfz$ mag.
DELVE DR1 is publicly available via the NOIRLab Astro Data Lab science platform.
\end{abstract}



\section{Introduction}
\label{sec:intro}

The standard model of cosmology (\LCDM) is strongly supported by observations at large spatial scales \citep[e.g.,][]{Planck:2018,DES:2018}. 
However, this fundamental model for the growth and evolution of our Universe remains challenging to test on scales smaller than our Milky Way.
Starting with the Sloan Digital Sky Survey \citep[SDSS;][]{York:2000}, large digital sky surveys have revolutionized our understanding of galaxies with stellar masses $\lesssim 10^5 \Msun$ (see recent reviews by \citealt{McConnachie:2012} and \citealt{Simon:2019}). 
We now know that our Milky Way is surrounded by scores (and likely hundreds) of faint galaxies, which span orders of magnitude in luminosity \citep[e.g.,][and references therein]{Drlica-Wagner:2020}.
Discoveries of faint satellites around our nearest galactic neighbors have begun to extend these studies beyond the Milky Way \citep[\eg,][]{Martin:2013,Chiboucas:2013,Muller:2015,Carlin:2016,Smercina:2018,Crnojevic:2019}.
Furthermore, we observe the tidal remnants of faint satellite galaxies traversing our own Galactic halo \citep[e.g.,][]{Ibata:2001a,Belokurov:2006,Shipp:2018} and the halos of other nearby galaxies \citep[e.g.,][]{Malin:1997,Ibata:2001b,Martinez-Delgado:2008,Mouhcine:2010}.
The discovery of the faintest galaxies and their remnants represents an observational frontier for large digital sky surveys, while the study of these systems continues to improve our understanding of \LCDM at the smallest observable scales.

The Dark Energy Camera \citep[DECam;][]{Flaugher:2015}, mounted on the 4\,m Blanco Telescope at Cerro Tololo Inter-American Observatory in Chile, is an exceptional instrument for exploring the faintest stellar systems.
The large field of view ($3 \deg^2$) and fast readout time ($27\second$) of DECam allow it to quickly image large areas of the sky.
DECam has been used by several large survey programs, including the Dark Energy Survey \citep[DES;][]{DES:2005,DES:2016}, the Survey of the Magellanic Stellar History \citep[SMASH;][]{Nidever:2017,Nidever:2021}, and the DECam Legacy Surveys \citep[DECaLS;][]{Dey:2019}.
Furthermore, the astronomical community has used DECam for targeted observing programs that have covered much of the remaining sky.

\begin{figure*}[t!]
\centering
\includegraphics[width=0.9\textwidth]{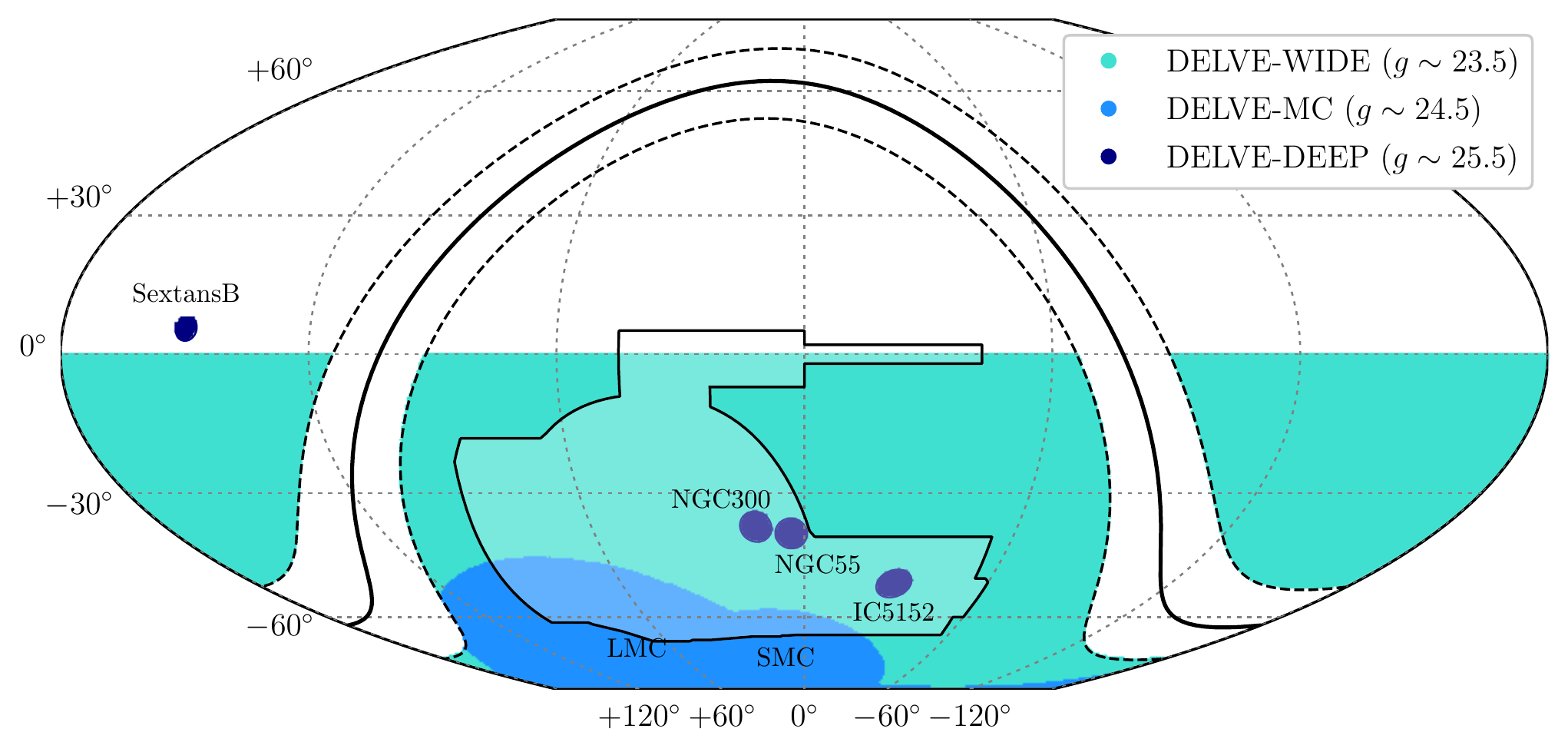}
\caption{\label{fig:delve}
DELVE combines 126 nights of allocated time with public archival data to cover the southern equatorial sky with DECam.
DELVE will provide contiguous multi-band imaging with a $5\sigma$ point-source depth of 
$g,r,i,z \gtrsim \widedepth$ mag over $\roughly \widearea \deg^2$ (turquoise region). 
In addition, a region of $\mcarea \deg^2$ around the Magellanic Clouds will be imaged to a depth of $g,r,i \geq \mcdepth$ mag (light-blue region).
Deep fields around four Magellanic analogs (Sextans B, NGC\,300, NGC\,55, and IC\,5152) will be imaged to a depth of $g,i \geq \deepdepth$ mag (dark-blue region).
The Galactic plane is indicated with a thick black line (dashed lines denote $b = {\pm}10\deg$), and the DES footprint is outlined in black. 
This figure uses an equal-area McBryde--Thomas flat polar quartic projection in celestial equatorial coordinates.
}
\end{figure*}

Here we present the {\bf DE}Cam {\bf L}ocal {\bf V}olume {\bf E}xploration survey (DELVE),\footnote{\url{https://delve-survey.github.io}} which is in the process of combining 126 nights of new DECam observations with existing public archival DECam data to assemble contiguous multi-band coverage of the entire high Galactic latitude ($|b| > 10 \deg$) southern sky (\figref{delve}).
The DELVE program consists of three main components:
(1) DELVE-WIDE seeks to complete DECam coverage of $\roughly \widearea \deg^2$ of the high-Galactic-latitude southern sky in $g,r,i,z$; 
(2) DELVE-MC provides deeper contiguous coverage of $\mcarea \deg^2$ in $g,r,i$ surrounding the Large and Small Magellanic Clouds (LMC and SMC); and 
(3) DELVE-DEEP performs deep imaging of $\deeparea \deg^2$ in $g,i$ around four Magellanic analogs in the Local Volume. 
Each survey component will combine new observations with archival data that have been self-consistently processed with state-of-the-art data management pipelines.
The wide-area DELVE data are processed with the DES Data Management pipeline \citep[DESDM;][]{Morganson:2018}, which includes point-spread function (PSF) fitting and source modeling.
The deeper DELVE data around the Magellanic Clouds and Magellanic analogs are processed with the multiepoch point-source fitting pipeline used by SMASH \citep{NideverDorta:2020} to enable deep, accurate photometry in these regions.
The DELVE data will be released on regular intervals with the first public data release (DELVE DR1) documented here.

This paper summarizes the DELVE science program, survey design, progress to date, and contents of the first data release.
We start in \secref{motivation} by describing the scientific motivation for DELVE. 
In \secref{progress} we document the survey strategy and observational progress to date.
In \secref{release} we describe DELVE DR1, including the input data set, processing, validation, and data access tools.
We provide several scientific demonstrations of the DELVE data in \secref{examples}, and we conclude in \secref{conclusion}.
Throughout this paper, all magnitudes are referenced in the AB system \citep{Oke:1974}, and all astronomical coordinates are provided in the Gaia-CRF2 reference frame \citep{Gaia:2018b} unless explicitly noted otherwise.


\section{Scientific Motivation}
\label{sec:motivation}

The \LCDM model predicts that galaxies like the Milky Way inhabit large dark matter halos that grow hierarchically by merging with and/or accreting smaller galaxies.
There is ample evidence for the \LCDM paradigm on large scales; however, small-scale tests are challenging due to the low luminosities of the faintest galaxies that inhabit low-mass dark matter halos.
In particular, ultra-faint galaxies with stellar masses $\lesssim 10^5 \Msun$ have only been identified out to distances of a few Mpc \citep[\eg,][]{McConnachie:2012, Martin:2013, Muller:2015, Carlin:2016, Smercina:2018, Crnojevic:2019}, while the census of the faintest dwarf galaxies (sometimes called ``hyper-faint'' dwarf galaxies; \citealt{Hargis:2014}) is incomplete even within the Milky Way halo \citep[\eg,][]{Tollerud:2008,Hargis:2014,Kim:2018,Simon:2019,Drlica-Wagner:2020}. 
Despite these observational challenges, the faintest galaxies provide crucial information about the role of environment and feedback on galaxy formation \citep[\eg,][]{Maschenko:2007,Wheeler:2015,Wheeler:2019,Munshi:2018,Agertz:2020,Karunakaran:2020}, reionization and the first stars \citep[\eg,][]{Bullock:2000,Shaprio:2004,Weisz:2014a,Weisz:2014b,Boylan-Kolchin:2015,Ishiyama:2016,Weisz:2017,Tollerud:2018,Graus:2019,Katz:2019}, and the nature of dark matter \citep[e.g.,][]{Bergstrom:1998,Spekkens:2013,Malyshev:2014,Ackermann:2015,Geringer-Sameth:2015,Brandt:2016,Bullock:2017,Nadler:2019b,Nadler:2020b}.
Ultra-faint galaxies also provide a unique opportunity to study the creation of heavy elements in some of the earliest star-forming environments \citep[e.g.,][]{Frebel:2015,Ji:2016,Roederer:2016}.

DELVE seeks to improve our understanding of dark matter and galaxy formation by studying the faintest satellite galaxies and their tidally disrupted remnants in a range of environments.
To accomplish this, DELVE consists of three survey programs, each with a specific observational aim.

\subsection{DELVE-WIDE}

The DELVE-WIDE program will complete DECam coverage over the entire high Galactic latitude ($|b|>10$) southern sky to provide a deep and accurate census of ultra-faint satellite galaxies around the Milky Way.
DELVE will reach a photometric depth in $g,r,i,z$ that is comparable to that of the first two years of DES.
Early DELVE-WIDE data have already resulted in the discovery of an ultra-faint satellite galaxy, Centaurus I \citep{Mau:2020}.
The full DELVE-WIDE survey will enable the detection of ultra-faint satellites similar to Centaurus I ($M_V = -5$ mag and $\mu = 27 \magasec^{-2}$) out to the virial radius of the Milky Way ($\roughly 300 \kpc$) with $>90\%$ efficiency \citep{Drlica-Wagner:2020}.
The combined model of the Milky Way and LMC satellite galaxy populations from \citet{Nadler:2020} predicts that the DELVE footprint (including the area covered by DES) contains $48 \pm 8$ satellite galaxies that are detectable by DELVE. 
Given the existing population of confirmed and candidate satellite galaxies \citep{Drlica-Wagner:2020}, this model predicts that DELVE could discover over a dozen ultra-faint satellite galaxies.

The DELVE-WIDE program also provides exceptional sensitivity to stellar streams, the remnants of tidally disrupted dwarf galaxies and globular clusters.
These resolved stellar structures provide insight into the formation and evolution of the Milky Way stellar halo \citep[e.g.,][]{Bullock:2005}.
The composition, morphology, and orbital properties of detected structures capture the recent accretion history of the Milky Way, including the masses, orbits, and metallicities of recently accreted satellites \citep[e.g.,][]{Bonaca:2020b}.
Stellar streams also probe both the large- and small-scale distribution of dark matter around the Milky Way: they trace the gravitational potential of the Milky Way over a large range of radii \citep[e.g.,][]{Johnston:1999,Bovy:2016,Erkal:2016,Bonaca:2018}, and they offer a promising way to test dark matter clustering below the threshold of galaxy formation \citep[e.g.,][]{Johnston:2002,Carlberg:2013,Erkal:2016b,Banik:2019}. 

The detection of stellar streams relies on deep, uniform coverage due to their low surface brightnesses ($\roughly32 \magasec^{-2}$) and large extents on the sky (tens of degrees).
Recent studies of stellar streams have emphasized the synergy between deep photometry with DECam, proper-motion measurements from {\it Gaia}, and radial velocities from massively multiplexed spectroscopic instruments \citep[e.g.,][]{Balbinot:2016,Shipp:2018,Jethwa:2018b,Shipp:2019,Li:2019,Li:2020,Shipp:2020,Bonaca:2020}.
DELVE-WIDE will extend the study of stellar streams by providing contiguous coverage across the southern hemisphere.

DELVE-WIDE will also enable a broad range of extragalactic science due to its wide, multi-band coverage.
In particular, DELVE-WIDE will enable extended southern-sky targeting for the Satellites Around Galactic Analogs (SAGA) program, which seeks to study more massive and luminous companions of Milky Way-like galaxies within 20--40\Mpc \citep{Geha:2017,Mao:2020}.
In addition, DELVE-WIDE will enable the search for strong gravitational lens systems, which can be used to probe the Hubble constant, dark energy, and the small-scale structure of dark matter \citep[e.g.,][]{Koopmans:2005,Treu:2010,Oguri:2010,Vegetti:2012,Treu:2018,Gilman:2019,Wong:2020}.
Furthermore, the DELVE-WIDE data can be used to study galaxies and galaxy clusters in a range of environments.
Several examples of extragalactic science with DELVE-WIDE can be found in \secref{examples}.

\subsection{DELVE-MC}

While it has long been hypothesized that the Magellanic Clouds (MCs) arrived with their own population of dwarf companions \citep[\eg,][]{Lynden-Bell:1976,DOnghia:2008}, the observational evidence for this model has been strengthened by the discovery of many ultra-faint satellites surrounding the MCs \citep[e.g.,][]{Bechtol:2015,Koposov:2015,Drlica-Wagner:2015,Drlica-Wagner:2016,Torrealba:2018,Koposov:2018,Cerny:2020}.
These discoveries have stimulated a flurry of interest in simulating and modeling the satellite populations of the MCs.
Simulations predict that up to a third of the satellites around the Milky Way originated with the MCs \citep[\eg,][]{Deason:2015,Wetzel:2015,Jethwa:2016,Dooley:2017b,Jahn:2019,Nadler:2020}.
Some studies suggest that the MCs should host more luminous satellites than are observed \citep[e.g.,][]{Dooley:2017b}, while others suggest that the MCs should host more faint satellites than are observed \citep[e.g.,][]{Jahn:2019}.
Furthermore, there is significant observational uncertainty in associating known ultra-faint galaxies with the MCs \citep{Kallivayalil:2018,Pardy:2020,Erkal:2020,Patel:2020}.

One issue in determining the satellite luminosity function of the MCs comes from the fact that the region around the MCs has only been observed by relatively shallow contiguous surveys \citep[\eg,][]{Drlica-Wagner:2016,Mackey:2018} and by deep surveys with low fill factors \citep[\eg,][]{Nidever:2017}.
The DELVE-MC program will provide deep, contiguous imaging of the MCs and their surrounding environment to robustly measure the population of faint satellites around the MCs.
DELVE has already started to probe this region with inhomogeneous early data, leading to the discovery an ultra-faint star cluster (DELVE 2; $M_V = -2.1$) located 12\kpc from the SMC and 28\kpc from the LMC \citep{Cerny:2020}.
The model of \citet{Nadler:2020} predicts that $\roughly 30\%$ of the ultra-faint galaxies contained within the DELVE footprint are associated with the MCs. 

The stellar masses, star formation histories, and interaction histories of the MCs are expected to influence the properties of their satellite populations \citep[e.g.,][]{Jethwa:2016,Dooley:2017b,Jahn:2019}.
SMASH has used DECam to study the main bodies and periphery of the MCs with a deep, partially filled survey \citep[][]{Nidever:2017}.
Photometric metallicities from SMASH suggest that the LMC periphery is not as metal-poor as would be expected in a ``classical'' halo produced by \LCDM-style hierarchical assembly.
This observation is consistent with the hypothesis that the stellar envelope of the LMC may be dominated by material from the outer LMC disk, likely stirred up through a recent interaction with the SMC \citep{Choi:2018a,Choi:2018b,Nidever:2019a}.
In parallel, recent observations of {\it Gaia}-selected red giant branch (RGB) stars suggest that even more structure exists in the periphery of the MCs \citep{Belokurov:2018,Gaia:2020b}.
The deep, contiguous imaging of DELVE-MC will extend below the oldest main-sequence turn off (MSTO) of the MCs and will be sensitive to faint substructures that can provide clues about their interaction history \citep{Massana:2020}. 
Comparisons between the stellar populations in the bodies, peripheries, and satellites of the MCs will help reconstruct the evolution of the MCs and their satellite system as they are accreted onto the Milky Way.

DELVE-MC is also well suited to study the gravitational wake of the LMC.
As the LMC moves through the Milky Way stellar halo, it pulls stars towards itself, creating an overdensity of stars along its past orbital path \citep[e.g.,][]{Garavito-Camargo:2019, Erkal:2020b}. 
Recently, it has been shown that the Pisces Overdensity \citep{Watkins:2009} matches the properties of the expected wake \citep{Belokurov:2019}. 
The wide coverage of DELVE will give a more complete view of the Milky Way's stellar halo close to the LMC, allowing us to better map the wake of the LMC and test the effect of dynamical friction. 
This, in turn, may constrain alternative dark matter models that modify dynamical friction \citep[\eg,][]{Lancaster:2020}.

\begin{figure*}[t!]
    \center
  \includegraphics[width=0.65\textwidth, trim=2.0cm 1.5cm 3.5cm 3.0cm, clip]{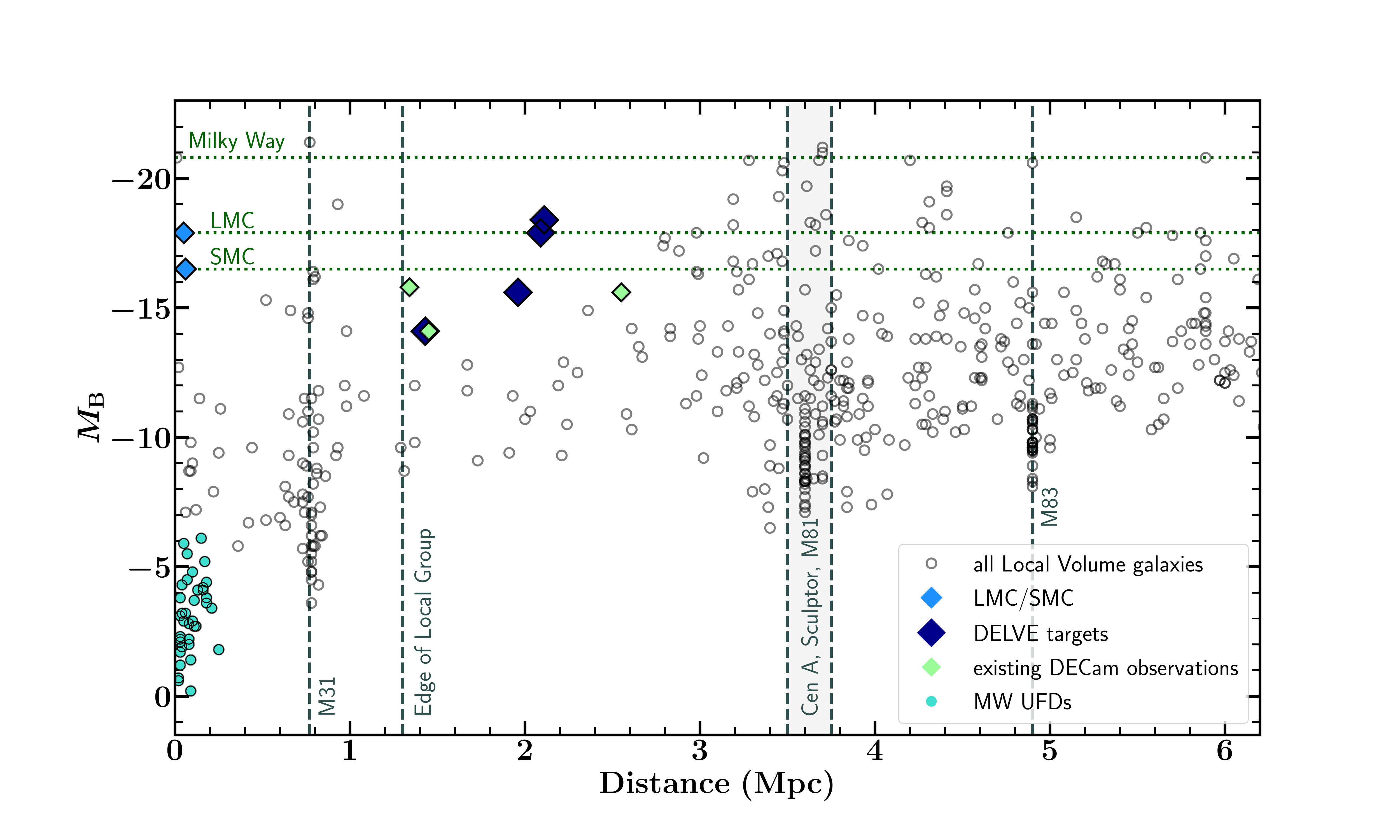}
\caption{$B$-band absolute magnitude vs.\ distance for nearby galaxies \citep{Karachentsev:2013}. 
We highlight systems targeted by DELVE: ultra-faint Milky Way satellites (turquoise circles), the LMC/SMC (blue diamonds), and MC analogs in the Local Volume (navy diamonds). Also highlighted as light-green diamonds are MC analogs with existing DECam data prior to DELVE. Little is known about the faint dwarf galaxy population ($M_{\rm B} \gtrsim -10$) beyond $\roughly 1 \Mpc$, with the exception of a handful of satellites around Milky Way-mass galaxies (vertical dashed lines). Dotted horizontal lines denote the luminosities of the Milky Way, LMC, and SMC.}
\label{fig:lumdist}
\end{figure*}

\subsection{DELVE-DEEP}

As our understanding of satellite galaxies and stellar substructures in the Local Group has improved, searches for faint stellar systems have extended to more distant galaxies. 
Within $\roughly 4\Mpc$, systematic imaging searches for dwarf companions and tidal debris have been undertaken for Centaurus\,A \citep[\eg,][]{Crnojevic:2016,Crnojevic:2019,Muller:2019}, M\,101 \citep[\eg,][]{Merritt:2014, Danieli:2017, Bennet:2019, Bennet:2020}, M\,94 \citep{Smercina:2018}, M\,81 \citep{Chiboucas:2013, Okamoto:2015, Okamoto:2019, Smercina:2020}, NGC\,253 \citep[e.g.,][]{Sand:2014, Toloba:2016, Romanowsky:2016}, and 10 other (approximately) Milky Way-mass hosts within the Local Volume \citep{Carlsten:2020a, Carlsten:2020b}. 
A complementary approach has been taken by the SAGA Survey, which identifies satellites of Milky Way-mass hosts located at 20--40\Mpc via spectroscopic verification of bright companions selected from SDSS and DECam imaging \citep{Geha:2017,Mao:2020}. 

The \LCDM model predicts that the abundance of satellite galaxies primarily depends on host halo mass \citep[e.g.,][]{Behroozi:2013,Moster:2013}. 
However, scatter in the stellar mass--halo mass relation \citep[e.g.,][]{Behroozi:2013,Garrison-Kimmel:2014,Garrison-Kimmel:2017,Munshi:2021}, or the effects of reionization, tides, ram pressure stripping, and host infall time, may be relatively more important for satellites of low-mass hosts than they are for satellites of more massive hosts \citep[e.g.,][]{Dooley:2017b}. 
Indeed, recent studies have shown that environmental effects of MC-mass hosts on their dwarf satellites are stronger than expected \citep[e.g.,][]{Garling:2020, Carlin:2019, Carlin:2020}. 
While the discovery of ultra-faint satellites associated with the MCs is broadly consistent with \LCDM, interpreting the population of MC satellites in a cosmological context is complicated by the current location of the MCs within the Milky Way's gravitational potential.

The DELVE-DEEP program targets nearby isolated galaxies with stellar masses similar to those of the MCs (``MC analogs'').
Observations of satellites around isolated MC analogs will allow studies of satellite/host demographics in environments outside the influence of a massive host.
DELVE-DEEP will complement existing DECam observations of the SMC-analog NGC\,3109 \citep{Sand:2015} and other similar surveys such as MADCASH \citep{Carlin:2016,Carlin:2019,Carlin:2020} and LBT-SONG \citep{Davis:2021}. 
By combining results from the DELVE-DEEP and other similar observing programs, it will be possible to perform a statistically robust analysis of the satellite populations of MC analogs.

DELVE-DEEP targets four isolated dwarf galaxies at 1.4--2.1\Mpc (\figref{lumdist} and \tabref{deep_targets}).
These galaxies were chosen as MC analogs: NGC\,55 and NGC\,300 have stellar masses roughly comparable to the LMC, while Sextans\,B and IC\,5152 are within an order of magnitude of the SMC ($\roughly 0.1-0.8\,M_{\rm \star, SMC}$). 
NGC 55 is an irregular, barred spiral galaxy \citep{deVaucouleurs:1991}; therefore, it closely resembles the LMC not only in stellar mass but also in morphology.  Unlike the LMC, NGC 55 is almost edge-on, which makes it well suited for stellar content studies. NGC 300 is an almost pure disk galaxy and is similar to M33, an LMC-mass satellite of M31. 
Both NGC 55 and NGC 300 belong to the nearby Sculptor Group, which is actually an unbound filamentary structure extended along our line of sight \citep{Karachentsev:2003}. 
Both Sextans B and IC 5152 are irregular dwarf galaxies. 
While Sextans B is a member of a very loose group of dwarf galaxies (the NGC 3109 association; \citealt{Tully:2006}), IC 5152 is an exceptionally isolated object, with the nearest neighbor being NGC 55 at a distance of 800 kpc \citep{Karachentsev:2002}. 
While all of our targets have been studied with deep imaging observations \citep[e.g.,][]{Tosi:1991,Zijlstra:1999,Tanaka:2011,Bellazzini:2014,Hillis:2016,Rodriguez:2016,Jang:2020}, systematic searches for dwarf satellites have not been possible due to limited sky coverage.

DELVE-DEEP will provide sensitivity to resolved stars $\gtrsim 1.5$ mag below the tip of the RGB of each target, enabling the detection of resolved satellite galaxies with $M_V \lesssim -7$ (comparable to the brightest ultra-faint satellites of the Milky Way).
DELVE-DEEP imaging covers the halos of each target \citep[$r_{\rm vir} \sim 110\kpc$ for an SMC-mass galaxy; e.g.,][]{Dooley:2017b} to provide a complete census of faint satellites with a total area of $\roughly \deeparea \deg^2$. 
According to predictions based on \LCDM, we expect to discover 5--17 satellites with $M_V \lesssim -7$ around the four targets \citep{Dooley:2017a,Dooley:2017b}. 
In addition to searches for faint satellites, the DELVE-DEEP data will enable detailed studies of the target galaxies, including searches for globular clusters and measurement of their stellar density profiles to large radii and low surface brightnesses \citep[e.g.,][]{Pucha:2019}.

\input{table_deep}


\section{Survey Design and Progress}
\label{sec:progress}

DELVE began observing in 2019 February and has collected $\roughly 12000$ new DECam exposures as of 2021 January.
The DELVE imaging data are released via the NOIRLab Astro Data Archive\footnote{\url{https://astroarchive.noirlab.edu/}} without any proprietary period.
Observations for the three DELVE programs are scheduled by an automated open-source tool that optimizes for field availability, air mass, sky brightness, and seeing.\footnote{\url{https://github.com/kadrlica/obztak}}
DELVE observing has been performed on site at CTIO and from remote stations in Tucson and at Fermilab.
Remote observing from home commenced in 2020 October due to the COVID-19 pandemic.
The observational strategy and status for each of the DELVE programs are described in more detail below.

\subsection{DELVE-WIDE}
\label{sec:survey_wide}

DELVE-WIDE seeks to assemble contiguous DECam coverage over the entire southern sky with $|b| > 10\deg$ (\figref{delve}).
Two-band photometry provides sufficient color--magnitude information to separate old, metal-poor halo populations from Milky Way foreground; however, additional color information is useful for a wide range of science topics.
Thus, DELVE-WIDE observes preferentially in the $g,i$ bands and coordinates with other DECam programs to process and assemble coverage in the $r,z$ bands (see \secref{release}).
DELVE-WIDE nominally performs $3 \times 90 \second$ dithered exposures in $g,i$ using the same icosahedral tiling scheme used by DECaLS \citep{Dey:2019},\footnote{Based on the scheme of Hardin, Sloane, and Smith (\url{http://neilsloane.com/icosahedral.codes})} but with larger dithered offsets (corresponding to roughly three times the DECam CCD dimensions). 
The DELVE tiling scheme allows the entire $\widearea \deg^2$ DELVE-WIDE footprint to be covered with three tilings in $g,i$ by $\roughly 43000$ DECam exposures.
However, a large fraction of the sky has already been covered by DES, DECaLS, and other DECam programs.
By preferentially targeting regions that lack sufficiently deep DECam data, the DELVE-WIDE footprint can be contiguously covered with 3 tilings in $g,i$ by $\roughly 20000$ new exposures.
DELVE-WIDE includes a fourth tiling when necessitated by observing constraints and existing DECam coverage.
As of 2021 January, DELVE-WIDE has collected \widenexp exposures.

When scheduling DELVE-WIDE, we use public metadata to assess the coverage and depth of archival DECam imaging. 
We calculate the effective exposure time for each archival exposure from the shutter-open time, $T_{\rm exp}$, and the effective exposure time scale factor, $t_{\rm eff}$, which combines the achieved seeing, sky brightness, and extinction due to clouds to assess the effective depth of an exposure \citep{Neilsen:2015}.
We create maps of the summed effective exposure time in each band as a function of sky position at a resolution of $\roughly 3\farcm4$ (\healpix $\nside=1024$).
Each DELVE-WIDE target exposure is compared to the existing summed effective exposure time at the target location of that exposure and is considered to have been covered if $>67\%$ of the exposure area exceeds a minimum depth threshold.
The minimum depth of each DELVE-WIDE target exposure is calculated from the tiling number, $N_{\rm tile} \in \{ 1, 2, 3, 4 \}$, the DELVE-WIDE shutter open time, $T_{\rm exp,0} = 90 \second$, and the minimum effective depth scale factor, $t_{\rm eff,min}$.
A DELVE-WIDE exposure in tiling $N_{\rm tile}$ is considered covered if the summed effective exposure time in that region of the sky is
\begin{equation}
\sum_{j} t_{{\rm eff},j} \times T_{{\rm exp}, j} > N_{\rm tile} \times t_{\rm eff, min} \times T_{\rm exp,0}.
\label{eqn:coverage}
\end{equation}
The cumulative exposure time from archival data (the left-hand side of the above equation) is calculated from archival exposures with $t_{{\rm eff},j} > 0.2$.
Motivated by experience from DES and other DECam programs, we set the minimum effective exposure time threshold as $t_{{\rm eff, min},g} = 0.4$ and $t_{{\rm eff, min},i} = 0.5$.
The DECam coverage map is updated with new archival data each semester, while completed DELVE exposures are included in real time during observing.

The DELVE-WIDE data are processed with the DESDM pipeline \citep{Morganson:2018} and serve as the basis for DELVE DR1. 
The DELVE-WIDE data processing is described in more detail in \secref{processing}.

\subsection{DELVE-MC}
\label{sec:survey_mc}

The DELVE-MC program seeks to map the periphery of the MCs by covering a contiguous region of $\roughly \mcarea \deg^2$ extending $25\deg$ around the LMC and $15\deg$ around the SMC to a depth comparable to that of SMASH, $g=24.8$, $r=24.5$, and $i=24.2$ mag \citep{Nidever:2017,Nidever:2019a}.
DELVE-MC observes in three bands to leverage $(g-r)$ and $(r-i)$ colors to help separate compact blue galaxies from Magellanic MSTO stars at $g \gtrsim 23$ mag \citep[][]{Nidever:2019a}.
Roughly half of the DELVE-MC footprint has already been covered to the desired depth by DES and SMASH, and DELVE is in the process of supplementing the remaining $\roughly 1100 \deg^2$ with $\roughly 2200$ new exposures.
DELVE-MC targets a total integrated exposure time of $800\second$ in $g,r$, and $1000\second$ in $i$ by using $3 \times 267\second$ dithered exposures in $g,r$ and $3 \times 333 \second$ dithered exposures in $i$.
The DELVE-MC tiling scheme is the same as the DELVE-WIDE survey, and regions of missing DECam coverage are determined following the procedure described in \secref{survey_wide}.
Due to the low elevation of the MCs as seen from CTIO, DECam observations of the MCs generally have larger-than-average PSF FWHM values and correspondingly lower-than-average $\teff$ values.
Based on SMASH observations, we set $t_{{\rm eff, min},g} = 0.3$, $t_{{\rm eff, min},r} = 0.3$, and $t_{{\rm eff, min},i} = 0.45$ when calculating the existing coverage.
DELVE-MC observations are scheduled when the PSF FWHM is $< 1\farcs1$ (as estimated for an $i$-band exposure taken at zenith) to help improve crowded-field photometry.
As of 2021 January, DELVE-MC has collected \mcnexp exposures.

The high stellar density in the DELVE-MC region motivates us to process the DELVE-MC data with a modified version of the \code{PHOTRED} pipeline \citep{NideverDorta:2020}.
This pipeline, called ``\code{DELVERED}'', was created to better handle large dithers between overlapping DELVE-MC exposures.  
\code{DELVERED} first performs PSF photometry for each exposure and then performs forced PSF photometry for overlapping exposures. 
At the exposure level, each night is processed separately.
This processing includes WCS correction for each CCD using \Gaia DR2, PSF photometry using DAOPHOT \citep{Stetson:1987,Stetson:1994}, photometric zeropoint determination, and aperture correction.  
The photometric zeropoint of each exposure is estimated in the same manner as for the NOIRLab Source Catalog \citep[NSC;][]{Nidever:2018,Nidever:2020a,Nidever:2020b} using all-sky catalogs and ``model magnitudes''.  
Forced PSF photometry is then performed on overlapping exposures that pass quality cuts on seeing and zeropoints.
\code{DAOPHOT}/\code{ALLFRAME} \citep{Stetson:1994} is run on all CCD images overlapping each $0.25 \deg \times 0.25 \deg$ patch of sky (referred to as a ``brick'' from the DECaLS tiling scheme) using a master source list generated from a multi-band stacked image and \SExtractor.  
The \code{ALLFRAME} results are then calibrated using the zeropoints and aperture corrections determined at the exposure level.  
Finally, weighted mean magnitudes, coordinates, and photometric variability indices are determined for each object using the measurements from the multiple exposures.

\subsection{DELVE-DEEP}
\label{sec:survey_deep}

DELVE-DEEP performs deep DECam imaging around four isolated dwarf galaxies: 
Sextans\,B, NGC\,55, NGC\,300, and IC\,5152.
DELVE-DEEP will achieve $5\sigma$ depths of $g = \deepdepthg \magn$ and $i = \deepdepthi \magn$ for each target with total integrated exposure times of $4500\second$ in $g$ and $3000\second$ in $i$. 
The nominal depth can be achieved in $15 \times 300\second$ exposures in $g$ and $10 \times 300\second$ exposures in $i$.
However, three of the four DELVE-DEEP targets (NGC\,55, NGC\,300, and IC\,5152) reside in the DES footprint; thus, we decrease the required exposure time by $900\second$ to account for the existing DES data and target $12 \times 300 \second$ $g$-band exposures and $7 \times 300 \second$ $i$-band exposures for these targets.

The DELVE-DEEP fields are chosen to roughly cover the angular region corresponding to the virial radius of the SMC \citep[$r_{\rm vir} \sim 110\kpc$;][]{Dooley:2017b} at the distance of each target (\tabref{deep_targets}).
This radius corresponds to $\roughly 3 \deg$ for NGC\,55, NGC\,300, and IC\,5152, but it is significantly larger for Sextans\,B due to its proximity ($\roughly 4.4\deg$).
Covering this large of a region around Sextans\,B is prohibitive given the DELVE-DEEP time allocation, and the coverage around Sextans\,B was reduced to $\roughly 3 \deg$.

Observations of each target start with the central pointing and progress outward.
Small dithers of $\roughly 2 \arcmin$ are applied in a hexagonal pattern to each tiling to cover the gaps between CCDs.
Star--galaxy separation is critical to the DELVE-DEEP science, and observations are performed only when the $i$-band zenith PSF is estimated to have ${\rm FWHM} < 0\farcs9$.
DELVE imaging for Sextans B was completed in 2019A, and imaging of NGC\,55 is $\roughly78\%$ complete as of 2020B.
 The DELVE-DEEP program has collected \deepnexp exposures as of 2021 January.

The DELVE-DEEP data are processed with the DESDM image co-addition pipeline in a manner similar to the DES deep fields \citep*{HartleyChoi:2020}. While this has allowed for early visual inspection and catalog analysis, the DESDM pipeline is not optimized for stellar photometry in the crowded fields near the DELVE-DEEP targets. We are exploring the use of \code{DELVERED} to perform crowded-field stellar photometry in the DELVE-DEEP fields.


\section{Data Release}
\label{sec:release}

The first DELVE data release, DELVE DR1, is based on new and archival DECam data collected as part of DELVE-WIDE. 
DELVE DR1 consists of a catalog of unique astronomical objects covering $\roughly 5000 \deg^2$ in each of $g,r,i,z$ and $\roughly 4000 \deg^2$ in all of $g,r,i,z$ simultaneously.
This section describes the DELVE DR1 data selection, processing, characterization, and validation.
The DELVE DR1 catalog can be accessed through the NOIRLab Astro Data Lab.\footnote{\url{https://datalab.noirlab.edu/delve}} 

\input{table_summary.tex}

\subsection{Data Set}
\label{sec:data}

DELVE DR1 consists of \nexp DECam exposures assembled from a combination of DELVE observations and archival DECam data.
The largest contributors to these data are DELVE itself, the DECam eROSITA Survey (DeROSITAS; PI Zenteno),\footnote{\url{http://astro.userena.cl/derositas}} DECaLS \citep[PI Schlegel;][]{Dey:2019}, and the Blanco Imaging of the Southern Sky Survey \citep[BLISS: PI Soares-Santos;][]{Mau:2019}.
However, over half of the exposures in DELVE DR1 come from $>150$ DECam community programs (\appref{propid}).
These data were downloaded from the NOIRLab Astro Data Archive.

\begin{figure}[t]
    \centering
    \includegraphics[width=\columnwidth]{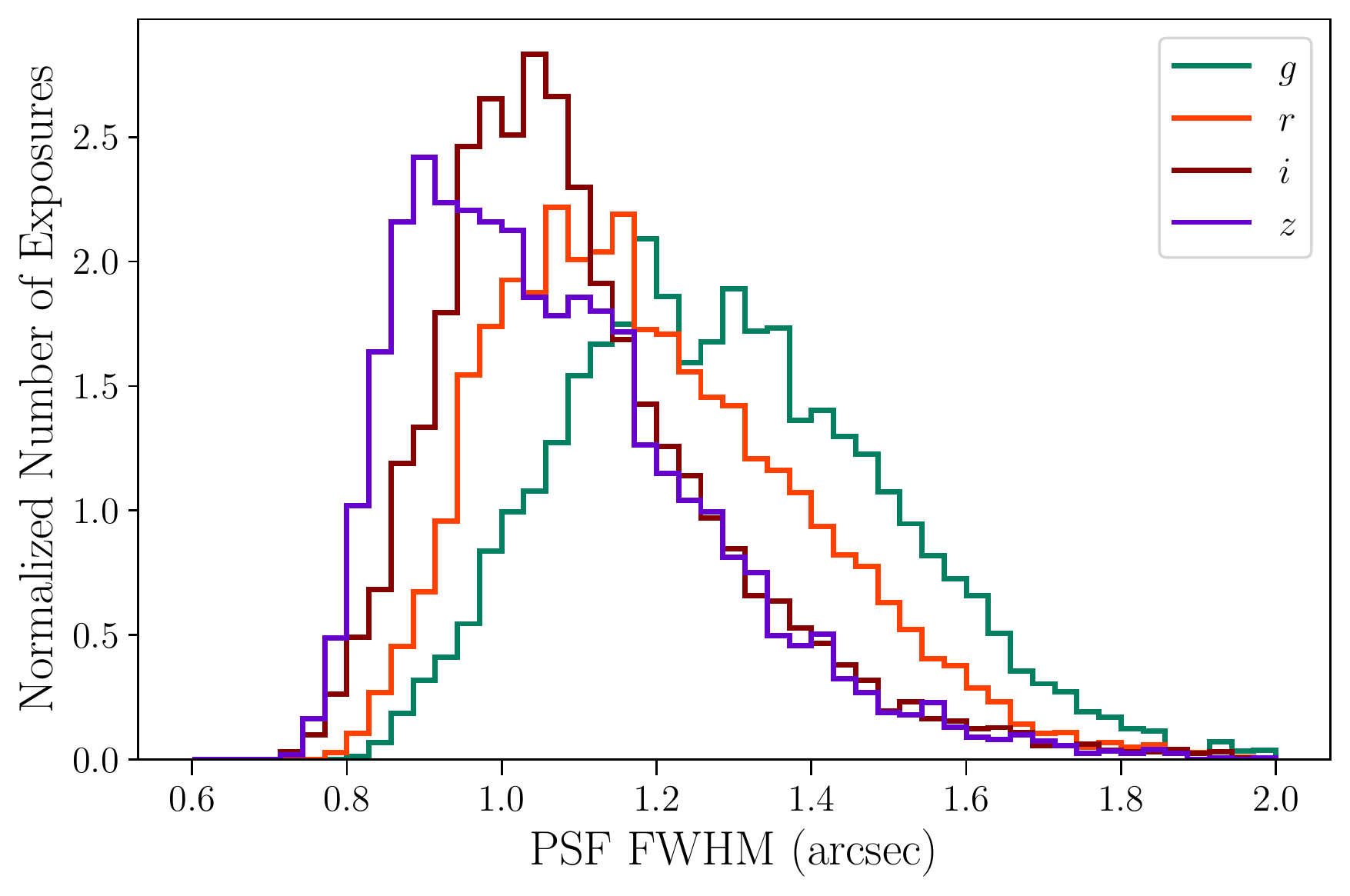}
    \caption{PSF FWHM distributions for DECam exposures included in DELVE DR1.}
    \label{fig:fwhm}
\end{figure}

The nominal DELVE DR1 region was defined as exposures having centroids with $\dec < 0\deg$ and $b > 10\deg$.
This region was extended to the Galactic plane in the region of $120\deg < \ra < 140\deg$ to enable an extended search for the Jet stream \citep{Jethwa:2018}.
Exposures were selected to have exposure time $30\second < \texp < 350\second$ and effective exposure time scale factor $\teff > 0.3$ (this is slightly higher than the $t_{{\rm eff},j} > 0.2$ requirement applied in \secref{survey_wide}). 
While no explicit cut was placed on the PSF FWHM, the cut on \teff removes exposures with very poor seeing (\figref{fwhm}).
Furthermore, exposures were required to have good astrometric solutions when matched to \Gaia DR2 by \SCAMP \citep{Bertin:2006} including $>250$ astrometric matches, $\chi^2_{\rm astrom} < 500$, $\Delta(\ra) < 150\mas$, and $\Delta(\dec) < 150\mas$.
The key properties of the DELVE DR1 are listed in \tabref{summary}.

\subsection{Data Processing}
\label{sec:processing}

All exposures in DELVE DR1 were consistently processed with the DESDM ``Final Cut'' pipeline \citep{Morganson:2018}.
Identical processing was performed at Fermilab and NCSA; DELVE DR1 is derived from the Fermilab processing. 
DELVE implements updates to the DESDM pipeline that were developed for DES DR2 (see Section 3 of \citealt{DES-DR2:2021}).
Data were reduced and detrended using custom, seasonally averaged bias and flat images, and full-exposure sky background subtraction was performed \citep{Bernstein:2018}. 
\SExtractor \citep{Bertin:1996} and \PSFEx \citep{Bertin:2011} were used to automate source detection and photometric measurement. 
Astrometric calibration was performed against \Gaia DR2 using \SCAMP \citep{Bertin:2006}.
Photometric zeropoints for each CCD were derived by performing a $1\arcsec$ match between the DELVE Final Cut \SExtractor catalogs and the ATLAS Refcat2 catalog \citep{Tonry:2018}, which places measurements from Pan-STARRS1 (PS1) DR1 \citep{Chambers:2016}, SkyMapper DR1 \citep{Wolf:2018}, and several other surveys onto the PS1 $g,r,i,z$-bandpass system.
Transformation equations from the ATLAS Refcat2 system to the DECam system were derived by comparing calibrated stars from DES DR1 (see \appref{transforms} for details).
Zeropoints derived from the DELVE processing and photometric calibration pipeline were found to agree with those derived by DES DR1 with an rms scatter of $\roughly 0.01 \magn$.

We built a multi-band catalog of unique sources by combining the \SExtractor catalogs from each individual CCD image following the procedure of \citet{Drlica-Wagner:2015}.
We took the set of \SExtractor detections with $\var{flags} < 4$, which allows neighboring and deblended sources, and $(\var{imaflags\_iso}\,\&\,2047) = 0$, which removes objects containing bad pixels within their isophotal radii \citep{Morganson:2018}.
We further required that each detection have a measured automatic aperture flux, a measured PSF flux, and a PSF magnitude error of $< 0.5$ mag.
We sorted \SExtractor detections into $\roughly 3 \deg^2$ \healpix pixels ($\nside=32$), and within each \healpix pixel we grouped detections into clusters by associating all detections within a $1\arcsec$ radius.
Occasionally, two real astronomical objects are located within $1 \arcsec$ of each other and were grouped into the same cluster.
We identified these ``double'' objects if they were contemporaneously detected on a single image.
If a double object was detected in two or more images, we split its parent cluster into two distinct clusters.
We did not split clusters when more than two objects were detected within $1\arcsec$. 
This splitting procedure means that the DELVE DR1 catalog should be treated with care when searching for multiple closely separated objects.

Each cluster of detections was associated with an object in the DELVE DR1 catalog.
The astrometric position of each object was calculated as the median of the individual single-epoch measurements of the object.
We track two sets of photometric quantities for each object: (1) measurements from the single exposure in each band that had the largest effective exposure time (i.e., the largest $\teff \times T_{\rm exp}$), and (2) the weighted average (\var{WAVG}) of the individual single-epoch measurements.
The weighted average and unbiased weighted standard deviation were calculated following the weighted sample prescriptions used by DES \citep{Gough:2009,DES-DR2:2021}.\footnote{\url{https://www.gnu.org/software/gsl/doc/html/statistics.html\#weighted-samples}}
In addition, we track cluster-level statistics such as the number of detections in each band.

We follow the procedure of DES DR1 \citep{DES-DR1:2018} to calculate the interstellar extinction from Milky Way foreground dust.
We calculate the value of $E(B-V)$ at the location of each catalog source by performing a bi-linear interpolation in $(\ra,\dec)$ to the maps of \citet{Schlegel:1998}.
The reddening correction for each source in each band, $A_b = R_b \times E(B-V)$, is calculated using the fiducial interstellar extinction coefficients from DES DR1: $R_g = 3.185$, $R_r = 2.140$, $R_i = 1.571$, and $R_z = 1.196$ \citep{DES-DR1:2018}.
Note that, following the procedure of DES DR1, the \citet{Schlafly:2011} calibration adjustment to the \citet{Schlegel:1998} maps is included in our fiducial reddening coefficients.
The $A_b$ values are included for each object in DELVE DR1 but are not applied to the magnitude columns by default.
The list of the photometric and astrometric properties provided in DELVE DR1 can be found in \appref{tables}.


\subsection{Sky Coverage}
\label{sec:coverage}

We quantify the area covered by DELVE DR1 in the form of \healpix maps with angular resolution of $\roughly 0\farcm86$ ($\nside = 4096$).
These maps were created by pixelizing the geometry of each DECam CCD using the \code{decasu}\footnote{\url{https://github.com/erykoff/decasu}} package built on \code{healsparse}.\footnote{\url{https://healsparse.readthedocs.io}}
This package calculates the geometry of each CCD using higher-resolution nested \healpix maps ($\nside = 16384$) and sums the resulting covered pixels to generate lower resolution maps containing the fraction of each pixel that is covered by the survey.
This bypasses the computationally intensive \code{mangle} processing done by DES \citep{Swanson:2008,Morganson:2018} while retaining the same accuracy at a resolution of $\nside=4096$.
We quantitatively estimate the covered area as the sum of the coverage fraction maps in each band independently, as well as the intersection of the maps in all four bands.
These values are reported in \tabref{summary} and visualized in \appref{depth}.
 

\subsection{Astrometry}
\label{sec:astrometry}

We assess the internal astrometric accuracy by comparing the distributions of angular separations of individual detections of the same objects over multiple exposures.
The median global astrometric spread is $\astrorepeatall \mas$ across all bands.
We find that this spread is fairly consistent within each band, with median offsets in $g,r,i,z$ of $\astrorepeatg, \astrorepeatr, \astrorepeati, \astrorepeatz\mas$.
Furthermore, we compare the DELVE DR1 catalog-coadd object locations to the locations of matched sources in the \Gaia DR2 catalog \citep{Gaia:2018}, and we find an overall astrometric agreement of $\astroabs \mas$. 
This comparison is somewhat circular, since the \Gaia DR2 catalog was used for the image-level astrometric calibration; however, the good agreement confirms that no significant astrometric offsets have been introduced by the catalog coaddition procedure.


\begin{figure*}
    \centering
    \includegraphics[width=0.98\textwidth]{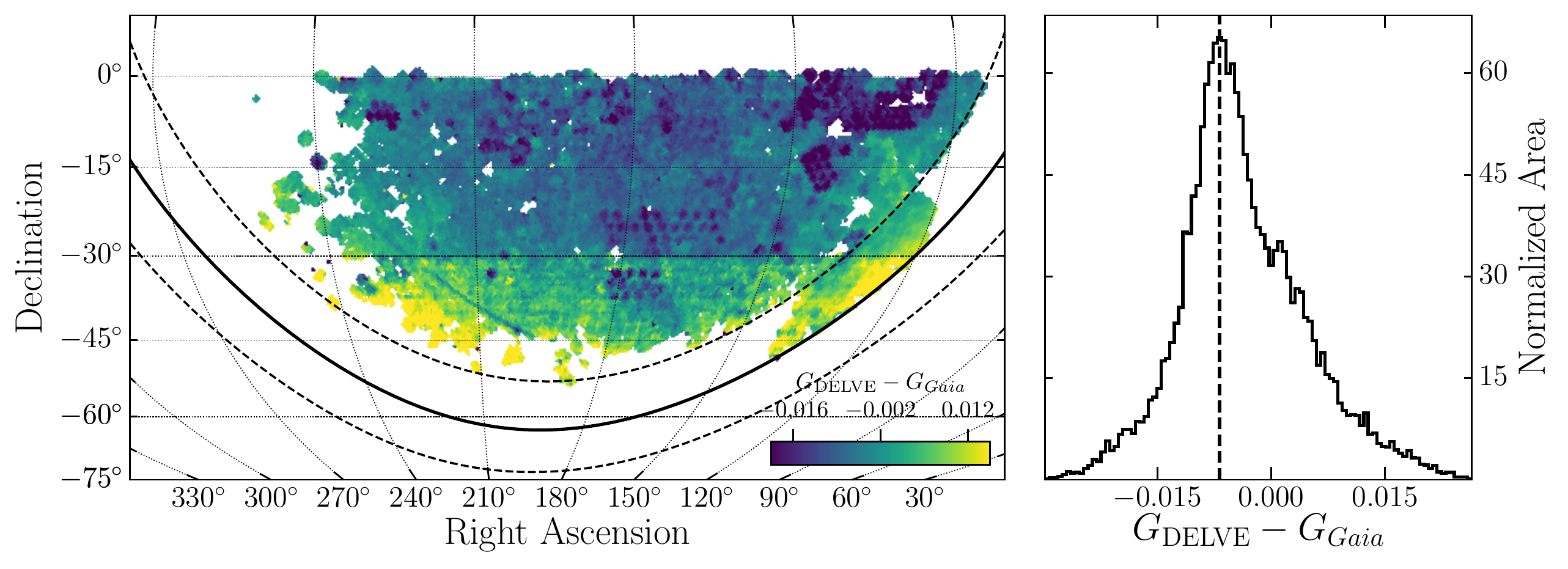}
    \caption{Median difference between the DELVE DR1 photometry transformed into the \Gaia $G$-band, $G_{\rm DELVE}$, and the measured \Gaia magnitude, $G_\Gaia$. The spatial distribution of the median difference in each pixel is shown in the left panel, while the right panel shows a histogram of the pixel values. A shift in the zeropoint can be seen at $\dec \sim -30\deg$, which corresponds to the boundary between the ATLAS Refcat2 use of PS1 and SkyMapper (\secref{photrel}).
This comparison is restricted to the area with overlapping DELVE DR1 coverage in all four bands ($g,r,i,z$).
}
    \label{fig:gaia}
\end{figure*}

\subsection{Relative Photometric Calibration}
\label{sec:photrel}

We assess the photometric repeatability in each band from the root-mean-square (rms) scatter between independent PSF measurements of bright stars. 
For each band, we select stars with $16 < \var{WAVG\_MAG\_PSF} < 18$ mag and calculate the median rms scatter in $\roughly 0.2 \deg^2$ \healpix pixels ($\nside=128$).
We estimate the median of the rms scatter over the entire footprint in each band.
This quantity is found to be $\roughly 5 \mmag$ and is listed for each band in \tabref{summary}.

We validate the photometric uniformity of DELVE DR1 by comparing to space-based photometry from \Gaia DR2 (\figref{gaia}).
We transform the $g,r,i,z$ photometry from DELVE to the \Gaia $G$ band following a set of transformations derived for DES \citep{Sevilla-Noarbe:2020,DES-DR2:2021}.
We compare the \Gaia DR2 $G$-band magnitude in the AB system to the $G$-band AB magnitude of stars in DELVE with $16 < r < 20 \magn$ and $0 < i-z < 1.5 \magn$.
We calculate the median difference, $G_{\rm DELVE} - G_\Gaia$, within each $\nside=128$ \healpix pixel (\figref{gaia}).
We find that the median offset between DELVE DR1 and \Gaia DR2 is $-4.8 \mmag$. 
We estimate the photometric uniformity across the DELVE DR1 data as the standard deviation of the median differences across pixels, which yields a value of \photgaia \mmag (\tabref{summary}).
Since the distribution of residuals is non-Gaussian, we also calculate the 68\% containment, which is $15.3 \mmag$.

Similar comparisons between DES and \Gaia DR2 have demonstrated that the nonuniformity of \Gaia observations can be the dominant contributor to photometric nonuniformity estimated using this technique \citep{Burke:2018,Sevilla-Noarbe:2020,DES-DR2:2021}.
However, it is clear from the spatial structure in \figref{gaia} that systematics in the DELVE DR1 calibration dominate the nonuniformity relative to \Gaia.
Furthermore, we observe a systematic shift of $11.2 \mmag$ relative to \Gaia at $\dec = -30\deg$. 
This is the declination at which ATLAS Refcat2 switches from using PS1 to SkyMapper, and a similar feature can be seen in the residuals of Figure 3 in \citet{Tonry:2018}. 
This offset is the dominant contributor to the broadening of the residuals between DELVE DR1 and \Gaia DR2 seen in the left panel of \figref{gaia}.
The relative photometric calibration of DELVE could be improved in the future by performing an internal relative calibration such as the ``ubercalibration'' of SDSS and PS1 \citep[][]{Padmanabhan:2008,Schlafly:2012}, or the  forward global calibration of DES \citep[i.e.,][]{Burke:2018}.

\begin{figure*}[t]
    \centering
    \includegraphics[width=0.49\textwidth]{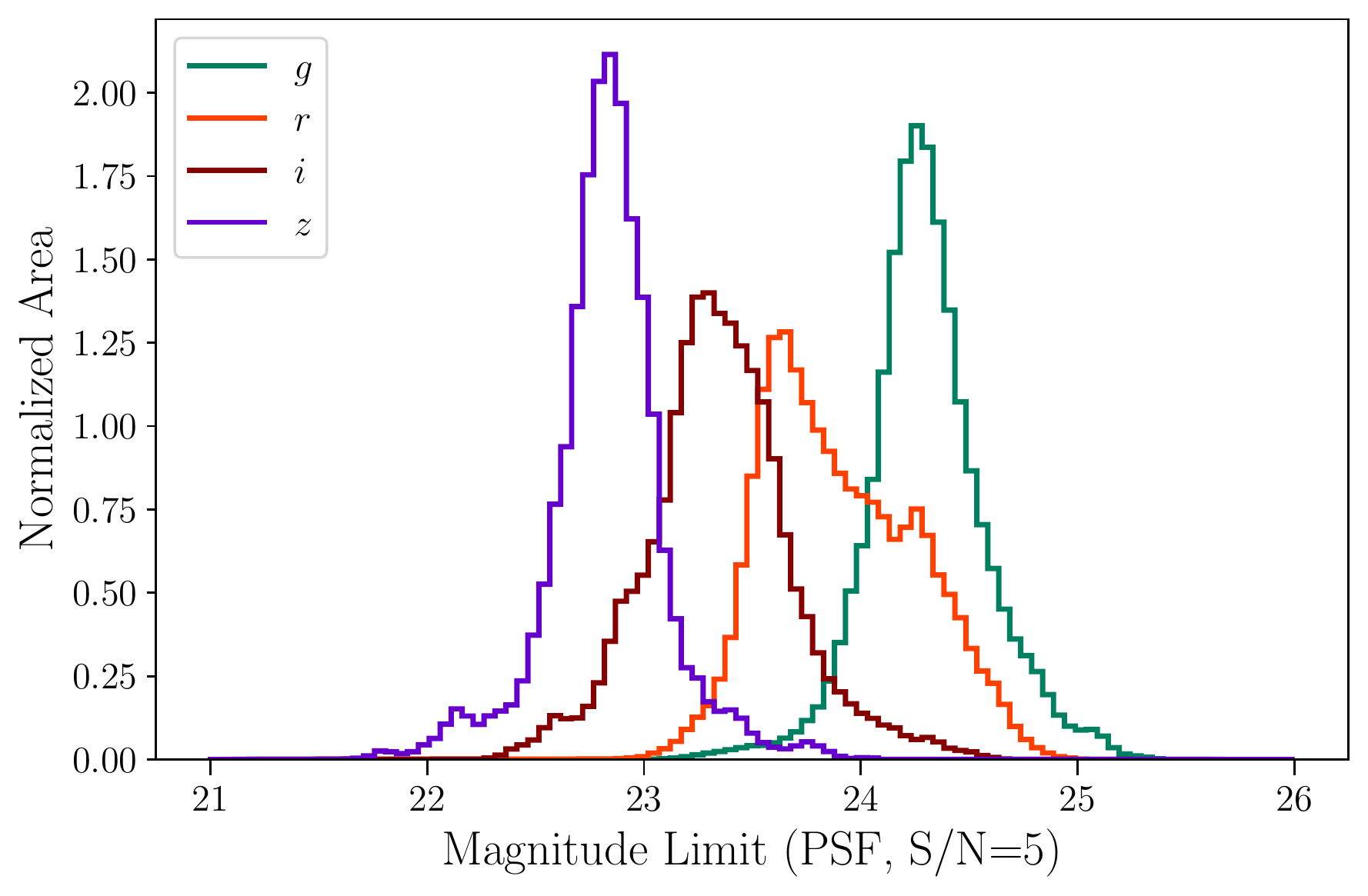}
    \includegraphics[width=0.49\textwidth]{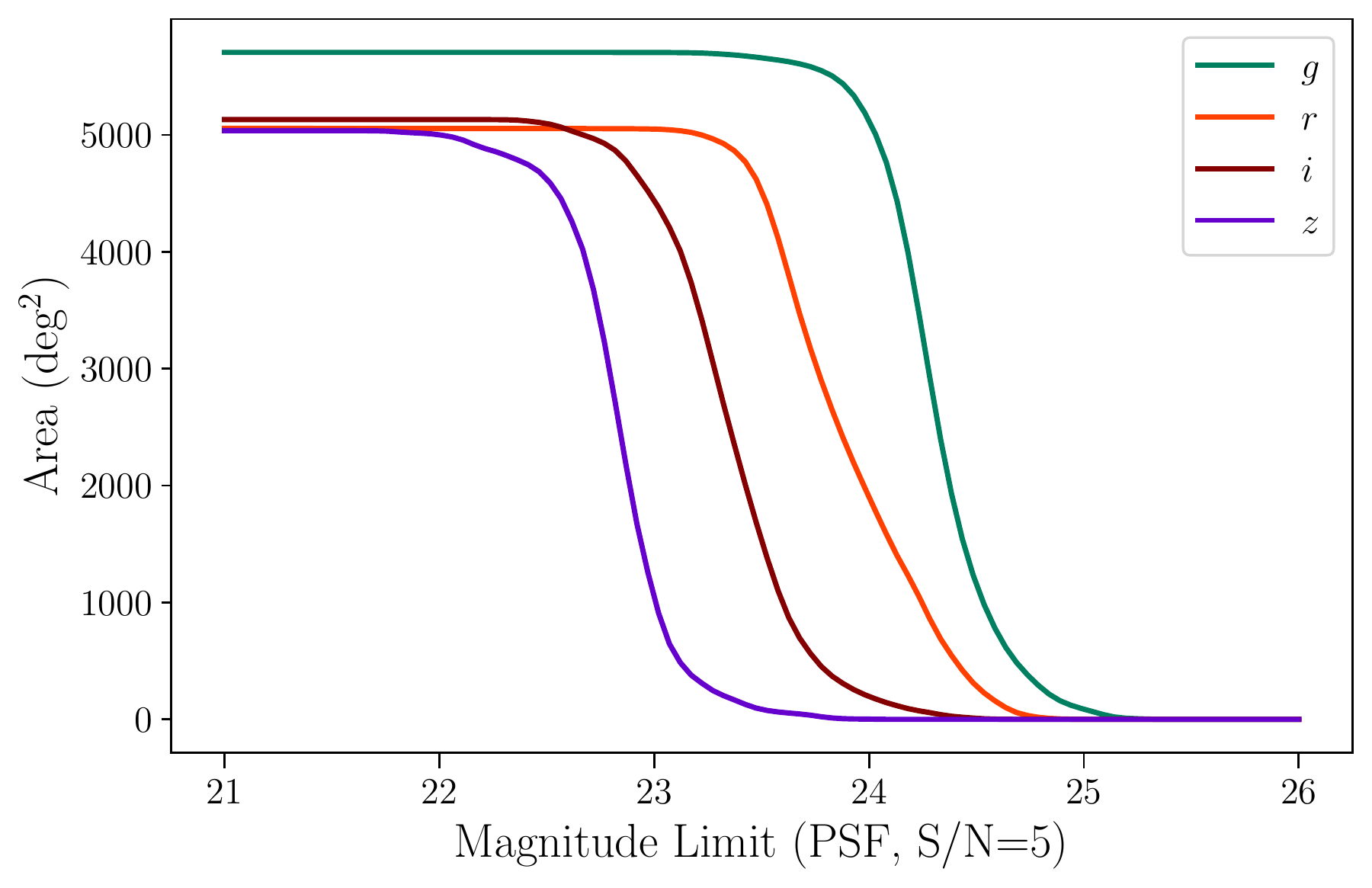}
    \caption{(Left) Distribution of PSF magnitude limits for point-like sources at S/N=5. (Right) DELVE DR1 survey area in each band as a function of the limiting PSF magnitude (S/N=5). These distributions look similar when calculated from the \magauto limiting magnitude for all sources but are shifted brighter by $\roughly 0.4 \magn$.
    }
    \label{fig:maglim}
\end{figure*}

\subsection{Absolute Photometric Calibration}
\label{sec:photabs}

We do not have a precise estimate of the absolute photometric calibration of DELVE DR1, but a rough estimate can be derived by comparing to other data sets.
The absolute photometry of DELVE DR1 is tied to the HST CalSpec standard star C26202 via DES DR2, which was used to adjust the zeropoints of the ATLAS Refcat2 transformation equations (\appref{transforms}).
Thus, DELVE DR1 cannot have a better absolute calibration than DES DR2, which sets a lower limit on the statistical uncertainty of $2.2 \mmag$ per band and a systematic uncertainty of $11$ to $12 \mmag$ per band (see Table 1 of \citealt{DES-DR2:2021}).
The global offset between DELVE DR1 and \Gaia DR2 (which is not seen in DES DR2) suggests that the absolute calibration cannot be better than $4.8 \mmag$.
Combining the maximum systematic uncertainty on the absolute calibration from DES DR2 and the DELVE DR1 offset relative to \Gaia, we conservatively estimate that the absolute photometric accuracy of DELVE DR1 is $\lesssim \photabs \mmag$.
This is comparable to the absolute photometric accuracy estimated for the DES first-year cosmology data set \citep{Drlica-Wagner:2018}.
In the future, the absolute photometric calibration of DELVE can be improved and verified directly following a similar procedure to DES DR2  using HST CalSpec standards and/or pure hydrogen atmosphere white dwarf stars \citep{DES-DR2:2021}.

\subsection{Photometric Depth}
\label{sec:depth}

The photometric depth of DELVE DR1 can be assessed in several ways.
One common metric is to determine the magnitude at which a fixed signal-to-noise ratio (S/N) is achieved \citep[e.g.,][]{Rykoff:2015}.
The statistical magnitude uncertainty is related to the S/N calculated from the flux, $F/\delta F$, via propagation of uncertainties and Pogson's law \citep{Pogson:1856},
\begin{equation}
\delta m = \frac{2.5}{\ln 10} \frac{\delta F}{F}.
\end{equation}
\noindent Using this equation, we estimate the magnitude at which DELVE DR1 achieves S/N=5 ($\delta m \approx 0.2171$) and S/N=10 ($\delta m \approx 0.1085$).
We calculate these magnitude limits for point-like sources using \magpsf and for all sources using \magauto.
For each magnitude and S/N combination, we select objects and interpolate the relationship between $m$ and ${\rm median}(\delta m)$ in $\roughly 12 \amin^2$ \healpix pixels ($\nside = 1024$).
The resulting median magnitude limits over the DELVE DR1 footprint are shown in \tabref{depth}.
We show histograms of the \magpsf magnitude limit for point-like sources at S/N=5 in the left panel of \figref{maglim}.
In the right panel of \figref{maglim} we show the DELVE DR1 area as a function of depth in each band.
The magnitude limits as a function of location on the sky are shown in \appref{depth}. 
The median S/N=10 point-source depth of DELVE DR1 is comparable to the point-source depth from the first two years of DES \citep{Drlica-Wagner:2015}, which is consistent with the goal of the DELVE-WIDE program.

\input{table_depth}

\subsection{Object Classification}
\label{sec:classification}

DELVE DR1 includes the \SExtractor \spreadmodel parameter, which can be used to separate spatially extended galaxies from point-like stars and quasars \citep[\eg,][]{Desai:2012}.
Following DES \citep[e.g.,][]{DES-DR1:2018,DES-DR2:2021}, we define \extclass parameters as a sum of several Boolean conditions,
{\par\footnotesize\begin{align}\begin{split}
\var{ext}&\var{ended\_class\_g} = \\
~&((\spreadmodel[g] + 3\, \spreaderrmodel[g]) > 0.005) \\
+&((\spreadmodel[g] +     \spreaderrmodel[g]) > 0.003) \\
+&((\spreadmodel[g] -     \spreaderrmodel[g]) > 0.003).
\end{split}\end{align}}

\noindent When true, each Boolean condition adds one unit to the classifier such that an \extclass value of 0 indicates high-confidence stars, 1 is likely stars, 2 is likely galaxies, and 3 is high-confidence galaxies.
Objects that lack coverage in a specific band or where the \spreadmodel fit failed are set to a sentinel value of $-9$.
We calculate \extclass values similarly for each band; however, we recommend the use of \extclass[G] since the $g$ band has the widest coverage and deepest limiting magnitude. 

\begin{figure}
    \centering
    \includegraphics[width=0.9\columnwidth]{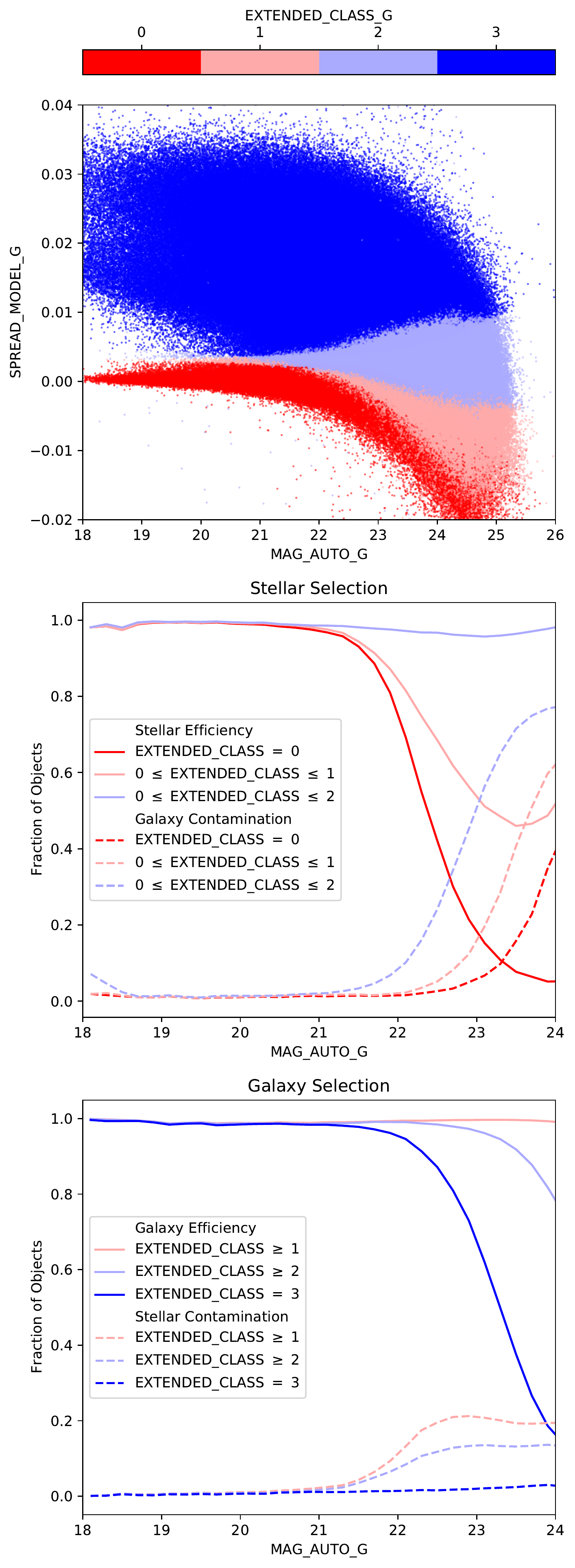}
    \caption{\emph{Top:} Stars and galaxies occupy distinct regions of \spreadmodel space. 
      At fainter magnitudes it becomes harder to distinguish between the two types of objects. 
    \emph{Middle:} DELVE DR1 stellar classification performance as estimated from matched HSC observations. 
    \emph{Bottom:} Similar to the middle panel, but showing DELVE DR1 galaxy classification performance.}
    \label{fig:stargal}
\end{figure}

We evaluate the performance of \extclass[g] by matching DELVE DR1 objects to data from the W04 (WIDE12H+GAMA15H) equatorial field of the wide layer of HSC-SSP PDR2 \citep{HSC-DR2}. 
The superior image quality ($i$-band PSF FWHM $\roughly 0\farcs6$) and depth ($i \sim 25.9 \magn$) of HSC-SSP PDR2 enable robust tests of star--galaxy separation in DELVE DR1.
The matched data set covers $\roughly 95 \deg^2$ and contains $\roughly 2.3$ million matched objects.
Following \citet{DES-DR1:2018}, we select point-like sources from HSC-SSP DR2 based on the difference between the $i$-band PSF and model magnitudes of sources,
{\par\footnotesize\begin{align}\begin{split}
\var{hsc}&\var{\_stars} = \\
& ((\var{i\_psfflux\_mag} - \var{i\_cmodel\_mag}) < 0.03) \\
& ||~ (~ ( (\var{i\_psfflux\_mag} - \var{i\_cmodel\_mag}) < 0.1) \\
& ~~~ \& ~ (\var{i\_psfflux\_mag < 22})~).
\end{split}\end{align}}
We require that the PSF and model magnitudes are very similar for fainter sources, while the agreement is relaxed for brighter sources.
This selection results in $\roughly 1.7$ million matched objects classified as galaxies and $\roughly 0.6$ million matched objects classified as stars.
We use these objects to evaluate the differential performance of DELVE DR1 \extclass[g] as a function of magnitude in \figref{stargal}.
We report the performance of nominal star ($0 \leq \extclass[g] \leq 1$) and galaxy ($2 \leq \extclass[g]$) classifications integrated over the magnitude range $19 \leq \magauto[g] \leq 22$ mag in \tabref{summary}.

\begin{figure*}[t!]
    \centering
    \includegraphics[width=0.49\textwidth]{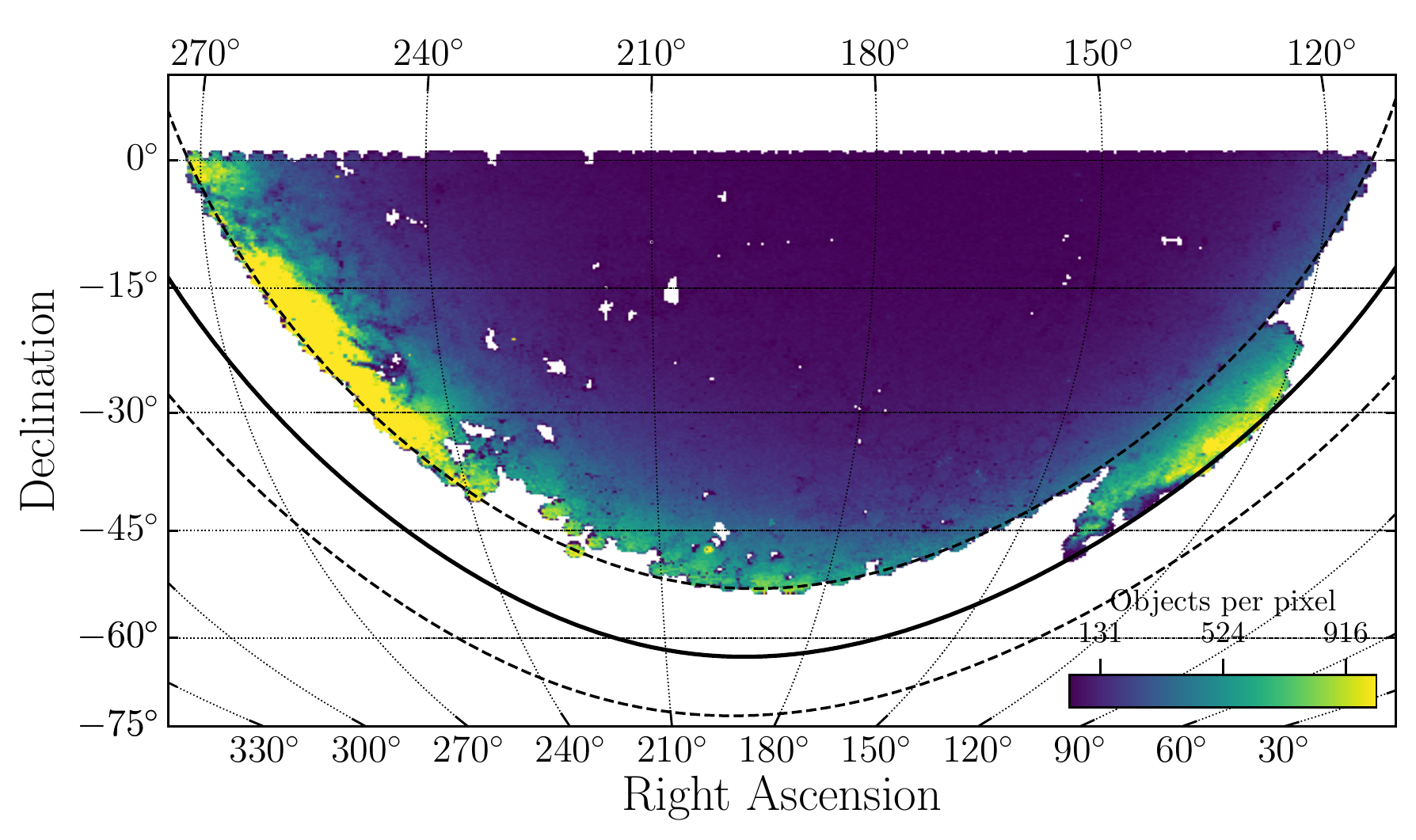}
    \includegraphics[width=0.49\textwidth]{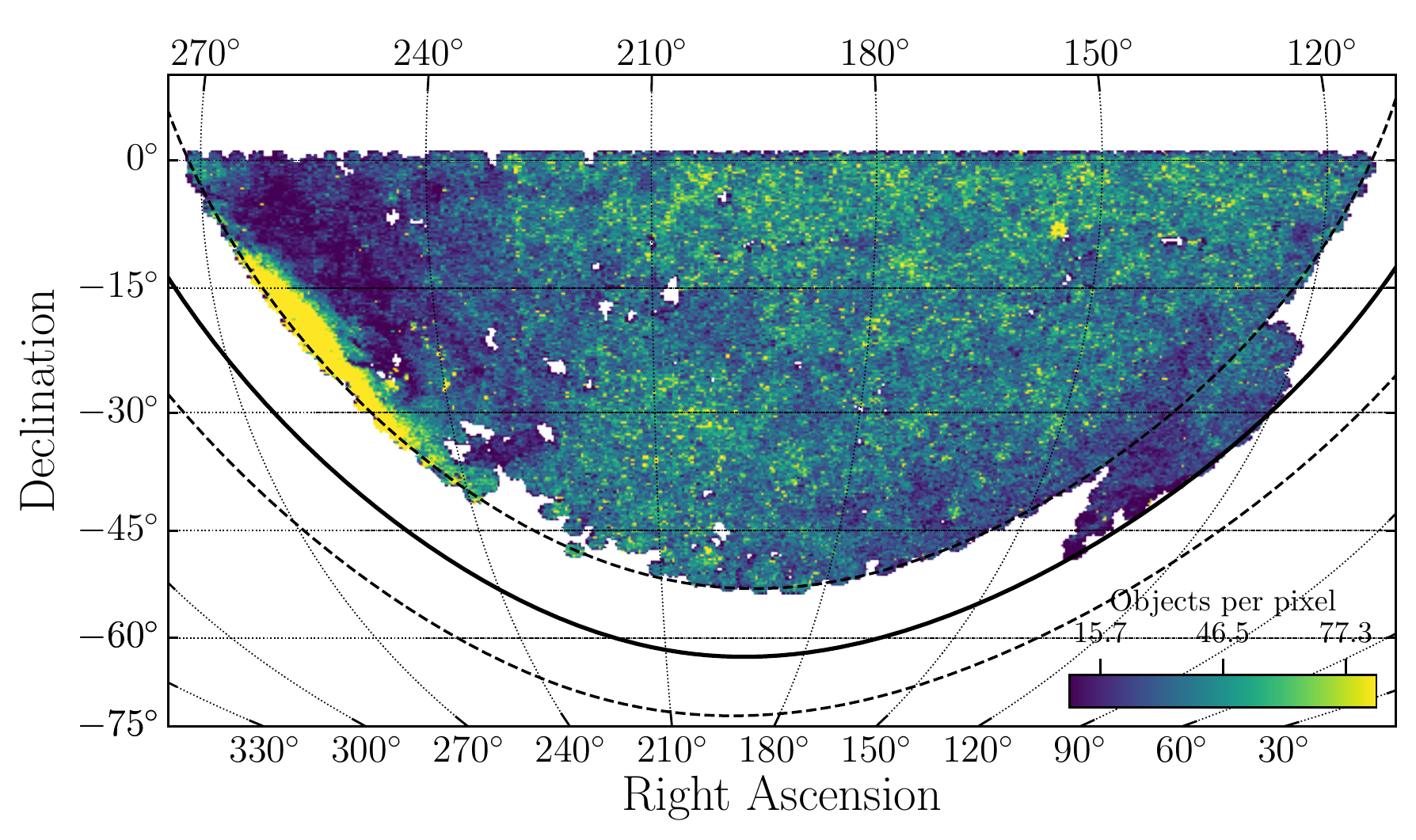}
    \caption{\label{fig:stargalmap} \textit{Left}: Stellar density map created with the $\extclass[g] = 0$ selection described in \secref{classification}. \textit{Right}: Analogous galaxy counts map created with the $\extclass[G] = 3$ selection. The region of lower galaxy density toward the northeast of the footprint can be attributed to higher interstellar extinction, which is not corrected for in this map. Color range units are number of objects per \healpix $\nside = 512$ pixel. Both maps use a magnitude threshold of $\magauto[g] < 22$.}
\end{figure*}

\begin{figure}[t!]
    \centering
    \includegraphics[width=\columnwidth]{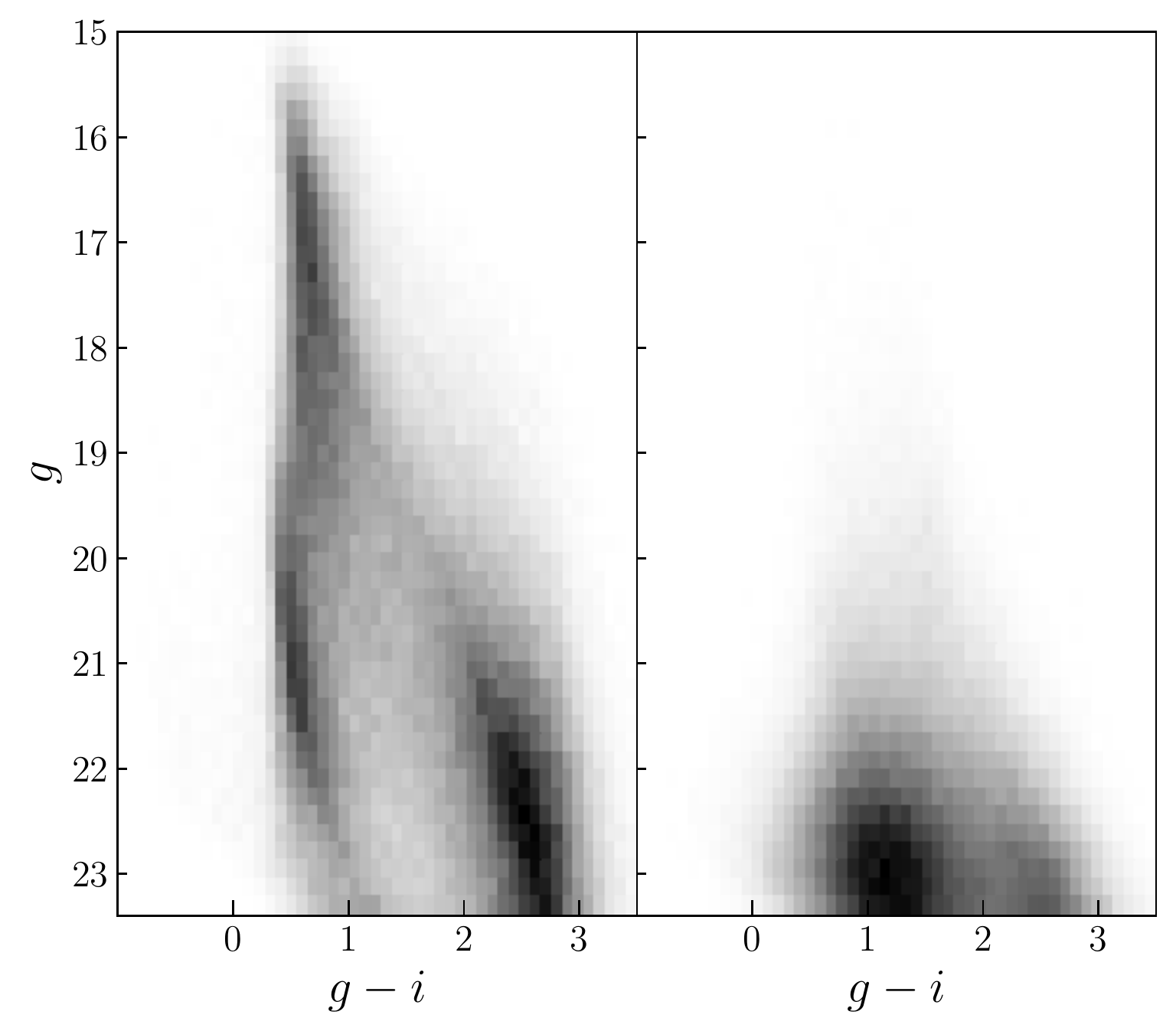}
    \caption{\label{fig:cmd} Color--magnitude diagrams for objects in a $\roughly 30\deg^2$ region centered on $\ra,\dec = (140,-20)\deg$. 
\textit{Left}: PSF magnitudes for objects classified as stars ($0 \leq \extclass[g] \leq 1$). 
\textit{Right}: Automatic aperture magnitude for objects classified as galaxies ($\extclass[g] \geq 2$).}
\end{figure}

Spatial maps of high-confidence stars and galaxies are shown in \figref{stargalmap}.
The stellar density map clearly shows increasing stellar density toward the Galactic plane, and the galaxy density map shows the large-scale clustering of galaxies.
Stellar contamination of the galaxy sample on the eastern edge of the footprint correlates with the Galactic bulge.
The underdense regions in the northeast and southwest of the galaxy map correlate with interstellar extinction, which has not been corrected for when creating this map.
Color--magnitude diagrams for looser selections of stars and galaxies are shown in \figref{cmd}.
 
\subsection{Known Issues}
\label{sec:issues}

The DELVE DR1 detrending and image reduction pipeline used flat-field images, bias images, bad pixel masks, and other image calibration products (i.e., ``supercals'') assembled during the processing of DES Y4.
These are believed to be the best available instrument calibrations for exposures taken prior to 2017 September (DECam exposure numbers $< 674970$).
However, newer calibrations are available from DES Y5 and Y6.
Applying these calibrations to exposures taken after 2017 September leads to a $\roughly 10 \mmag$ shift in the photometric zeropoints. 
While the DELVE DR1 calibration to ATLAS Refcat2 removes these shifts, we note that more recent processing of the DELVE data uses the new DES calibration products and thus the derived zeropoints will change slightly in future releases. 
Deriving updated calibration products extending beyond the end of DES in 2019 February is a topic of ongoing work.

As noted in \secref{photrel}, there is a residual photometric offset of $11.2\mmag$ that is apparent when comparing DELVE DR1 to \Gaia DR2 across the boundary of $\dec = -30\deg$.
This likely arises because ATLAS Refcat2 switches from PS1 to SkyMapper at this declination. 
Internal relative photometric calibration of DELVE is expected to reduce inhomogeneities in the photometric calibration in the future.

Regions around bright stars suffer from scattered and reflected light artifacts \citep[e.g.,][]{DES-DR1:2018}. 
Unlike DES, no effort was made to identify and remove affected CCDs prior to DELVE processing.
This can lead to regions of higher spurious object rates, biased object colors, and incorrect object sizes.
These regions can often be identified at catalog level by examining object colors \citep[i.e., Section 7.4 of][]{Drlica-Wagner:2018}, and most affected exposures were removed during DELVE DR1 validation.

Several problematic regions can be seen as spurious overdensities in the galaxy map shown in \figref{stargalmap}.
A prominent overdensity at $(\ra,\dec) \sim (155.5,-8)\deg$ has been associated with a PSF fitting failure that leads to incorrect star/galaxy separation.
Another spurious overdensity at $(\ra,\dec) \sim (190.5,-45)\deg$ (\healpix 10491) corresponds to a partial failure in the object matching procedure.
Each of these issues has been corrected in subsequent processing of the DELVE data.

The DESDM pipeline used to produce DELVE DR1 is not optimized for crowded-field photometry. This leads to severe blending and incompleteness in regions of high stellar density (see Section 4.6 of \citealt{DES-DR1:2018}). Blending is likely responsible for some of the increased stellar contamination in the galaxy sample close to the Galactic bulge seen in \figref{stargalmap}.

In general, the catalog creation process merged object detections located $< 1\arcsec$ from each other on different exposures (\secref{processing}).
The process for splitting two physical objects located within $1\arcsec$ required that these objects were detected on more than one exposure. 
This splitting process was found to be implemented suboptimally in DELVE DR1.
Furthermore, two objects were not split if they were only resolved in one exposure due to lack of observations or reduced image quality in other observations.
Finally, higher-multiplicity systems (i.e., three or more objects within 1\arcsec) were merged into a single object.
Therefore, the DELVE DR1 catalog should be used with caution when analyzing multiple sources separated by $<1\arcsec$.
The process of associating individual detections into a unique object catalog will be improved in future data releases.


\subsection{Data Access}
\label{sec:access}

Access to DELVE DR1 is provided through the Astro Data Lab \citep{Fitzpatrick:2016,Nikutta:2020},\footnote{\url{https://datalab.noirlab.edu}} part of the Community Science and Data Center (CSDC) hosted by NOIRLab.
The Astro Data Lab is a science platform that provides online access and analysis services for large survey data sets.
It provides database access to catalog data from both a Table Access Protocol (TAP)\footnote{\url{http://ivoa.net/documents/TAP}} service and from direct PostgreSQL queries via web-based, command-line, and programmatic query interfaces.
The Astro Data Lab also provides the ability to cross-match catalogs against other tables held by Astro Data Lab; a cross-match has already been performed between DELVE DR1 and \Gaia EDR3 \citep[\appref{tables};][]{Gaia:2020a}.
The Astro Data Lab also provides an image cutout service, built on the Simple Image Access (SIA) protocol, that points at the petabyte-scale holdings of NOIRLab's Astro Data Archive, including the DELVE DR1 images. 
Furthermore, the Astro Data Lab provides a JupyterHub-based notebook analysis environment capable of accessing all of its data sets and services, as well as remote storage space for user databases and files. 
We used templates provided by the Astro Data Lab to build the DELVE DR1 web pages\footnote{\url{https://datalab.noirlab.edu/delve}} and Jupyter notebooks demonstrating DELVE DR1 data access.\footnote{\url{https://github.com/noaodatalab/notebooks-latest/blob/master/05_Contrib/Galactic/DELVE_DR1/DELVE_access_dr1.ipynb}}  
More detailed information on using the data services for DELVE DR1 science can be found on the Astro Data Lab website.


\section{Science Examples}
\label{sec:examples}

Below we present a few examples of scientific investigations that are possible with the DELVE DR1 data. This list is not intended to be comprehensive, but instead provides a glimpse at the wide range of topics that can be studied with DELVE DR1.

\subsection{Faint Milky Way Satellites}

\begin{figure*}[t]
    \centering
    \includegraphics[width=0.98\textwidth]{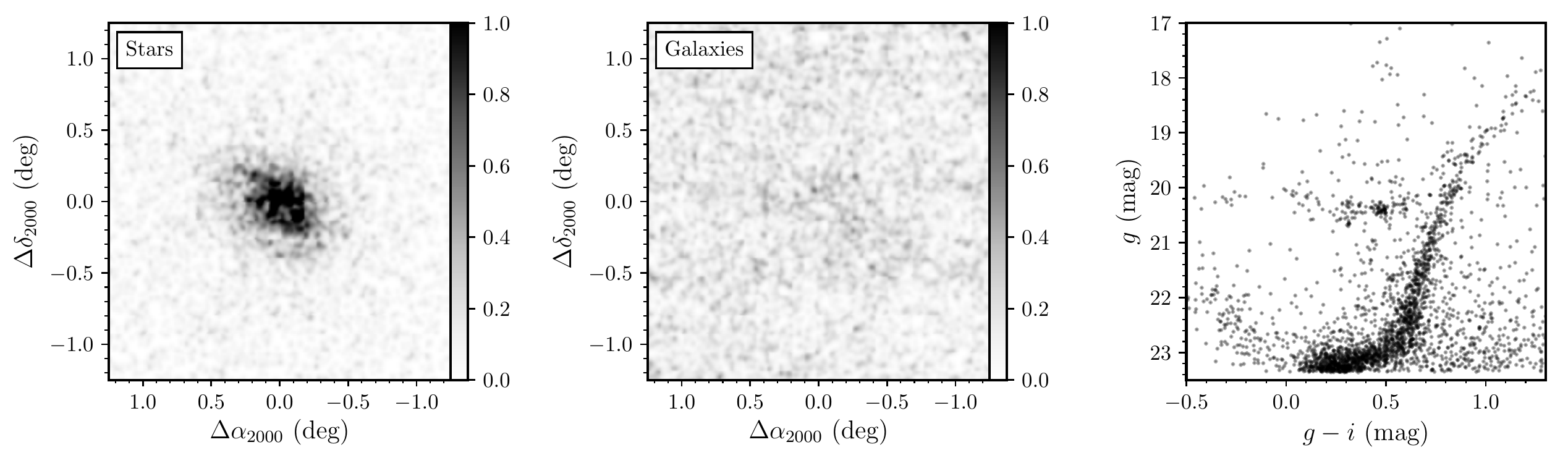}
    \caption{The Sextans dwarf spheroidal galaxy observed with DELVE DR1. \textit{Left:} Normalized spatial density of stellar sources ($0 \leq \extclass[g] \leq 1$ using the prescription from \secref{classification}) smoothed with a $1\arcmin$ Gaussian kernel. \textit{Middle:} Normalized spatial density of galaxies ($\extclass[g] \geq 2$) smoothed by the same kernel. \textit{Right:} Color--magnitude diagram of all stellar sources within $10\arcmin$ of the system centroid. }
    \label{fig:sextans}
\end{figure*}

One of the primary science objectives of DELVE is the discovery of ultra-faint satellites in the halo of the Milky Way.
Early DELVE data have already resulted in the discovery of three systems: Centaurus I, DELVE 1, and DELVE 2 \citep{Mau:2020,Cerny:2020}.
Additional satellites are expected to be discovered as the coverage, quality, homogeneity, and depth of the DELVE data continue to improve.

DELVE DR1 includes several known stellar systems spanning a range of luminosities.
The DELVE DR1 footprint contains the Milky Way dwarf spheroidal satellite galaxies Centaurus I \citep{Mau:2020}, Hydra II \citep{Martin:2015}, Leo IV \citep{2007ApJ...654..897B}, and Sextans \citep{1990MNRAS.244P..16I}. 
Furthermore, DELVE DR1 contains several faint outer halo star clusters including BLISS 1 \citep{Mau:2019}, DELVE 1 \citep{Mau:2020}, Kim 3 \citep{2016ApJ...820..119K}, and Laevens 1/Crater 1 \citep{Laevens:2014a,2014MNRAS.441.2124B}.
The DELVE DR1 data can be used to study the extended stellar populations of known dwarf galaxies and star clusters. 
Such studies can detect signatures of tidal disturbance, which can help inform the evolutionary history of these systems \citep[e.g.,][]{2019ApJ...885...53M}.

In \figref{sextans}, we show an example of the DELVE DR1 data surrounding the Sextans dwarf spheroidal galaxy, which is located at a distance of $84.7 \pm 0.4$ \kpc \citep{Vivas:2019}. 
In the left (middle) panel, we plot the smoothed spatial distribution of stars (galaxies) after applying a selection based on \extclass[g] as presented in \secref{classification}.
We note similar variations in the central stellar density of Sextans as were reported by \citet{Roderick:2016}.
In the right panel, we present a color--magnitude diagram for the stars located within $10\arcmin$ of Sextans, demonstrating that the DELVE DR1 $g$-band depth provides high-precision photometry down to the subgiant branch of this system.

\subsection{Stellar Streams}
\label{sec:streams}

\begin{figure}[t!]
    \centering
    \includegraphics[width=0.98\columnwidth]{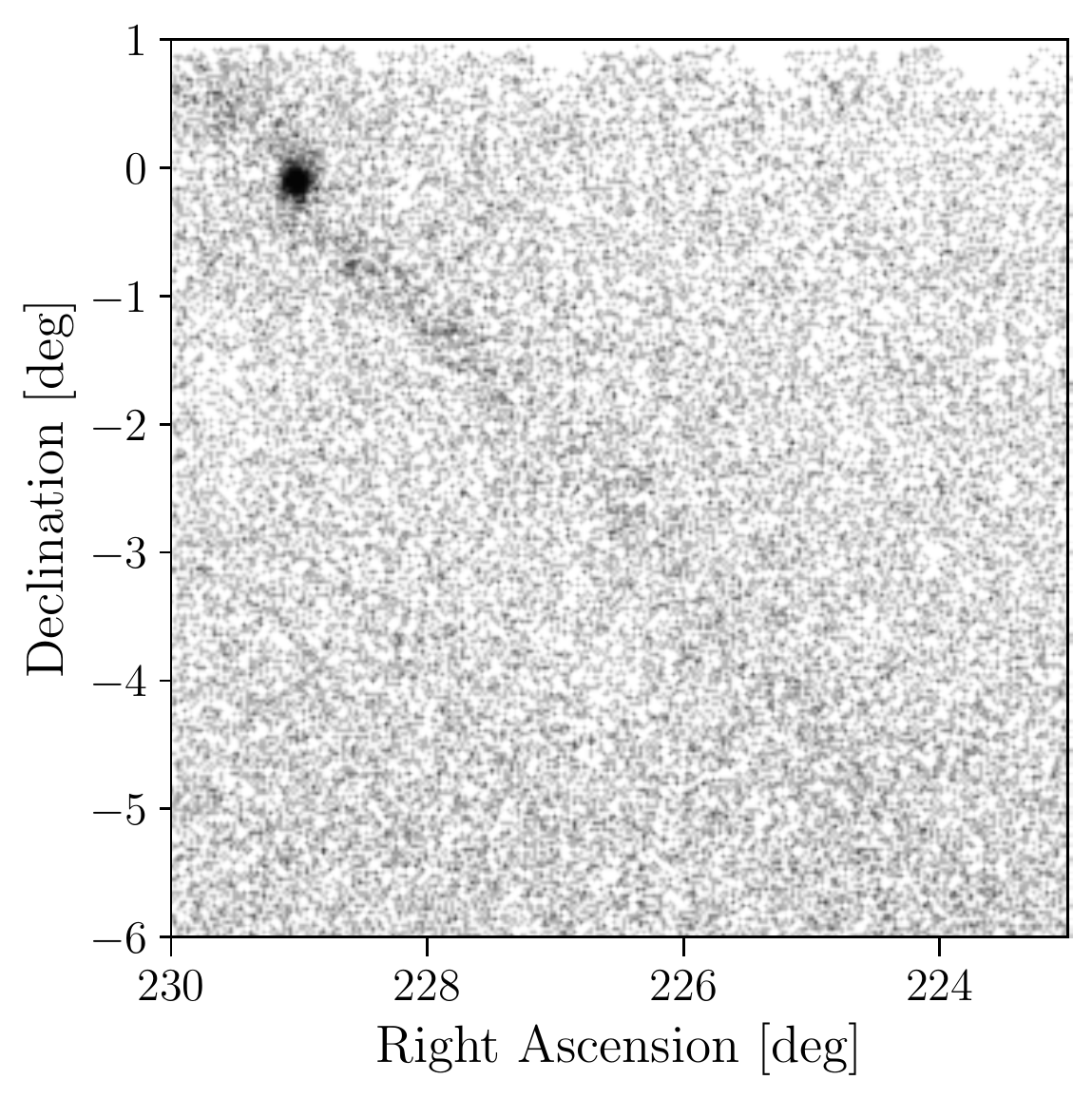}
    \caption{Leading tail of the disrupting globular cluster Palomar 5. Stellar objects were selected from DELVE DR1 with $0 \leq \extclass[g] \leq 1$ and a color--magnitude filter consistent with the stellar population of the Palomar 5 cluster (\secref{streams}).
}
    \label{fig:pal5}
\end{figure}

The contiguous, deep coverage of DELVE DR1 enables the discovery and study of stellar streams at large Galactocentric distances \citep[\eg,][]{Odenkirchen:2001,Newberg:2002,Belokurov:2006,Grillmair:2006,Bonaca:2012,Koposov:2014,Bernard:2014,Bernard:2016,Grillmair:2017,Drlica-Wagner:2015,Balbinot:2016,Shipp:2018,Shipp:2020}.
Following on other recent analyses \citep[\eg,][]{Shipp:2019, Shipp:2020, Bonaca:2020, Li:2020}, DELVE DR1 photometry can be used to characterize known streams in tandem with complementary proper-motion measurements from \Gaia and targeted spectroscopic follow-up.
The phase-space distribution of the population of stellar streams can be leveraged to construct a global map of the Milky Way's gravitational field \citep[e.g.,][]{Bovy:2016,Erkal:2016,Bonaca:2018}, while their internal structure may hold clues to the particle nature of dark matter \citep{Carlberg:2013,Erkal:2016b,Banik:2019}.

In \figref{pal5}, we show an example of the tidal disruption of the Palomar 5 (Pal 5) globular cluster \citep{Odenkirchen:2001}.
We select likely stars ($0 \leq \extclass[G] \leq 1$; \secref{classification}) consistent with the main sequence of Pal 5 in a de-reddened color--magnitude diagram.
Specifically, we select stars consistent with an old ($11.5\Gyr$), metal-poor (${\rm [Fe/H]} = -1.3$) isochrone from \citet{Dotter:2008} shifted to a distance of $22.5\kpc$.
To mitigate effects of nonuniform depth, we only include stars that are brighter than $g = 23.1$ mag.
The leading arm of Pal 5 resides in the DELVE DR1 footprint and is prominently detected extending to $\dec \lesssim -5 \deg$, confirming similar results from \citet{Bonaca:2020}.

\subsection{Strong Lensing}

\begin{figure}[t!]
    \centering
    \includegraphics[width=0.98\columnwidth]{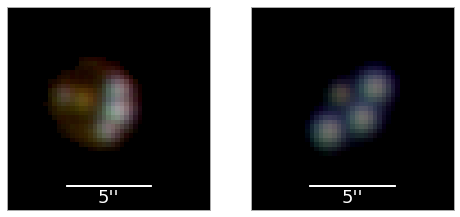}
    \caption{DELVE composite color images ($g, r, i$) of the quadruply lensed quasar systems J1131GL \citep[left;][]{Sluse:2003} and 2M J1134-2103 \citep[right;][]{Lucey:2018}. 
    \label{fig:delve_quads}}
\end{figure}

Strong gravitational lens systems are excellent probes of galaxy structure, dark matter, and dark energy \citep[e.g.,][]{Treu:2010}.
The total catalog of confirmed strong lens systems numbers only in the hundreds \citep[e.g.,][and references therein]{Jacobs:2019a}, 
but recent wide-area surveys have rapidly increased the number of candidate systems through both visual inspection of image data and the use of machine-learning techniques.
Thousands of new lens candidates have been identified in DES \citep[\eg,][]{Nord:2016,Diehl:2017, Jacobs:2019b,Jacobs:2019a}, DECaLS \citep[\eg,][]{Huang:2020, Huang:2021}, and PS1 \citep[\eg,][]{Canameras:2020}.
DELVE is expected to contain several thousand galaxy--galaxy lenses and $\roughly 100$ lensed quasars brighter than $i \sim 18.5 \magn$ based on the estimates of \citet{Collett:2015} and \citet{Treu:2018}, respectively.
In \figref{delve_quads}, we show DELVE images of two quadruply lensed quasars: J1131GL \citep[left;][]{Sluse:2003} and 2M J1134-2103 \citep[right;][]{Lucey:2018}.
DELVE has partnered with the {\it STRong lensing Insights into the Dark Energy Survey} \citep[STRIDES;][]{Treu:2018} to initiate searches for previously undiscovered galaxy--galaxy lenses and lensed quasars using the DELVE data.

\subsection{Galaxies in Different Environments}

While DELVE primarily focuses on faint galaxies in the Local Volume, the DELVE DR1 data can also be used to study more luminous galaxies at larger distances.
The wide coverage and depth of DELVE enable the investigation of galaxy populations spanning a range of different environments (from isolated galaxies to galaxy clusters).
The study of galaxies in different environments can help reveal the mechanisms responsible for variations in the structure and physical properties of galaxies as they form and evolve over cosmic time \citep[e.g.,][]{Darvish:2018,Chartab:2020}.
As an example, early community data processed by DELVE have been used to perform environmental studies of local analogs to high-redshift galaxies \citep{Santana-Silva:2020}.

Tidal streams around massive galaxies are a ubiquitous aspect of galaxy formation that has not yet been fully explored due to the deep imaging required to detect these systems beyond the Local Group \citep{Bullock:2005, Cooper:2010, Martinez-Delgado:2010, Morales:2018}.  
A systematic survey for galactic remnants in a large sample of nearby galaxies is needed to  understand whether the recent merger histories of the Milky Way and Andromeda are ``typical.'' 
DELVE has partnered with the {\it Stellar Stream Legacy Survey} \citep{Martinez-Delgado:2021}, a systematic imaging survey of tidal features around nearby galaxies that has reached surface brightnesses as faint as $\roughly 29 \magasec^{-2}$ using public data from the DESI Legacy Imaging Surveys \citep{Dey:2019}. 
The DELVE data will enable a search for extragalactic stellar streams around several thousand galaxies in the southern hemisphere.  
A catalog of extragalactic stellar streams will probe the current mass assembly rate of nearby galaxies, the stellar content and orbits of satellites, and the formation of stellar halos.


\subsection{Galaxy Clusters}

\begin{figure*}[t!]
  \centering
  \includegraphics[width=0.35\textwidth, trim=0cm -1.5cm 0cm 0cm]{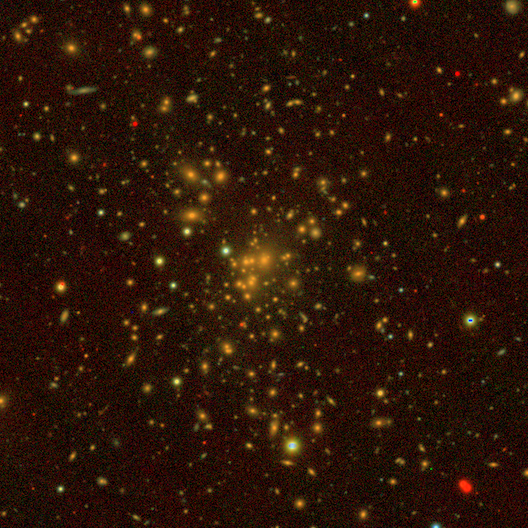}
  \includegraphics[width=0.55\textwidth, trim=0cm 0.5cm 0cm 0cm, clip]{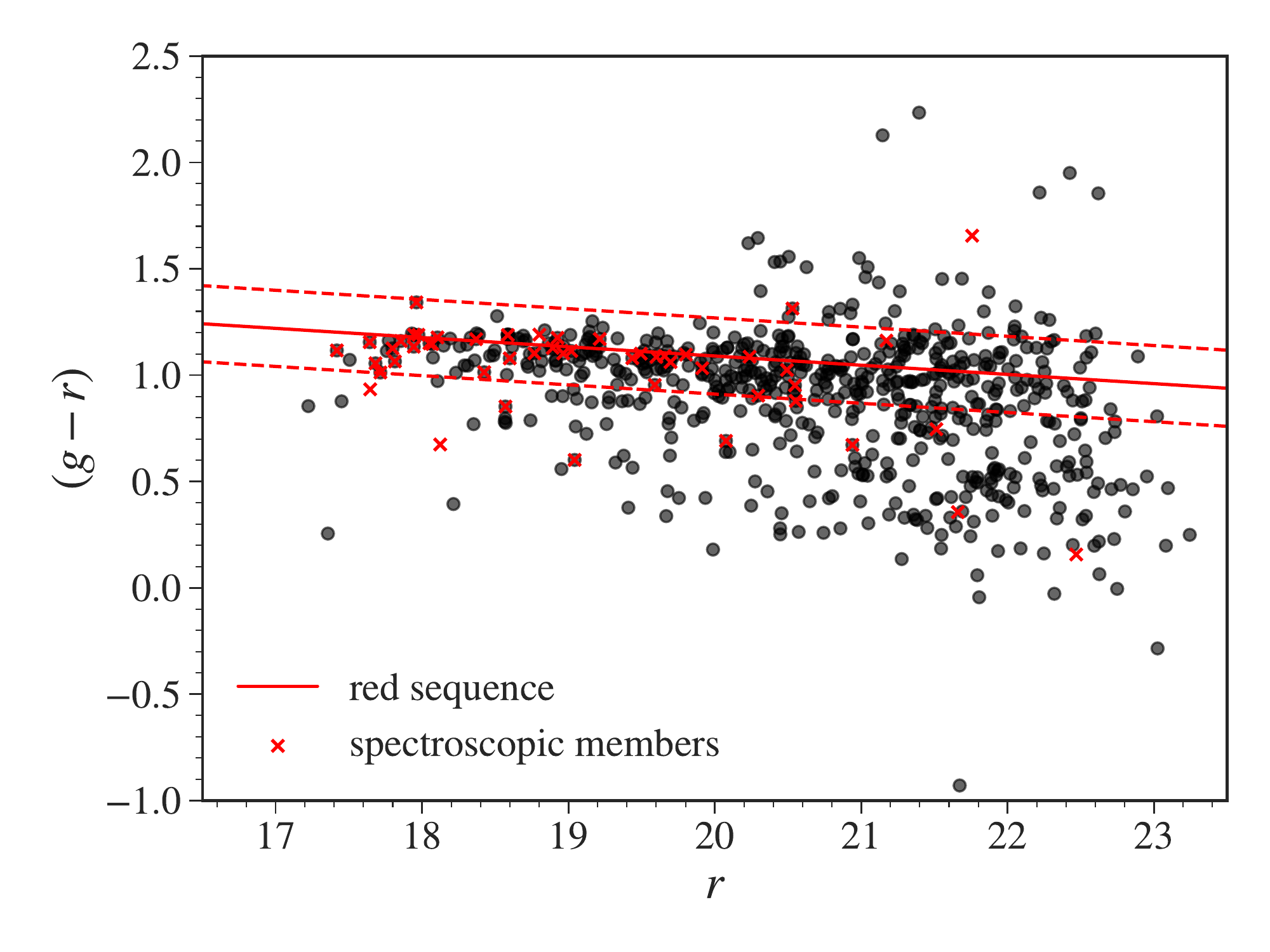}
  \caption{DELVE DR1 data around the rich galaxy cluster Abell 1689 at a redshift of $z=0.18$. (Left) False-color image assembled from $g,r,z$ images covering a $5\farcm8 \times 5\farcm8$ region around Abell 1689. (Right) Color--magnitude diagram of galaxies in Abell 1689 after a statistical background subtraction based on the field galaxy population. The red solid line is a linear red sequence relationship derived from the DES Y1 redMaPPer cluster catalog \citep{McClintock2019}, and the red dashed lines represent the expected 95\% containment region for red sequence galaxies. The red markers are spectroscopically confirmed members of the system.
}
\label{fig:a1689}
\end{figure*}

Galaxy clusters are a powerful tool for studying cosmology and galaxy evolution in dense environments \citep[][]{Allen2011,Zenteno16,Hennig2017,Bocquet2018,DESY1ClusterCosmology}. 
Assembling a large population of galaxy clusters is essential for improving measurements of cosmological parameters \citep[][]{kidsviking,DES:2018,McClintock2019}.
DELVE collaborates closely with DeROSITAS (PI: Zenteno) to assemble contiguous imaging in $g,r,i,z$ and enable wide-area optical studies of galaxy clusters. 
The optical data can be combined with data from large surveys in the submillimeter \citep[\eg,][]{Bleem:2020,Hilton:2020} and X-ray \citep[\eg, eROSITA;][]{Predehl:2020} regimes. 
Multiwavelength studies of galaxy clusters can help mitigate systemics in optical studies \citep{Grandis2021} and provide information about the dynamical state of galaxy clusters, which is important for cosmological measurements \citep[\eg,][]{Andrade-Santos:2012,Andrade-Santos:2017,Lovisari:2020a,Lovisari:2020b} and studies of galaxy evolution \citep[\eg,][]{Zenteno20}. 

As a proof of concept, \figref{a1689} shows DELVE DR1 data associated with the rich galaxy cluster Abell 1689 \citep[\eg,][]{Limousin2007}. 
We use a kernel-density estimate to subtract likely field galaxies from the color--magnitude diagram \citep{Pimbblet2002}, and we recover a prominent red sequence without the need for spectroscopic membership information or photometric redshifts. 
The colors and magnitudes of DELVE DR1 galaxies (black circles) follow the linear red sequence relation derived from the DES Y1 redMaPPer catalog \citep[red line;][]{McClintock2019} demonstrating consistency between these data sets in the galaxy cluster regime. 
In the future, we plan to assemble a catalog of galaxy clusters by using a cluster finder algorithm such as redMaPPer \citep{2014ApJ...785..104R} or WAZP \citep{Aguena:2021}. 
These algorithms have been validated on DES data and can be naturally adapted to DELVE data.

\section{Summary}
\label{sec:conclusion}

DELVE seeks to study the fundamental physics of dark matter and galaxy formation through rigorous systematic studies of dwarf galaxies and stellar substructures in the Local Volume.
To do so, DELVE aims to complete contiguous coverage of the southern, high Galactic latitude sky in $g,r,i,z$ to a depth comparable to two years of DES observations.
As of 2021 January, DELVE has completed slightly less than half of the 126 nights of scheduled DECam observing. 
The first major public release of DELVE DR1 combines observations from the DELVE-WIDE program with archival DECam data to cover $\roughly 4000 \deg^2$ of the northern Galactic cap to a $5\sigma$ depth of $g=\maglimpsfg, r=\maglimpsfr, i=\maglimpsfi, z=\maglimpsfz \magn$ (\tabref{summary}).
The DELVE DR1 catalog contains PSF and automatic aperture measurements for $\roughly 520$ million astronomical objects produced by the DESDM pipeline and accessible through the NOIRLab Astro Data Lab.

DELVE DR1 has utility for a broad range of scientific investigations in the Local Volume and beyond.
Future DELVE data releases will increase the coverage, uniformity, and depth of the DELVE catalogs.
Furthermore, these releases will include products from the DELVE-MC and DELVE-DEEP programs, which are processed with the multiepoch point-source pipeline from SMASH.
We anticipate that DELVE DR1 and future DELVE data releases will be a valuable resource for the community in advance of the Vera C.\ Rubin Observatory Legacy Survey of Space and Time.


\input{ack.tex}

\facilities{Blanco (DECam), Astro Data Lab, \Gaia} 

\software{
\code{astropy} \citep{astropy:2018},
\code{fitsio},\footnote{\url{https://github.com/esheldon/fitsio}}
\healpix \citep{Gorski:2005},\footnote{\url{http://healpix.sourceforge.net}}
\code{healpy} \citep{Zonca:2019},\footnote{\url{https://github.com/healpy/healpy}}
\code{healsparse},\footnote{\url{https://healsparse.readthedocs.io/en/latest/}}
\code{matplotlib} \citep{Hunter:2007},
\code{numpy} \citep{NumPy:2020},
\PSFEx \citep{Bertin:2011},
\code{scipy} \citep{Scipy:2020},
\scamp \citep{Bertin:2006}, 
\code{skymap},\footnote{\url{https://github.com/kadrlica/skymap}}
\SExtractor \citep{Bertin:1996},
\swarp \citep{Bertin:2002, Bertin:2010}
}


\appendix


\section{DECam Data}
\label{app:propid}

DELVE DR1 combines DECam observations taken by DELVE with archival DECam data from 157 other programs. These programs and their contributions to the DELVE DR1 data set are listed in \tabref{propid}.

\input{table_propid.tex}


\section{Transformation Equations}
\label{app:transforms}

Initial transformation equations were derived between ATLAS Refcat2 \citep{Tonry:2018} and the DES Standard Bandpasses \citep{DES-DR1:2018} using a set of matched stars from ATLAS Refcat2 and DES DR1 with $16.0 < i < 20.0$ mag located within a $\roughly 20\deg^2$ region centered at $(\ra,\dec) = (17, -30) \deg$.
We performed a straight-line fit to the difference in the DES DR1 \var{WAVG\_PSF} magnitudes and the ATLAS Refcat2 magnitudes vs.\ color over a color range typical of the main-sequence stellar locus. 
We further refined the zeropoints of these initial transformation equations using a set of exposures processed by both DELVE and DES. 
We compare zeropoints derived by the DES global calibration \citep{Burke:2018} to zeropoints derived from the DELVE processing by matching to ATLAS Refcat2 using the initial transformation equations. 
We adjusted the zeropoints of the ATLAS Refcat2 transformation equations so that there was no mean offset between the DES zeropoints and the DELVE zeropoints for these exposures (these zeropoint adjustments were $\lesssim 0.01$ mag).

\begin{figure*}[t]
    \centering
    \includegraphics[width=0.98\textwidth]{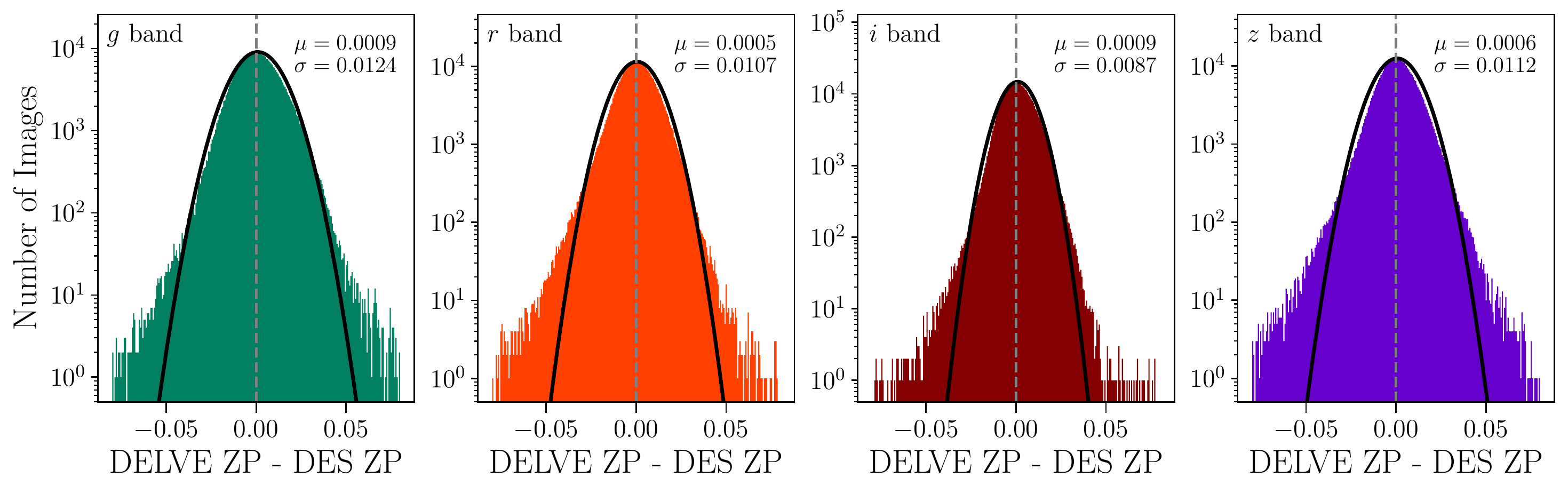}
    \caption{Comparison of CCD image zeropoints for DES exposures processed by DES and DELVE. The DELVE zeropoints are derived from ATLAS Refcat2, while the DES zeropoints derived from the DES global calibration \citep{Burke:2018}. 
    The spread between these two techniques has a standard deviation of  $< 0.01 \magn$ (black curve), with a small fraction of outliers with larger deviations.
    }
    \label{fig:zps}
\end{figure*}

The resulting transformation equations in $g, r, i, z$ between ATLAS Refcat2 (in the PS1 system) and DECam are
\begin{align*}
g_{\rm DECam} &= g_{\rm PS1} + 0.0994(g_{\rm PS1}-r_{\rm PS1}) - 0.0319 \\
r_{\rm DECam} &= r_{\rm PS1} -0.1335(g_{\rm PS1}-r_{\rm PS1}) + 0.0215 \\
i_{\rm DECam} &= i_{\rm PS1} -0.3407(i_{\rm PS1}-z_{\rm PS1}) - 0.0013 \\
z_{\rm DECam} &= r_{\rm PS1} -0.2575(r_{\rm PS1}-z_{\rm PS1}) - 0.0201.
\end{align*}\normalsize
\noindent We used these transformation equations to derive DELVE zeropoints for a larger set of DES exposures and to compare  with the DES zeropoints in \figref{zps}.
The mean offsets between the DELVE and DES zeropoints are $<1\mmag$ in all bands.
The DELVE zeropoints have an rms scatter of $\roughly 0.01$ mag per CCD when compared to the DES zeropoints for these exposures.
The transformations in $g,r$ are valid for stars with $-0.2 < (g_{\rm PS1} - r_{\rm PS1}) < 1.2 \magn$, and the transformations in $i,z$ are valid for stars with $-0.2 < (i_{\rm PS1} - z_{\rm PS1}) < 0.3 \magn$. 
Transformations between the DES/DECam system and other systems can be found in Appendix A of \citet{DES-DR2:2021}.

\section{DELVE DR1 Tables}
\label{app:tables}
 
The DELVE DR1 catalog data are accessible through the \code{DELVE\_DR1.OBJECTS} table hosted by the Astro Data Lab.
This table includes the photometric properties assembled from a catalog-level co-add of the individual single-epoch measurements.
The table columns are described in \tabref{dr1_main}. A cross-match between DELVE DR1 and \Gaia EDR3 is also provided in \code{DELVE\_DR1.{\allowbreak}X1P5\_\_OBJECTS\_\_{\allowbreak}GAIA\_EDR3\_\_GAIA\_SOURCE}. The columns of this table are described in \tabref{dr1_gaia}.

\input{table_dr1_main.tex}
\input{table_dr1_gaia.tex}


\section{Depth}
\label{app:depth}

\begin{figure*}[t]
\centering
\includegraphics[width=0.8\textwidth]{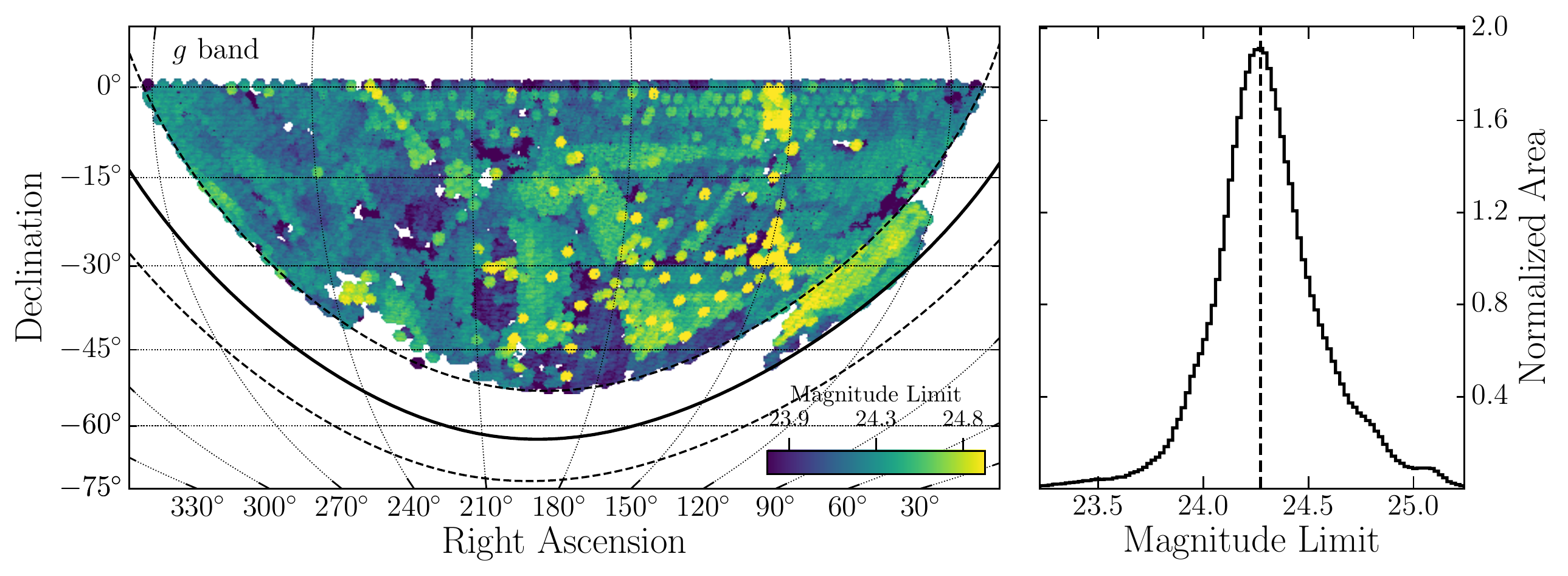}\\
\includegraphics[width=0.8\textwidth]{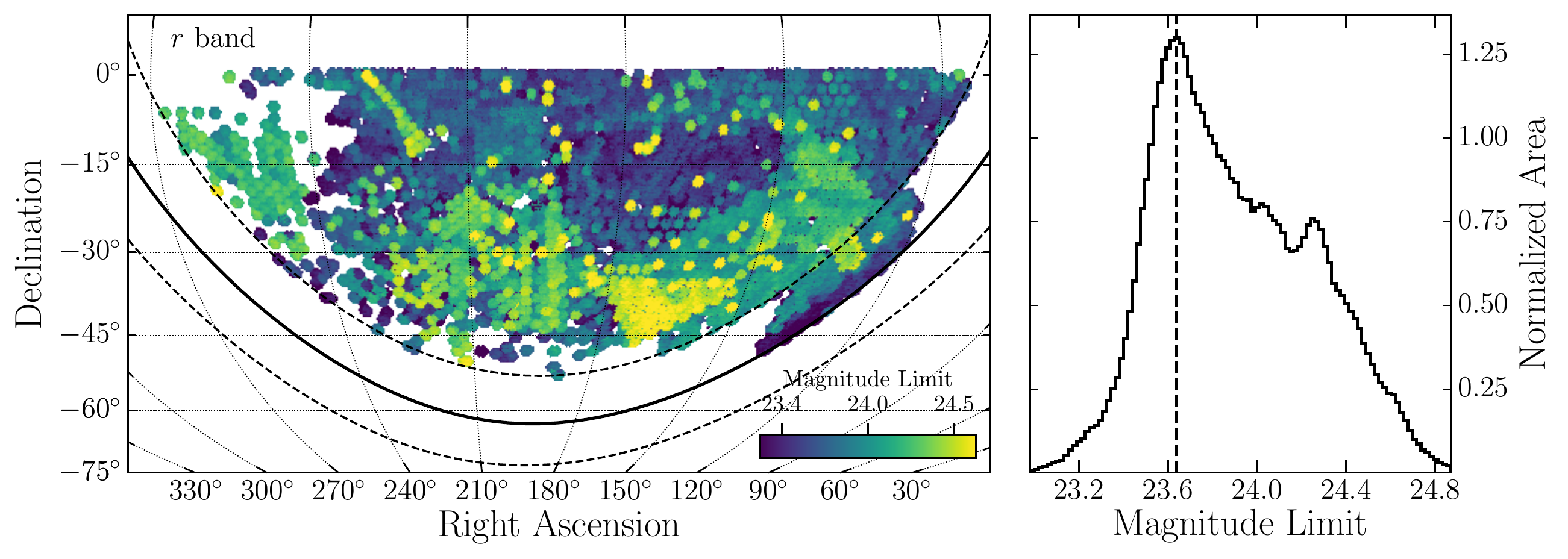}\\
\includegraphics[width=0.8\textwidth]{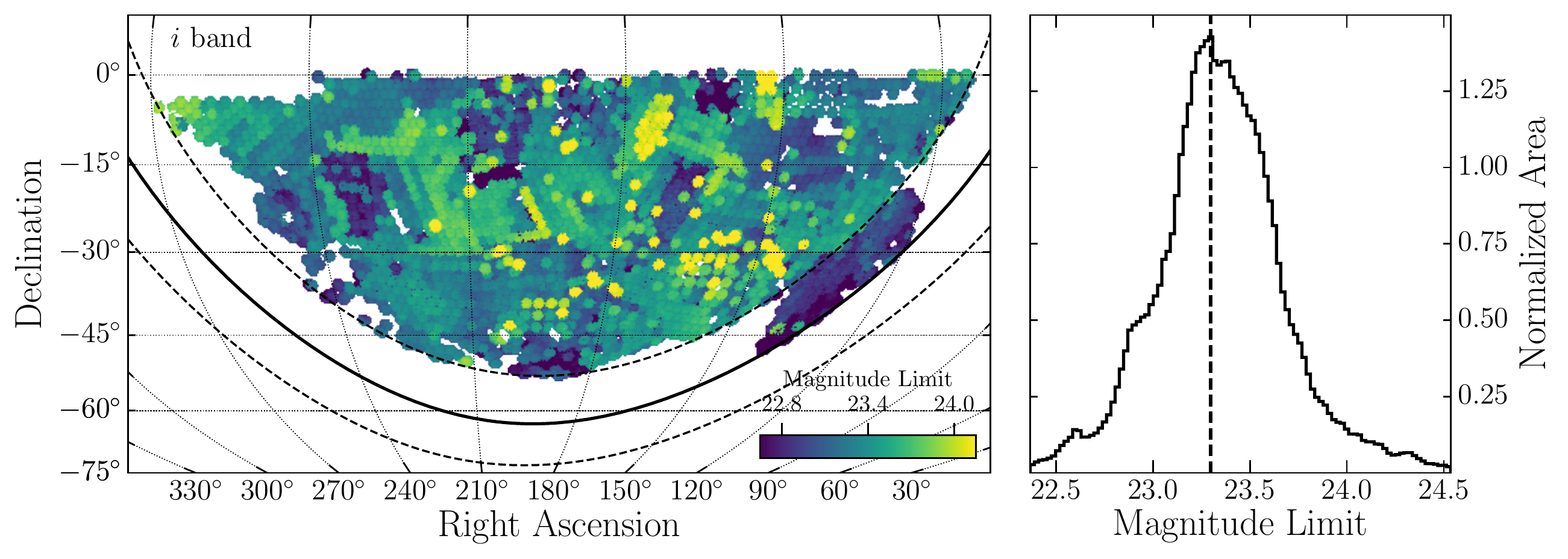}\\
\includegraphics[width=0.8\textwidth]{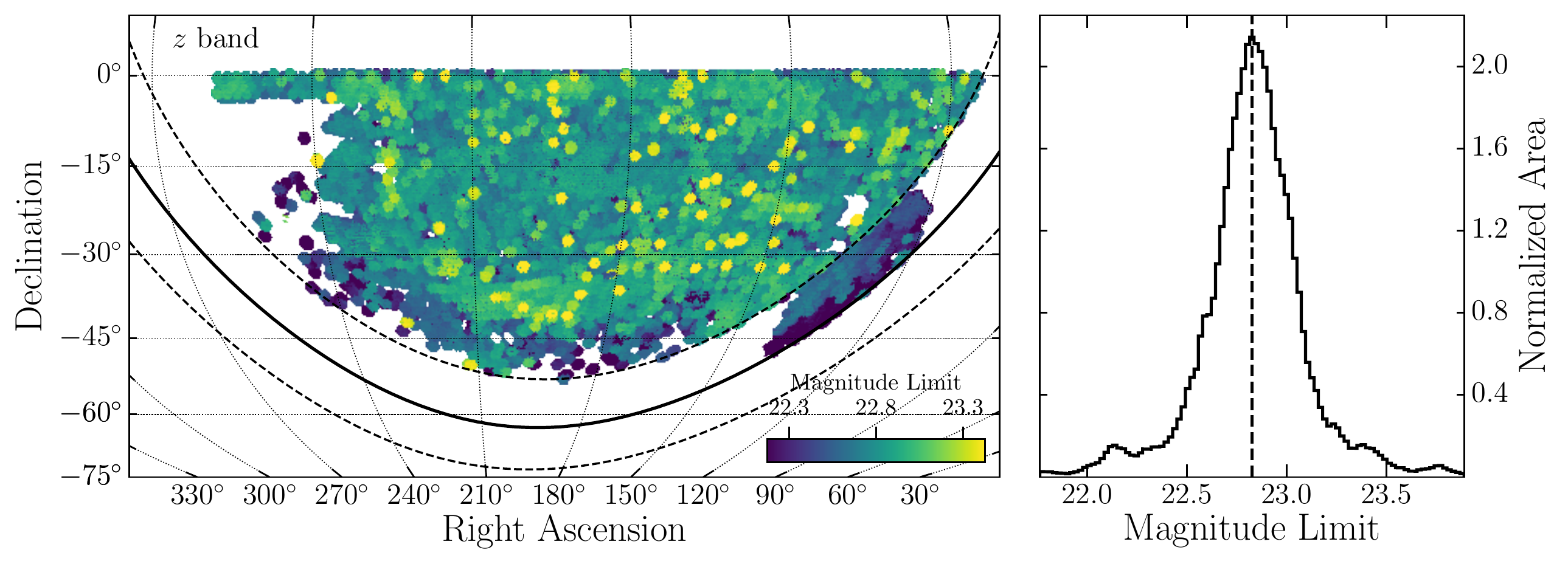}
\caption{Sky maps and histograms of the S/N=5 magnitude limit computed from the statistical uncertainty in \var{MAG\_PSF}. Sky maps are plotted using an equal-area McBryde--Thomas flat polar quartic projection in celestial equatorial coordinates.
}
\label{fig:maglim_map}
\end{figure*}

This appendix includes sky maps showing variations in the S/N=5 depth of DELVE DR1 in the $g,r,i,z$ bands. The S/N=5 depth was derived from the magnitude at which the median magnitude uncertainty was $\delta m = 0.2171$ mag (\secref{depth}). These values were derived in  $\roughly 12 \amin^2$ \healpix pixels ($\nside = 1024$) and are shown in \figref{maglim_map}. 
A region of deeper $r$-band imaging can be seen around ($\ra,\dec) \sim (170,-40)\deg$. 
This corresponds to archival data from the SLAMS survey \citep[][]{Jethwa:2018b} that were taken in excellent conditions. 

\clearpage

\bibliographystyle{yahapj_twoauthor_arxiv_amp}
\bibliography{main}

\end{document}

%% file: commands.tex
\usepackage{amsmath}
\usepackage{amssymb}
\usepackage{xspace}
\usepackage{xifthen}
\usepackage{eso-pic}

\newcommand{\Gaia}{{\it Gaia}\xspace}

\definecolor{forestgreen}{HTML}{228B22}
\definecolor{urlblue}{HTML}{000000}

\newcommand{\eg}{e.g.\xspace}

\mathchardef\mhyphen="2D

\newcommand{\roughly}{\ensuremath{ {\sim}\,} }

\newlength{\dhatheight}

\newcommand{\code}[1]{\texttt{#1}\xspace}

\newcommand{\unit}[1]{\ensuremath{\mathrm{\,#1}}\xspace}

\newcommand{\Gyr}{\unit{Gyr}}

\newcommand{\mas}{\unit{mas}}
\newcommand{\amin}{\unit{arcmin}}
\newcommand{\asec}{\unit{arcsec}}

\newcommand{\kpc}{\unit{kpc}}
\newcommand{\Mpc}{\unit{Mpc}}
\newcommand{\second}{\unit{s}}

\newcommand{\Msun}{\unit{M_\odot}}

\newcommand{\magn}{\unit{mag}}
\newcommand{\mmag}{\unit{mmag}}
\newcommand{\magasec}{\magn \asec}
\newcommand{\e}{\unit{e^{-}}}

\newcommand{\secref}[1]{Section~\ref{sec:#1}}
\newcommand{\appref}[1]{Appendix~\ref{app:#1}}
\newcommand{\tabref}[1]{Table~\ref{tab:#1}}

\newcommand{\figref}[1]{Figure~\ref{fig:#1}}

\newcommand{\var}[1]{%
 \ensuremath{\texttt{\MakeUppercase{#1}}}\xspace
}
\newcommand{\bandvar}[2][]{%
  \ifthenelse{\isempty{#1}}{\var{#2}}{\var{#2\_#1}}%
}
\newcommand{\spreadmodel}[1][]{\bandvar[#1]{spread\_model}}
\newcommand{\spreaderrmodel}[1][]{\bandvar[#1]{spreaderr\_model}}

\newcommand{\extclass}[1][]{\bandvar[#1]{extended\_class}}
\newcommand{\magauto}[1][]{\bandvar[#1]{mag\_auto}}
\newcommand{\magpsf}[1][]{\bandvar[#1]{mag\_psf}}

\newcommand{\LCDM}{\ensuremath{\rm \Lambda CDM}\xspace}

\newcommand{\ra}{{\ensuremath{\rm RA}}\xspace}
\newcommand{\dec}{{\ensuremath{\rm Dec}}\xspace}

\newcommand{\SWARP}{\code{SWarp}}
\newcommand{\swarp}{\SWARP}
\newcommand{\SExtractor}{\code{SourceExtractor}}

\newcommand{\PSFEx}{\code{PSFEx}}

\newcommand{\SCAMP}{\code{SCAMP}}
\newcommand{\scamp}{\SCAMP}

\newcommand{\HEALPix}{\code{HEALPix}}
\newcommand{\healpix}{\HEALPix}

\newcommand{\nside}{\code{nside}}
\newcommand{\teff}{\ensuremath{t_{\rm eff}}\xspace}
\newcommand{\texp}{\ensuremath{T_{\rm exp}}\xspace}

\providecommand\physrep{\ref@jnl{Phys.~Rep.}}%
\providecommand\apjs{\ref@jnl{ApJS}}%
\providecommand{\jcap}{\ref@jnl{JCAP}}%

%% file: numbers.tex
\newcommand{\widearea}{15000\xspace}
\newcommand{\mcarea}{2200\xspace}
\newcommand{\deeparea}{135\xspace}

\newcommand{\widedepth}{23.5\xspace}
\newcommand{\mcdepth}{24.5\xspace}
\newcommand{\deepdepth}{25.5\xspace}

\newcommand{\deepdepthg}{26.0\xspace}
\newcommand{\deepdepthi}{25.0\xspace}

\newcommand{\nobjects}{519,737,142 \xspace}

\newcommand{\areanimagesg}{5599}
\newcommand{\areanimagesr}{5106}
\newcommand{\areanimagesi}{5065}
\newcommand{\areanimagesz}{5153}
\newcommand{\areanimagesgriz}{4075}

\newcommand{\nexp}{29929\xspace}
\newcommand{\nexpg}{8336}
\newcommand{\nexpr}{7713}
\newcommand{\nexpi}{6840}
\newcommand{\nexpz}{7040}

\newcommand{\widenexp}{9778\xspace}
\newcommand{\mcnexp}{1467\xspace}
\newcommand{\deepnexp}{419\xspace}

\newcommand{\astroabs}{22} 

\newcommand{\astrorepeatg}{24}
\newcommand{\astrorepeatr}{26}
\newcommand{\astrorepeati}{25}
\newcommand{\astrorepeatz}{28}
\newcommand{\astrorepeatall}{26}

\newcommand{\photabs}{\ensuremath{20}} 

\newcommand{\photrepeatg}{4.9}
\newcommand{\photrepeatr}{5.0}
\newcommand{\photrepeati}{4.5}
\newcommand{\photrepeatz}{5.4}

\newcommand{\photgaia}{9.1}

\newcommand{\maglimpsfg}{24.3}
\newcommand{\maglimpsfr}{23.9}
\newcommand{\maglimpsfi}{23.3}
\newcommand{\maglimpsfz}{22.8}
\newcommand{\maglimpsfteng}{23.5}
\newcommand{\maglimpsftenr}{23.0}
\newcommand{\maglimpsfteni}{22.6}
\newcommand{\maglimpsftenz}{22.1}

\newcommand{\maglimautog}{23.9}
\newcommand{\maglimautor}{23.4}
\newcommand{\maglimautoi}{22.9}
\newcommand{\maglimautoz}{22.4}
\newcommand{\maglimautoteng}{22.9}
\newcommand{\maglimautotenr}{22.5}
\newcommand{\maglimautoteni}{22.0}
\newcommand{\maglimautotenz}{21.5}

\newcommand{\medfwhmg}{1.28} 
\newcommand{\medfwhmr}{1.16} 
\newcommand{\medfwhmi}{1.07} 
\newcommand{\medfwhmz}{1.04} 

\newcommand{\starefficiency}{96} 
\newcommand{\starcontamination}{2} 
\newcommand{\galefficiency}{99} 
\newcommand{\galcontamination}{3}

%% file: authors.tex

\author[0000-0001-8251-933X]{A.~Drlica-Wagner}
\affiliation{Fermi National Accelerator Laboratory, P.O.\ Box 500, Batavia, IL 60510, USA}
\affiliation{Kavli Institute for Cosmological Physics, University of Chicago, Chicago, IL 60637, USA}
\affiliation{Department of Astronomy and Astrophysics, University of Chicago, Chicago IL 60637, USA}
\author[0000-0002-3936-9628]{J.~L.~Carlin}
\affiliation{Rubin Observatory/AURA, 950 North Cherry Avenue, Tucson, AZ, 85719, USA}
\author[0000-0002-1793-3689]{D.~L.~Nidever}
\affiliation{Department of Physics, Montana State University, P.O. Box 173840, Bozeman, MT 59717-3840}
\affiliation{NSF's National Optical-Infrared Astronomy Research Laboratory, 950 N. Cherry Ave., Tucson, AZ 85719, USA}
\author[0000-0001-6957-1627]{P.~S.~Ferguson}
\affiliation{George P. and Cynthia Woods Mitchell Institute for Fundamental Physics and Astronomy,\\ Texas A\&M University, College Station, TX 77843, USA}
\affiliation{Department of Physics and Astronomy, Texas A\&M University, College Station, TX 77843, USA}
\author[0000-0003-2511-0946]{N.~Kuropatkin}
\affiliation{Fermi National Accelerator Laboratory, P.O.\ Box 500, Batavia, IL 60510, USA}
\author[0000-0002-6904-359X]{M.~Adam\'ow}
\affiliation{National Center for Supercomputing Applications, University of Illinois, 1205 West Clark St., Urbana, IL 61801, USA}
\affiliation{Center for Astrophysical Surveys, National Center for Supercomputing Applications, Urbana, IL, 61801, USA}
\author[0000-0003-1697-7062]{W.~Cerny}
\affiliation{Kavli Institute for Cosmological Physics, University of Chicago, Chicago, IL 60637, USA}
\affiliation{Department of Astronomy and Astrophysics, University of Chicago, Chicago IL 60637, USA}
\author[0000-0003-1680-1884]{Y.~Choi}
\affiliation{Space Telescope Science Institute, 3700 San Martin Drive, Baltimore, MD 21218, USA}
\author{J.~H.~Esteves}
\affiliation{Department of Physics, University of Michigan, Ann Arbor, MI 48109, USA}
\author[0000-0002-9144-7726]{C.~E.~Mart\'inez-V\'azquez}
\affiliation{Cerro Tololo Inter-American Observatory, NSF's National Optical-Infrared Astronomy Research Laboratory,\\ Casilla 603, La Serena, Chile}
\author[0000-0003-3519-4004]{S.~Mau}
\affiliation{Department of Physics, Stanford University, 382 Via Pueblo Mall, Stanford, CA 94305, USA}
\affiliation{Kavli Institute for Particle Astrophysics \& Cosmology, P.O.\ Box 2450, Stanford University, Stanford, CA 94305, USA}
\author{A.~E.~Miller}
\affiliation{Leibniz-Institut f\"ur Astrophysik Potsdam (AIP), An der Sternwarte 16, D-14482 Potsdam, Germany}
\affiliation{Institut f\"{u}r Physik und Astronomie, Universit\"{a}t Potsdam, Haus 28, Karl-Liebknecht-Str. 24/25, D-14476 Golm (Potsdam), Germany}
\author[0000-0001-9649-4815]{B.~Mutlu-Pakdil}
\affiliation{Kavli Institute for Cosmological Physics, University of Chicago, Chicago, IL 60637, USA}
\affiliation{Department of Astronomy and Astrophysics, University of Chicago, Chicago IL 60637, USA}
\author[0000-0002-7357-0317]{E.~H.~Neilsen}
\affiliation{Fermi National Accelerator Laboratory, P.O.\ Box 500, Batavia, IL 60510, USA}
\author[0000-0002-7134-8296]{K.~A.~G.~Olsen}
\affiliation{NSF's National Optical-Infrared Astronomy Research Laboratory, 950 N. Cherry Ave., Tucson, AZ 85719, USA}
\author[0000-0002-6021-8760]{A.~B.~Pace}
\affiliation{McWilliams Center for Cosmology, Carnegie Mellon University, 5000 Forbes Ave, Pittsburgh, PA 15213, USA}
\author[0000-0002-7134-8296]{A.~H.~Riley}
\affiliation{George P. and Cynthia Woods Mitchell Institute for Fundamental Physics and Astronomy,\\ Texas A\&M University, College Station, TX 77843, USA}
\affiliation{Department of Physics and Astronomy, Texas A\&M University, College Station, TX 77843, USA}
\author[0000-0002-1594-1466]{J.~D.~Sakowska}
\affiliation{Department of Physics, University of Surrey, Guildford GU2 7XH, UK}
\author[0000-0003-4102-380X]{D.~J.~Sand}
\affiliation{Department of Astronomy/Steward Observatory, 933 North Cherry Avenue, Room N204, Tucson, AZ 85721-0065, USA}
\author[0000-0003-3402-6164]{L.~Santana-Silva}
\affiliation{NAT-Universidade Cruzeiro do Sul / Universidade Cidade de S{\~a}o Paulo, Rua Galv{\~a}o Bueno, 868, 01506-000, S{\~a}o Paulo, SP, Brazil}
\author[0000-0002-7134-8296]{E.~J.~Tollerud}
\affiliation{Space Telescope Science Institute, 3700 San Martin Drive, Baltimore, MD 21218, USA}
\author[0000-0001-7211-5729]{D.~L.~Tucker}
\affiliation{Fermi National Accelerator Laboratory, P.O.\ Box 500, Batavia, IL 60510, USA}
\author[0000-0003-4341-6172]{A.~K.~Vivas}
\affiliation{Cerro Tololo Inter-American Observatory, NSF's National Optical-Infrared Astronomy Research Laboratory,\\ Casilla 603, La Serena, Chile}
\author{E.~Zaborowski}
\affiliation{Kavli Institute for Cosmological Physics, University of Chicago, Chicago, IL 60637, USA}
\author[0000-0001-6455-9135]{A.~Zenteno}
\affiliation{Cerro Tololo Inter-American Observatory, NSF's National Optical-Infrared Astronomy Research Laboratory,\\ Casilla 603, La Serena, Chile}
\author[0000-0003-1587-3931]{T.~M.~C.~Abbott}
\affiliation{Cerro Tololo Inter-American Observatory, NSF's National Optical-Infrared Astronomy Research Laboratory,\\ Casilla 603, La Serena, Chile}
\author[0000-0002-7069-7857]{S.~Allam}
\affiliation{Fermi National Accelerator Laboratory, P.O.\ Box 500, Batavia, IL 60510, USA}
\author[0000-0001-8156-0429]{K.~Bechtol}
\affiliation{Department of Physics, University of Wisconsin-Madison, Madison, WI 53706, USA}
\affiliation{LSST, 933 North Cherry Avenue, Tucson, AZ 85721, USA}
\author[0000-0003-0642-6558]{C.~P.~M.~Bell}
\affiliation{Leibniz-Institut f\"ur Astrophysik Potsdam (AIP), An der Sternwarte 16, D-14482 Potsdam, Germany}
\author[0000-0002-5564-9873]{E.~F.~Bell }
\affiliation{Department of Astronomy, University of Michigan, 1085 S.\ University Ave., Ann Arbor, MI, 48109-1107, USA}
\author{P.~Bilaji}
\affiliation{Kavli Institute for Cosmological Physics, University of Chicago, Chicago, IL 60637, USA}
\affiliation{Department of Astronomy and Astrophysics, University of Chicago, Chicago IL 60637, USA}
\author[0000-0003-4383-2969]{C.~R.~Bom}
\affiliation{Centro Brasileiro de Pesquisas F\'isicas, Rua Dr. Xavier Sigaud 150, 22290-180 Rio de Janeiro, RJ, Brazil}
\author[0000-0002-3690-105X]{J.~A.~Carballo-Bello}
\affiliation{Instituto de Alta Investigaci\'on, Sede Esmeralda, Universidad de Tarapac\'a, Av. Luis Emilio Recabarren 2477, Iquique, Chile}
\author[0000-0002-1763-4128]{D.~Crnojevi\'{c}}
\affiliation{University of Tampa, Department of Chemistry, Biochemistry, and Physics, 401 West Kennedy Boulevard, Tampa, FL 33606, USA}
\author[0000-0002-6797-696X]{M.-R.~L. Cioni}
\affiliation{Leibniz-Institut f\"ur Astrophysik Potsdam (AIP), An der Sternwarte 16, D-14482 Potsdam, Germany}
\author{A.~Diaz-Ocampo}
\affiliation{Departamento de F\'isica y Astronom\'ia, Facultad de Ciencias, Universidad de La Serena. Av. Juan Cisternas 1200, La Serena, Chile}
\author[0000-0001-5486-2747]{T.~J.~L.~de~Boer}
\affiliation{Institute for Astronomy, University of Hawai\'{i}, 2680 Woodlawn Drive, Honolulu, HI 96822, USA}
\author[0000-0002-8448-5505]{D.~Erkal}
\affiliation{Department of Physics, University of Surrey, Guildford GU2 7XH, UK}
\author[0000-0002-4588-6517]{R.~A.~Gruendl}
\affiliation{Department of Astronomy, University of Illinois, 1002 W. Green Street, Urbana, IL 61801, USA}
\affiliation{National Center for Supercomputing Applications, 1205 West Clark St., Urbana, IL 61801, USA}
\author{D.~Hernandez-Lang}
\affiliation{University of La Serena, La Serena, Chile}
\affiliation{Cerro Tololo Inter-American Observatory, NSF's National Optical-Infrared Astronomy Research Laboratory,\\ Casilla 603, La Serena, Chile}
\affiliation{Gemini Observatory, La Serena, Chile}
\author{A.~K.~Hughes}
\affiliation{Department of Astronomy/Steward Observatory, 933 North Cherry Avenue, Room N204, Tucson, AZ 85721-0065, USA}
\author[0000-0001-5160-4486]{D.~J.~James}
\affiliation{ASTRAVEO, LLC, PO Box 1668, Gloucester, MA 01931}
\author[0000-0001-6421-0953]{L.~C.~Johnson}
\affiliation{Center for Interdisciplinary Exploration and Research in Astrophysics (CIERA) and Department of Physics and Astronomy, Northwestern University,\\ 1800 Sherman Ave, Evanston, IL 60201 USA}
\author[0000-0002-9110-6163]{T.~S.~Li}
\affiliation{Observatories of the Carnegie Institution for Science, 813 Santa Barbara St., Pasadena, CA 91101, USA}
\affiliation{Department of Astrophysical Sciences, Princeton University, Princeton, NJ 08544, USA}
\affiliation{NHFP Einstein Fellow}
\author[0000-0002-1200-0820]{Y.-Y.~Mao}
\affiliation{Department of Physics and Astronomy, Rutgers, The State University of New Jersey, Piscataway, NJ 08854, USA}
\affiliation{NHFP Einstein Fellow}
\author[0000-0003-3835-2231]{D.~Mart\'{i}nez-Delgado}
\affiliation{Instituto de Astrof\'{i}sica de Andaluc\'{i}a, CSIC, E-18080 Granada, Spain}
\author{P.~Massana}
\affiliation{Department of Physics, University of Surrey, Guildford GU2 7XH, UK}
\affiliation{Isaac Newton Group of Telescopes, Apartado 321, E-38700 Santa Cruz de La Palma, Canary Islands, Spain}
\author[0000-0001-5435-7820]{M.~McNanna}
\affiliation{Department of Physics, University of Wisconsin-Madison, Madison, WI 53706, USA}
\author[0000-0002-7016-5471]{R.~Morgan}
\affiliation{Department of Physics, University of Wisconsin-Madison, Madison, WI 53706, USA}
\author[0000-0002-1182-3825]{E.~O.~Nadler}
\affiliation{Department of Physics, Stanford University, 382 Via Pueblo Mall, Stanford, CA 94305, USA}
\affiliation{Kavli Institute for Particle Astrophysics \& Cosmology, P.O.\ Box 2450, Stanford University, Stanford, CA 94305, USA}
\author[0000-0002-8282-469X]{N.~E.~D.~No\"el}
\affiliation{Department of Physics, University of Surrey, Guildford GU2 7XH, UK}
\author[0000-0002-6011-0530]{A.~Palmese}
\affiliation{Fermi National Accelerator Laboratory, P.O.\ Box 500, Batavia, IL 60510, USA}
\affiliation{Kavli Institute for Cosmological Physics, University of Chicago, Chicago, IL 60637, USA}
\author[0000-0002-8040-6785]{A.~H.~G.~Peter }
\affiliation{CCAPP, Department of Physics, and Department of Astronomy, The Ohio State University,\\ 191 W. Woodruff Ave., Columbus, OH 43210}
\author[0000-0001-9376-3135]{E.~S.~Rykoff}
\affiliation{Kavli Institute for Cosmological Physics, University of Chicago, Chicago, IL 60637, USA}
\affiliation{SLAC National Accelerator Laboratory, Menlo Park, CA 94025, USA}
\author[0000-0003-3136-9532]{J. S\'{a}nchez}
\affiliation{Fermi National Accelerator Laboratory, P.O.\ Box 500, Batavia, IL 60510, USA}
\author[0000-0002-7052-6900]{N.~Shipp}
\affiliation{Department of Astronomy and Astrophysics, University of Chicago, Chicago IL 60637, USA}
\affiliation{Kavli Institute for Cosmological Physics, University of Chicago, Chicago, IL 60637, USA}
\author[0000-0002-4733-4994]{J.~D.~Simon}
\affiliation{Observatories of the Carnegie Institution for Science, 813 Santa Barbara St., Pasadena, CA 91101, USA}
\author[0000-0003-2599-7524]{A.~Smercina}
\affiliation{Astronomy Department, University of Washington, Box 351580, Seattle, WA 98195-1580, USA}
\author[0000-0002-6904-359X]{M.~Soares-Santos}
\affiliation{Department of Physics, University of Michigan, Ann Arbor, MI 48109, USA}
\author[0000-0003-1479-3059]{G.~S.~Stringfellow}
\affiliation{Center for Astrophysics and Space Astronomy, University of Colorado, 389 UCB, Boulder, CO 80309-0389, USA}
\author[0000-0002-7052-6900]{K.~Tavangar}
\affiliation{Kavli Institute for Cosmological Physics, University of Chicago, Chicago, IL 60637, USA}
\affiliation{Department of Astronomy and Astrophysics, University of Chicago, Chicago IL 60637, USA}
\author[0000-0002-7052-6900]{R.~P.~van der Marel}
\affiliation{Space Telescope Science Institute, 3700 San Martin Drive, Baltimore, MD 21218, USA}
\affiliation{Center for Astrophysical Sciences, Department of Physics \& Astronomy, Johns Hopkins University, Baltimore, MD 21218, USA}
\author[0000-0002-7123-8943]{A.~R.~Walker}
\affiliation{Cerro Tololo Inter-American Observatory, NSF's National Optical-Infrared Astronomy Research Laboratory,\\ Casilla 603, La Serena, Chile}
\author[0000-0002-5077-881X]{R.~H.~Wechsler}
\affiliation{Department of Physics, Stanford University, 382 Via Pueblo Mall, Stanford, CA 94305, USA}
\affiliation{Kavli Institute for Particle Astrophysics \& Cosmology, P.O.\ Box 2450, Stanford University, Stanford, CA 94305, USA}
\affiliation{SLAC National Accelerator Laboratory, Menlo Park, CA 94025, USA}
\author[0000-0002-5077-881X]{J.~F.~Wu}
\affiliation{Space Telescope Science Institute, 3700 San Martin Drive, Baltimore, MD 21218, USA}
\author[0000-0002-7134-8296]{B.~Yanny}
\affiliation{Fermi National Accelerator Laboratory, P.O.\ Box 500, Batavia, IL 60510, USA}
\collaboration{66}{(DELVE Collaboration)}

\author[0000-0002-9080-0751]{M.~Fitzpatrick}
\affiliation{NSF's National Optical-Infrared Astronomy Research Laboratory, 950 N. Cherry Ave., Tucson, AZ 85719, USA}
\author[0000-0003-0952-5789]{L.~Huang}
\affiliation{NSF's National Optical-Infrared Astronomy Research Laboratory, 950 N. Cherry Ave., Tucson, AZ 85719, USA}
\author[0000-0001-9631-831X]{A.~Jacques}
\affiliation{NSF's National Optical-Infrared Astronomy Research Laboratory, 950 N. Cherry Ave., Tucson, AZ 85719, USA}
\author[0000-0002-7052-6900]{R.~Nikutta}
\affiliation{NSF's National Optical-Infrared Astronomy Research Laboratory, 950 N. Cherry Ave., Tucson, AZ 85719, USA}
\author[0000-0002-1140-5463]{A.~Scott}
\affiliation{NSF's National Optical-Infrared Astronomy Research Laboratory, 950 N. Cherry Ave., Tucson, AZ 85719, USA}

\collaboration{5}{(Astro Data Lab)}

%% file: table_deep.tex
\begin{deluxetable*}{ l c c c c c c c c}
\tablewidth{0pt}
\tabletypesize{\footnotesize}
\tablecaption{DELVE-DEEP isolated dwarf galaxy targets.} 
\label{tab:deep_targets}
\tablehead{
Target & R.A.    & Decl.   & Distance & $M_B$ & $M_*$       & $R_{110}$ & $N_{\rm sat, exp}$ & References \\
       & (deg) & (deg) & (Mpc) & (mag) & ($M_\odot$) & (deg)     & &
}
\startdata
NGC\,55  & 3.79  & -39.22 & 2.1  & -18.4 & $3.0\times10^9$    & 3.0 & 2--6 & (1)(2)(3) \\
NGC\,300 & 13.72 & -37.68 & 2.0 & -17.9 & $2.6 \times 10^9$   & 3.0 & 2--5 & (4)(2)(3) \\
Sextans\,B & 150.00 & 5.33 & 1.4 & -14.1 & $6.8 \times 10^7$  & 4.4 & 0--2 & (1)(5)(6) \\
IC\,5152 & 330.67 & -51.30 & 2.0 & -15.6 & $5.1 \times 10^8$  & 3.2 & 1--4 & (7)(5)(8) \\
\enddata
\tablecomments{Stellar masses are derived from the $K$-band luminosity reported by \citet{Karachentsev:2013} assuming a $K$-band mass-to-light ratio of unity. $R_{110}$ denotes the angular size of a 110 kpc region (roughly the virial radius) at the distance of each target. $N_{\rm sat, exp}$ is the number of expected satellites with $M_{\rm V} < -7$ according to \citet{Dooley:2017a,Dooley:2017b}.
}
\tablerefs{Values come from the compilation of \citet{Karachentsev:2013}. For each galaxy, we provide the original references for distance, absolute magnitude, and $K$-band luminosity: (1) \citet{Dalcanton:2009}, (2) \citet{daCosta:1998}, (3) \citet{Huchra:2012}, (4) \citet{Freedman:2001}, (5) \citet{deVaucouleurs:1991}, (6) \citet{Karachentsev:2013}, (7) \citet{Jacobs:2009}, (8) \citet{Fingerhut:2010}}
\end{deluxetable*}

%% file: table_summary.tex
\begin{deluxetable*}{l c c c c c}
\tablewidth{0pt}
\tabletypesize{\footnotesize}
\tablecaption{ DELVE DR1 key numbers and data quality summary. } 
\label{tab:summary}
\tablehead{
Survey Characteristic & \multicolumn{4}{c}{Band} & \colhead{Reference} \\[-0.5em]
 & $g$ & $r$ & $i$ & $z$ &
}
\startdata
Number of exposures & \nexpg & \nexpr & \nexpi & \nexpz &  \secref{data} \\
Median PSF FWHM (arcsec) & \medfwhmg & \medfwhmr & \medfwhmi & \medfwhmz &  \secref{data} \\
Sky coverage (individual bands, deg$^{2}$)  & \areanimagesg & \areanimagesr & \areanimagesi & \areanimagesz &  \secref{coverage} \\ 
Sky coverage ($g,r,i,z$ intersection, deg$^{2}$) & \multicolumn{4}{c}{\areanimagesgriz} & \secref{coverage} \\ 
Astrometric repeatability (angular distance, \mas) & \astrorepeatg & \astrorepeatr & \astrorepeati & \astrorepeatz &  \secref{astrometry} \\
Astrometric accuracy vs.\ \Gaia (angular distance, \mas) & \multicolumn{4}{c}{\astroabs} & \secref{astrometry} \\ 
Photometric repeatability (mmag)  & \photrepeatg & \photrepeatr & \photrepeati & \photrepeatz & \secref{photrel} \\
Photometric uniformity vs.\ \Gaia (mmag)  & \multicolumn{4}{c}{\photgaia} & \secref{photrel} \\
Absolute photometric uncertainty (mmag) & \multicolumn{4}{c}{$\lesssim \photabs$} & \secref{photabs} \\ 
Magnitude limit (PSF, ${\rm S/N} = 5$) & \maglimpsfg & \maglimpsfr & \maglimpsfi & \maglimpsfz & \secref{depth} \\
Magnitude limit (AUTO, ${\rm S/N} = 5$) & \maglimautog & \maglimautor & \maglimautoi & \maglimautoz & \secref{depth}\\
Galaxy selection ($\var{EXTENDED\_COADD} \geq 2$, $19 \leq \magauto[g] \leq 22$) & \multicolumn{4}{c}{Eff. $>\galefficiency\%$; Contam. $<\galcontamination\%$} & \secref{classification} \\
Stellar selection ($\var{EXTENDED\_COADD} \leq 1$, $19 \leq \magauto[g] \leq 22$) & \multicolumn{4}{c}{Eff. $>\starefficiency\%$; Contam. $<\starcontamination\%$} & \secref{classification} \\
\enddata
\end{deluxetable*}

%% file: table_depth.tex
\begin{deluxetable}{l c c c c c}
\tablewidth{0pt}
\tabletypesize{\footnotesize}
\tablecaption{ DELVE DR1 median depth estimates. } 
\label{tab:depth}
\tablehead{
\colhead{Measurement} & &\multicolumn{4}{c}{Magnitude Limit}  \\[-0.5em]
                      & & $g$ & $r$ & $i$ & $z$   \\[-0.25em]
                      & & (mag) & (mag) & (mag) & (mag)
}
\startdata
\var{MAG\_PSF~} (S/N=5)  & & \maglimpsfg & \maglimpsfr & \maglimpsfi & \maglimpsfz  \\
\var{MAG\_PSF~} (S/N=10) & & \maglimpsfteng & \maglimpsftenr & \maglimpsfteni & \maglimpsftenz  \\
\var{MAG\_AUTO} (S/N=5)  & & \maglimautog & \maglimautor & \maglimautoi & \maglimautoz  \\
\var{MAG\_AUTO} (S/N=10) & & \maglimautoteng & \maglimautotenr & \maglimautoteni & \maglimautotenz \\
\enddata
\tablecomments{\var{MAG\_PSF} depth is estimated using point-like sources while \var{MAG\_AUTO} depth measurements are derived from all DELVE DR1 sources (see \secref{depth}).}
\end{deluxetable}

%% file: ack.tex
\section{Acknowledgments}
The DELVE project is partially supported by Fermilab LDRD project L2019-011 and the NASA Fermi Guest Investigator Program Cycle 9 No. 91201.
This work is supported by the Visiting Scholars Award Program of the Universities Research Association.
ABP acknowledges support from NSF grant AST-1813881.
This research received support from the National Science Foundation (NSF) under grant no. NSF DGE-1656518 through the NSF Graduate Research Fellowship received by SM.
JLC acknowledges support from NSF grant AST-1816196.
JDS acknowledges support from NSF grant AST-1714873.
SRM acknowledges support from NSF grant AST-1909497.
DJS acknowledges support from NSF grants AST-1821967 and 1813708.
DMD acknowledges financial support from the State Agency for Research
of the Spanish MCIU through the ``Centre of Excellence Severo Ochoa''
award for the Instituto de Astrofísica de Andaluc\'ia (SEV-2017-0709).
CPMB and MRLC acknowledge support from the European Research Council (ERC) under the European Union's Horizon 2020 research and innovation programme (grant agreement no.\ 682115).

This project used data obtained with the Dark Energy Camera (DECam), which was constructed by the Dark Energy Survey (DES) collaboration.
Funding for the DES Projects has been provided by 
the DOE and NSF (USA),   
MISE (Spain),   
STFC (UK), 
HEFCE (UK), 
NCSA (UIUC), 
KICP (U. Chicago), 
CCAPP (Ohio State), 
MIFPA (Texas A\&M University),  
CNPQ, 
FAPERJ, 
FINEP (Brazil), 
MINECO (Spain), 
DFG (Germany), 
and the collaborating institutions in the Dark Energy Survey, which are
Argonne Lab, 
UC Santa Cruz, 
University of Cambridge, 
CIEMAT-Madrid, 
University of Chicago, 
University College London, 
DES-Brazil Consortium, 
University of Edinburgh, 
ETH Z{\"u}rich, 
Fermilab, 
University of Illinois, 
ICE (IEEC-CSIC), 
IFAE Barcelona, 
Lawrence Berkeley Lab, 
LMU M{\"u}nchen, and the associated Excellence Cluster Universe, 
University of Michigan, 
NSF's National Optical-Infrared Astronomy Research Laboratory, 
University of Nottingham, 
Ohio State University, 
OzDES Membership Consortium
University of Pennsylvania, 
University of Portsmouth, 
SLAC National Lab, 
Stanford University, 
University of Sussex, 
and Texas A\&M University.

This work has made use of data from the European Space Agency (ESA) mission {\it Gaia} (\url{https://www.cosmos.esa.int/gaia}), processed by the {\it Gaia} Data Processing and Analysis Consortium (DPAC, \url{https://www.cosmos.esa.int/web/gaia/dpac/consortium}).
Funding for the DPAC has been provided by national institutions, in particular the institutions participating in the {\it Gaia} Multilateral Agreement.

Based on observations at Cerro Tololo Inter-American Observatory, NSF's National Optical-Infrared Astronomy Research Laboratory (2019A-0305; PI: Drlica-Wagner), which is operated by the Association of Universities for Research in Astronomy (AURA) under a cooperative agreement with the National Science Foundation.

This manuscript has been authored by Fermi Research Alliance, LLC, under contract No.\ DE-AC02-07CH11359 with the US Department of Energy, Office of Science, Office of High Energy Physics. The United States Government retains and the publisher, by accepting the article for publication, acknowledges that the United States Government retains a non-exclusive, paid-up, irrevocable, worldwide license to publish or reproduce the published form of this manuscript, or allow others to do so, for United States Government purposes.

%% file: table_propid.tex
\begin{deluxetable}{| c c c | c c c | c c c |}
\tablecolumns{3}
\tabletypesize{\scriptsize}
\tablecaption{\label{tab:propid}
DECam community data included in DELVE DR1}
\tablehead{
\colhead{Prop.ID}  & \colhead{PI} & \colhead{$N_{\rm exp}$} &\colhead{Prop.ID}  & \colhead{PI} & \colhead{$N_{\rm exp}$} & \colhead{Prop.ID}  & \colhead{PI} & \colhead{$N_{\rm exp}$}}
\startdata
2019A-0305 & Alex Drlica-Wagner & 3087 &  2015A-0631 & Alfredo Zenteno & 108 &  2018A-0137 & Jeffrey Cooke & 20\\
2014B-0404 & David Schlegel & 2882 &  2017B-0312 & Bryan Miller & 108 &  2015B-0307 & Armin Rest & 20 \\
2018A-0386 & Alfredo Zenteno & 1798 &  2020A-0402 & \ldots & 107 &  2013B-0612 & Julio Chaname & 20 \\
2019A-0272 & Alfredo Zenteno & 1513 &  2015A-0107 & Claudia Belardi & 106 &  2014A-0348 & Haojing Yan & 20\\
2017A-0260 & Marcelle Soares-Santos & 1257 &  2013B-0614 & Ricardo Munoz & 106 &  2018B-0122 & Armin Rest & 19\\
2017A-0388 & Alfredo Zenteno & 1187 &  2017A-0918 & Alexandra Yip & 101 &  2018A-0059 & Herve Bouy & 18\\
2018A-0913 & Brad Tucker & 1039  &  2014A-0327 & Armin Rest & 99  &  2014A-0339 & Jonathan Hargis & 18\\
2015A-0608 & Francisco Forster & 760 &  2016A-0397 & Anja von der Linden & 98 &  2018A-0907 & Ricardo Munoz & 18\\
2013A-0214 & Maureen Van Den Berg & 716 &  2018A-0206 & Abhijit Saha & 94 &  2019A-0337 & David E Trilling & 18\\
2013B-0440 & David Nidever & 715 &  2013B-0421 & Armin Rest & 92  &  2013B-0627 & Gastao B Lima Neto & 18\\
2013A-0724 & Lori Allen & 615 &  2019A-0265 & Douglas P Finkbeiner & 90 &  2018A-0371 & Sangeeta Malhotra & 16 \\
2018B-0271 & Douglas P Finkbeiner & 527 &  2017A-0909 & Jeffrey Cooke & 90 &  2013A-0737 & Scott Sheppard & 16 \\
2013A-0614 & Sarah Sweet & 510 &  2014B-0146 & Mark Sullivan & 86 &  2014A-0496 & Aren Heinze & 16 \\
2013A-0327 & Armin Rest & 464 & 2014A-0623 & Ken Freeman & 83 &  2013A-0386 & Paul Thorman & 16\\
2016B-0909 & Camila Navarrete & 463 &  2014A-0321 & Marla Geha & 80 &  2014A-0386 & Ian dell'Antonio & 15\\
2020A-0399 & Alfredo Zenteno & 452 &  2017A-0913 & Luidhy Santana da Silva & 77 & 2013A-2101& \ldots & 15\\
2014A-0624 & Helmut Jerjen & 428 &  2015A-0306 & Eduardo Balbinot & 71 &  2014B-0610 & Julio Chaname & 14\\
2013A-0360 & Anja von der Linden & 425 &  2013A-0529 & R Michael Rich & 69 &  2014A-0634 & David James & 13\\
2015A-0616 & Helmut Jerjen & 425 &  2013A-0719 & Abhijit Saha & 68 &  2015A-0610 & Cesar Fuentes & 12 \\
2014A-0270 & Carl J Grillmair & 368 &  2019A-0308 & Ian Dell'Antonio & 65  &  2015A-0617 & David M Nataf & 11\\
2013A-0741 & David Schlegel & 323 &  2014A-0622 & Iraklis Konstantopoulos & 65 &  2014A-0480 & R Michael Rich & 10\\
2015A-0630 & Thomas H. Puzia & 309 &  2017B-0285 & Armin Rest & 65 &  2014A-0073 & Mukremin Kilic & 10 \\
2014A-0035 & Herve Bouy & 306 &  2019A-0101 & Patrick M Hartigan & 64 &  2018A-0380 & Armin Rest & 10 \\
2014A-0608 & Francisco Forster & 286 &  2015A-0205 & Eric Mamajek & 64 &  2018B-0904 & Lee Splitter & 10\\
2014A-0415 & Anja von der Linden & 282 &  2013A-0615 & Joss Bland-Hawthorn & 63 &  2014B-0064 & Mukremin Kilic & 10 \\
2015A-0620 & Ana Bonaca & 268 &  2020A-0238 & Clara Martinez-Vazquez & 62 &  2015A-0609 & Julio Carballo-Bello & 9\\
2014A-0412 & Armin Rest & 266 &  2017A-0911 & Ana Chies Santos & 62  &  2012B-0001 & Josh Frieman & 9 \\
2015A-0110 & Thomas De Boer & 264  &  2016A-0384 & Jacqueline McCleary & 61 &  2014A-0621 & Dougal Mackey & 7\\
2019B-0323 & Alfredo Zenteno & 254  &  2014A-0613 & David Rodriguez & 59 &  2013A-0455 & Scott Sheppard & 7 \\
2017A-0914 & Grant Tremblay & 250  &  2014A-0313 & Kathy Vivas & 55 &  2013A-0621 & Matias Gomez & 6 \\
2017A-0916 & Julio Carballo-Bello & 240 &  2017B-0163 & Prashin Jethwa & 55  &  2015A-0151 & Annalisa Calamida & 6\\
2019A-0205& Daniel Goldstein & 225 &  2013A-0612 & Yun-Kyeong Sheen & 52 &  2015B-0607 & Jeffrey Cooke & 5 \\ 
2016B-0279 & Douglas P Finkbeiner & 208  &  2014A-0610 & Matthew Taylor & 51 &  2017B-0078 & Herve Bouy & 5\\
2018A-0273 & William Dawson & 205 &  2015A-0615 & Brendan McMonigal & 50 &  2015B-0314 & Brad Tucker & 5  \\
2015A-0163 & Carl J Grillmair & 186 &  2018A-0912 & Attila Popping & 49 &  2012B-0624 & Aaron Robotham & 4 \\
2016A-0189 & Armin Rest & 184 & 2012B-0003 & \ldots & 48 &  2013B-0531 & Eric Mamajek & 4 \\
2016A-0327 & Douglas P Finkbeiner & 183 &  2016A-0614 & Thomas H. Puzia & 48 &  2017A-0366 & Sangeeta Malhotra & 3\\
2017A-0060 & Denija Crnojevic & 172 &  2014A-0239 & Mark Sullivan & 48  &  2013A-0613 & Ricardo Munoz & 3 \\
2015A-0130 & Denija Crnojevic & 172 &  2015A-0371 & Armin Rest & 47 &  2012B-0448 & Paul Thorman & 3\\
2015A-0121 & Anja von der Linden & 160 &  2016A-0190 & Arjun Dey & 44 &  2013A-0608 & Ricardo Demarco & 2 \\
2014A-0306 & Xinyu Dai & 159 &  2012B-0569 & Lori Allen & 39  &  2014A-0255 & Anton Koekemoer & 2 \\
2015A-0619 & Thiago Goncalves & 156 &  2017B-0103 & Wayne Barkhouse & 38  &  2016A-0610 & Leopoldo Infante & 2\\
2016B-0910 & Thomas H. Puzia & 153  &  2012B-0363 & Josh Bloom & 37 &  2014B-0375 & Armin Rest & 2 \\
2018A-0911 & Francisco Forster & 152 &  2018A-0369 & Armin Rest & 34 &  2013A-0609 & Douglas P Geisler & 1\\
2013A-0411 & David Nidever & 140  &  2012B-0506 & Daniel D Kelson & 32 &  2013A-0610 & Mario Hamuy & 1 \\
2018A-0276 & Ian Dell'Antonio & 139 &  2018A-0159 & Kathy Vivas & 32 &  2016A-0386 & Sangeeta Malhotra & 1 \\
2017B-0279 & Armin Rest & 138 &  2017A-0210 & Alistair Walker & 31 &  2013A-0351 & Arjun Dey & 1 \\
2018A-0215 & Jeffrey Carlin & 135 &  2013A-0611 & Dougal Mackey & 30 &  2017A-0308 & Annalisa Calamida & 1\\
2016B-0301 & Armin Rest & 134 &  2013A-0723 & Eric Mamajek & 25 &  2017B-0307 & Scott Sheppard & 1 \\
2018A-0251 & Douglas P Finkbeiner & 131 &  2017B-0110 & Edo Berger & 24 &  2017B-0951 & Kathy Vivas & 1 \\
2014A-0620 & Andrew Casey & 121 &  2014A-0429 & Douglas P Finkbeiner & 23 &   2019A-0315 & Matthew Penny & 1\\
2015B-0606 & Katharine Lutz & 113 &  2016A-0622 & Paulo Lopes & 22 &  2016A-0191 & Armin Rest & 1 \\
2018A-0909 & Thomas H. Puzia & 110  &  2015A-0397 & Armin Rest & 21 & \textbf{TOTAL} & \textbf{DELVE DR1} & 29929\\
\enddata
\tablecomments{Programs are ordered by the number of exposures contributed. The largest single contributor to the DELVE DR1 data set is the DELVE program itself. Programs with no principal investigator (PI) listed are generally Target-of-Opportunity (ToO) programs.}
\end{deluxetable}

%% file: table_dr1_main.tex
\begin{deluxetable}{l l c}
    	\centering
	\tablewidth{0pt}
	\tabletypesize{\scriptsize}
	\tablecaption{\code{DELVE.DR1\_MAIN} table description: \nobjects rows; 215 columns \label{tab:dr1_main}}
	\tablehead{
		\colhead{Column Name} & \colhead{Description} & \colhead{Columns}
	}
        \startdata
        QUICK\_OBJECT\_ID & Unique identifier for each object & 1 \\
	RA & Right ascension derived from the median position of each detection (deg) & 1 \\
	DEC & Declination derived from the median position of each detection (deg) & 1 \\
	A\_IMAGE\_{G,R,I,Z} & Semi-major axis of adaptive aperture in image coordinates (pix) & 4 \\
	B\_IMAGE\_{G,R,I,Z} & Semi-minor axis of adaptive aperture in image coordinates (pix) & 4 \\
	CCDNUM\_{G,R,I,Z} & CCD number for best exposure in each band & 4 \\
	CLASS\_STAR\_{G,R,I,Z} & Neural-network-based star--galaxy classifier (see \SExtractor manual for details) & 4 \\
	EBV & $E(B-V)$ value at the object location interpolated from the map of \citet{Schlegel:1998} & 1 \\
	EXPNUM\_{G,R,I,Z} & Exposure number for best exposure in each band & 4 \\
	EXPTIME\_{G,R,I,Z} & Shutter-open exposure time for best exposure in each band & 4 \\
	EXTENDED\_CLASS\_{G,R,I,Z} & Spread-model-based morphology class (see \secref{classification}) & 4\\
                                   & $-9$ unknown, 0 high-confidence star, 1 likely star, 2 likely galaxy, 3 high-confidence galaxy &  \\
	EXTINCTION\_{G,R,I,Z} & Interstellar extinction from \citet{Schlegel:1998} $E(B-V)$ and DES $A_b$ (see \secref{processing}) & 4 \\
	FLAGS\_{G,R,I,Z} & \SExtractor flags for the best detection in each band & 4 \\
	HPX2048 & HEALPix index for each object in RING format at resolution $\nside=2048$ & 1 \\
	HTM9 & HTM Level-9 index & 1 \\
	MAG\_AUTO\_{G,R,I,Z} & Automatic aperture magnitude in each band (see \SExtractor manual) & 4 \\
	MAGERR\_AUTO\_{G,R,I,Z} & Automatic aperture magnitude uncertainty in each band (see \SExtractor manual) & 4 \\
	MAG\_PSF\_{G,R,I,Z} & PSF magnitude in each band (see \SExtractor manual) & 4 \\
	MAGERR\_PSF\_{G,R,I,Z} & PSF magnitude uncertainty in each band (see \SExtractor manual) & 4 \\
	NEPOCHS\_{G,R,I,Z} & Number of single-epoch detections for this object & 4 \\
    NEST4096 & \healpix index for each object in NEST format at resolution $\nside=4096$ & 1 \\
    RANDOM\_ID & Random ID in the range 0.0 to 100.0 for subsampling & 1 \\
    RING256 & \healpix index for each object in RING format at resolution $\nside=256$ & 1 \\
    RING32 & \healpix index for each object in RING format at resolution $\nside=32$ & 1 \\
    SPREAD\_MODEL\_{G,R,I,Z} & Likelihood-based star--galaxy classifier \citep[see][]{Desai:2012} & 4 \\
    SPREADERR\_MODEL\_{G,R,I,Z} & Likelihood-based star--galaxy classifier uncertainty \citep[see][]{Desai:2012} & 4 \\
    T\_EFF\_{G,R,I,Z} & Effective exposure time scale factor for best exposure in each band \citep[see][]{Neilsen:2015} & 4 \\
    THETA\_IMAGE\_{G,R,I,Z} & Position angle of automatic aperture in image coordinates (deg) & 4\\
    WAVG\_FLAGS\_{G,R,I,Z} & OR of \SExtractor flags from all detections in each band & 4 \\
    WAVG\_MAG\_AUTO\_{G,R,I,Z} & Weighted average of automatic aperture magnitude measurements in each band & 4 \\
    WAVG\_MAGERR\_AUTO\_{G,R,I,Z} & Sum in quadrature of the automatic aperture magnitude uncertainties in each band & 4 \\
    WAVG\_MAGRMS\_AUTO\_{G,R,I,Z} & Unbiased weighted standard deviation of the automatic aperture magnitude in each band & 4 \\
    WAVG\_MAG\_PSF\_{G,R,I,Z} & Weighted average of PSF magnitude measurements in each band & 4 \\
    WAVG\_MAGERR\_PSF\_{G,R,I,Z} & Sum in quadrature of the PSF magnitude uncertainties in each band & 4 \\
    WAVG\_MAGRMS\_PSF\_{G,R,I,Z} &  Unbiased weighted standard deviation of the PSF magnitude in each band & 4 \\
    WAVG\_SPREAD\_MODEL\_{G,R,I,Z} & Weighted average spread model in each band & 4 \\
    WAVG\_SPREADERR\_MODEL\_{G,R,I,Z} & Sum in quadrature of the spread model uncertainties in each band & 4 \\
    WAVG\_SPREADRMS\_MODEL\_{G,R,I,Z} & Unbiased weighted standard deviation of \var{spread\_model} in each band & 4 \\
\enddata
\end{deluxetable}

%% file: table_dr1_gaia.tex
\begin{deluxetable}{l l c}
    	\centering
	\tablewidth{0pt}
	\tabletypesize{\scriptsize}
	\tablecaption{\code{DELVE\_DR1.X1P5\_\_OBJECTS\_\_GAIA\_EDR3\_\_GAIA\_SOURCE} table description: 143,701,359 rows; 7 columns \label{tab:dr1_gaia}}
	\tablehead{
		\colhead{Column Name} & \colhead{Description} & \colhead{Columns}
	}
        \startdata
	DEC1 & Declination in DELVE DR1 (deg) & 1\\
	DEC2 & Declination in \Gaia EDR3 (deg) & 1\\
        DISTANCE & Distance between RA1,DEC1 and RA2,DEC2 (arcsec) & 1 \\
        ID1 & ID in DELVE DR1 (\var{quick\_object\_id}) & 1 \\
        ID2 & ID in \Gaia EDR3 (\var{source\_id}) & 1 \\
        RA1 & Right ascension from DELVE DR1 (deg) & 1 \\
        RA2 & Right ascension from \Gaia EDR3 (deg) & 1 \\
\enddata
\end{deluxetable}

%% file: main.bbl
\begin{thebibliography}{}
\providecommand\natexlab[1]{#1}
\providecommand\JournalTitle[1]{#1}
\providecommand{\eprint}[1][]{\url{#1}}

\bibitem[{{Ackermann} {et~al.}(2015){Ackermann} \& {Albert}
  {et~al.}}]{Ackermann:2015}
{Ackermann}, M., {Albert}, A., {Anderson}, B., {et~al.} 2015,
  \href{http://dx.doi.org/10.1103/PhysRevLett.115.231301}{\JournalTitle{\prl},
  115, 231301}, \eprint arXiv:{1503.02641}

\bibitem[{{Agertz} {et~al.}(2020){Agertz} \& {Pontzen} {et~al.}}]{Agertz:2020}
{Agertz}, O., {Pontzen}, A., {Read}, J.~I., {et~al.} 2020,
  \href{http://dx.doi.org/10.1093/mnras/stz3053}{\JournalTitle{\mnras}, 491,
  1656}, \eprint arXiv:{1904.02723}

\bibitem[{{Aguena} {et~al.}(2021){Aguena} \& {Benoist} {et~al.}}]{Aguena:2021}
{Aguena}, M., {Benoist}, C., {da Costa}, L.~N., {et~al.} 2021,
  \href{http://dx.doi.org/10.1093/mnras/stab264}{\JournalTitle{\mnras}, 502,
  4435}, \eprint arXiv:{2008.08711}

\bibitem[{{Aihara} {et~al.}(2019){Aihara} \& {AlSayyad} {et~al.}}]{HSC-DR2}
{Aihara}, H., {AlSayyad}, Y., {Ando}, M., {et~al.} 2019,
  \href{http://dx.doi.org/10.1093/pasj/psz103}{\JournalTitle{\pasj}, 71, 114},
  \eprint arXiv:{1905.12221}

\bibitem[{{Allen} {et~al.}(2011){Allen} \& {Evrard} \& {Mantz}}]{Allen2011}
{Allen}, S.~W., {Evrard}, A.~E., \& {Mantz}, A.~B. 2011,
  \href{http://dx.doi.org/10.1146/annurev-astro-081710-102514}{\JournalTitle{\araa},
  49, 409}, \eprint arXiv:{1103.4829}

\bibitem[{{Andrade-Santos} {et~al.}(2012){Andrade-Santos} \& {Lima Neto} \&
  {Lagan{\'a}}}]{Andrade-Santos:2012}
{Andrade-Santos}, F., {Lima Neto}, G.~B., \& {Lagan{\'a}}, T.~F. 2012,
  \href{http://dx.doi.org/10.1088/0004-637X/746/2/139}{\JournalTitle{\apj},
  746, 139}, \eprint arXiv:{1112.1971}

\bibitem[{{Andrade-Santos} {et~al.}(2017){Andrade-Santos} \& {Jones}
  {et~al.}}]{Andrade-Santos:2017}
{Andrade-Santos}, F., {Jones}, C., {Forman}, W.~R., {et~al.} 2017,
  \href{http://dx.doi.org/10.3847/1538-4357/aa7461}{\JournalTitle{\apj}, 843,
  76}, \eprint arXiv:{1703.08690}

\bibitem[{{Astropy Collaboration}(2018){Astropy Collaboration} \&
  {Price-Whelan} {et~al.}}]{astropy:2018}
{Astropy Collaboration}. 2018,
  \href{http://dx.doi.org/10.3847/1538-3881/aabc4f}{\JournalTitle{\aj}, 156,
  123}, \eprint arXiv:{1801.02634}

\bibitem[{{Balbinot} {et~al.}(2016){Balbinot} \& {Yanny} \& {Li} \& {Santiago}
  \& {Marshall} {et~al.}}]{Balbinot:2016}
{Balbinot}, E., {Yanny}, B., {Li}, T.~S., {et~al.} 2016,
  \href{http://dx.doi.org/10.3847/0004-637X/820/1/58}{\JournalTitle{\apj}, 820,
  58}, \eprint arXiv:{1509.04283}

\bibitem[{{Banik} {et~al.}(2019){Banik} \& {Bovy} \& {Bertone} \& {Erkal} \&
  {de Boer}}]{Banik:2019}
{Banik}, N., {Bovy}, J., {Bertone}, G., {Erkal}, D., \& {de Boer}, T.~J.~L.
  2019, \JournalTitle{arXiv e-prints}, arXiv:1911.02663, \eprint
  arXiv:{1911.02663}

\bibitem[{{Bechtol} {et~al.}(2015){Bechtol} \& {Drlica-Wagner}
  {et~al.}}]{Bechtol:2015}
{Bechtol}, K., {Drlica-Wagner}, A., {Balbinot}, E., {et~al.} 2015,
  \href{http://dx.doi.org/10.1088/0004-637X/807/1/50}{\JournalTitle{\apj}, 807,
  50}, \eprint arXiv:{1503.02584}

\bibitem[{{Behroozi} {et~al.}(2013){Behroozi} \& {Wechsler} \&
  {Conroy}}]{Behroozi:2013}
{Behroozi}, P.~S., {Wechsler}, R.~H., \& {Conroy}, C. 2013,
  \href{http://dx.doi.org/10.1088/0004-637X/770/1/57}{\JournalTitle{\apj}, 770,
  57}, \eprint arXiv:{1207.6105}

\bibitem[{{Bellazzini} {et~al.}(2014){Bellazzini} \& {Beccari}
  {et~al.}}]{Bellazzini:2014}
{Bellazzini}, M., {Beccari}, G., {Fraternali}, F., {et~al.} 2014,
  \href{http://dx.doi.org/10.1051/0004-6361/201423659}{\JournalTitle{\aap},
  566, A44}, \eprint arXiv:{1404.1697}

\bibitem[{{Belokurov} {et~al.}(2019){Belokurov} \& {Deason}
  {et~al.}}]{Belokurov:2019}
{Belokurov}, V., {Deason}, A.~J., {Erkal}, D., {et~al.} 2019,
  \href{http://dx.doi.org/10.1093/mnrasl/slz101}{\JournalTitle{\mnras}, 488,
  L47}, \eprint arXiv:{1904.07909}

\bibitem[{{Belokurov} {et~al.}(2014){Belokurov} \& {Irwin}
  {et~al.}}]{2014MNRAS.441.2124B}
{Belokurov}, V., {Irwin}, M.~J., {Koposov}, S.~E., {et~al.} 2014,
  \href{http://dx.doi.org/10.1093/mnras/stu626}{\JournalTitle{\mnras}, 441,
  2124}, \eprint arXiv:{1403.3406}

\bibitem[{{Belokurov} {et~al.}(2006){Belokurov} \& {Zucker}
  {et~al.}}]{Belokurov:2006}
{Belokurov}, V., {Zucker}, D.~B., {Evans}, N.~W., {et~al.} 2006,
  \href{http://dx.doi.org/10.1086/504797}{\JournalTitle{\apjl}, 642, L137},
  \eprint arXiv:{astro-ph/0605025}

\bibitem[{{Belokurov} {et~al.}(2007){Belokurov} \& {Zucker}
  {et~al.}}]{2007ApJ...654..897B}
{Belokurov}, V., {Zucker}, D.~B., {Evans}, N.~W., {et~al.} 2007,
  \href{http://dx.doi.org/10.1086/509718}{\JournalTitle{\apj}, 654, 897},
  \eprint{astro-ph/0608448}

\bibitem[{{Belokurov} \& {Erkal}(2019)}]{Belokurov:2018}
{Belokurov}, V.~A. \& {Erkal}, D. 2019,
  \href{http://dx.doi.org/10.1093/mnrasl/sly178}{\JournalTitle{\mnras}, 482,
  L9}, \eprint arXiv:{1808.00462}

\bibitem[{{Bennet} {et~al.}(2019){Bennet} \& {Sand} {et~al.}}]{Bennet:2019}
{Bennet}, P., {Sand}, D.~J., {Crnojevi{\'c}}, D., {et~al.} 2019,
  \href{http://dx.doi.org/10.3847/1538-4357/ab46ab}{\JournalTitle{\apj}, 885,
  153}, \eprint arXiv:{1906.03230}

\bibitem[{{Bennet} {et~al.}(2020){Bennet} \& {Sand} {et~al.}}]{Bennet:2020}
{Bennet}, P., {Sand}, D.~J., {Crnojevi{\'c}}, D., {et~al.} 2020,
  \href{http://dx.doi.org/10.3847/2041-8213/ab80c5}{\JournalTitle{\apjl}, 893,
  L9}, \eprint arXiv:{2002.11126}

\bibitem[{{Bergstr{\"o}m} {et~al.}(1998){Bergstr{\"o}m} \& {Ullio} \&
  {Buckley}}]{Bergstrom:1998}
{Bergstr{\"o}m}, L., {Ullio}, P., \& {Buckley}, J.~H. 1998,
  \href{http://dx.doi.org/10.1016/S0927-6505(98)00015-2}{\JournalTitle{Astroparticle
  Physics}, 9, 137}, \eprint arXiv:{astro-ph/9712318}

\bibitem[{{Bernard} {et~al.}(2014){Bernard} \& {Ferguson}
  {et~al.}}]{Bernard:2014}
{Bernard}, E.~J., {Ferguson}, A.~M.~N., {Schlafly}, E.~F., {et~al.} 2014,
  \href{http://dx.doi.org/10.1093/mnrasl/slu089}{\JournalTitle{\mnras}, 443,
  L84}, \eprint arXiv:{1405.6645}

\bibitem[{{Bernard} {et~al.}(2016){Bernard} \& {Ferguson}
  {et~al.}}]{Bernard:2016}
{Bernard}, E.~J., {Ferguson}, A. M.~N., {Schlafly}, E.~F., {et~al.} 2016,
  \href{http://dx.doi.org/10.1093/mnras/stw2134}{\JournalTitle{\mnras}, 463,
  1759}, \eprint arXiv:{1607.06088}

\bibitem[{{Bernstein} {et~al.}(2018){Bernstein} \& {Abbott}
  {et~al.}}]{Bernstein:2018}
{Bernstein}, G.~M., {Abbott}, T.~M.~C., {Armstrong}, R., {et~al.} 2018,
  \href{http://dx.doi.org/10.1088/1538-3873/aaa753}{\JournalTitle{\pasp}, 130,
  054501}, \eprint arXiv:{1710.10943}

\bibitem[{{Bertin}(2006)}]{Bertin:2006}
{Bertin}, E. 2006, in Astronomical Society of the Pacific Conference Series,
  Vol. 351, Astronomical Data Analysis Software and Systems XV, ed.
  C.~{Gabriel}, C.~{Arviset}, D.~{Ponz}, \& S.~{Enrique}, 112

\bibitem[{{Bertin}(2010)}]{Bertin:2010}
{Bertin}, E. 2010, {SWarp: Resampling and Co-adding FITS Images Together},
  Astrophysics Source Code Library, \eprint ascl:{1010.068}

\bibitem[{{Bertin}(2011)}]{Bertin:2011}
{Bertin}, E. 2011, in Astronomical Society of the Pacific Conference Series,
  Vol. 442, Astronomical Data Analysis Software and Systems XX, ed. I.~N.
  {Evans}, A.~{Accomazzi}, D.~J. {Mink}, \& A.~H. {Rots}, San Francisco, CA,
  435

\bibitem[{{Bertin} \& {Arnouts}(1996)}]{Bertin:1996}
{Bertin}, E. \& {Arnouts}, S. 1996, \JournalTitle{\aaps}, 117, 393

\bibitem[{{Bertin} {et~al.}(2002){Bertin} \& {Mellier} \& {Radovich} \&
  {Missonnier} \& {Didelon} \& {Morin}}]{Bertin:2002}
{Bertin}, E., {Mellier}, Y., {Radovich}, M., {et~al.} 2002, in Astronomical
  Society of the Pacific Conference Series, Vol. 281, Astronomical Data
  Analysis Software and Systems XI, ed. D.~A. {Bohlender}, D.~{Durand}, \&
  T.~H. {Handley}, 228

\bibitem[{{Bleem} {et~al.}(2020){Bleem} \& {Bocquet} {et~al.}}]{Bleem:2020}
{Bleem}, L.~E., {Bocquet}, S., {Stalder}, B., {et~al.} 2020,
  \href{http://dx.doi.org/10.3847/1538-4365/ab6993}{\JournalTitle{\apjs}, 247,
  25}, \eprint arXiv:{1910.04121}

\bibitem[{Bocquet {et~al.}(2019)}]{Bocquet2018}
Bocquet, S. {et~al.} 2019,
  \href{http://dx.doi.org/10.3847/1538-4357/ab1f10}{\JournalTitle{\apj}, 878,
  55}, \eprint arXiv:{1812.01679}

\bibitem[{{Bonaca} {et~al.}(2012){Bonaca} \& {Geha} \&
  {Kallivayalil}}]{Bonaca:2012}
{Bonaca}, A., {Geha}, M., \& {Kallivayalil}, N. 2012,
  \href{http://dx.doi.org/10.1088/2041-8205/760/1/L6}{\JournalTitle{\apjl},
  760, L6}, \eprint arXiv:{1209.5391}

\bibitem[{{Bonaca} \& {Hogg}(2018)}]{Bonaca:2018}
{Bonaca}, A. \& {Hogg}, D.~W. 2018,
  \href{http://dx.doi.org/10.3847/1538-4357/aae4da}{\JournalTitle{\apj}, 867,
  101}, \eprint arXiv:{1804.06854}

\bibitem[{{Bonaca} {et~al.}(2020){Bonaca} \& {Pearson} {et~al.}}]{Bonaca:2020}
{Bonaca}, A., {Pearson}, S., {Price-Whelan}, A.~M., {et~al.} 2020,
  \href{http://dx.doi.org/10.3847/1538-4357/ab5afe}{\JournalTitle{\apj}, 889,
  70}, \eprint arXiv:{1910.00592}

\bibitem[{{Bonaca} {et~al.}(2021){Bonaca} \& {Naidu} {et~al.}}]{Bonaca:2020b}
{Bonaca}, A., {Naidu}, R.~P., {Conroy}, C., {et~al.} 2021,
  \href{http://dx.doi.org/10.3847/2041-8213/abeaa9}{\JournalTitle{\apjl}, 909,
  L26}, \eprint arXiv:{2012.09171}

\bibitem[{{Bovy} {et~al.}(2016){Bovy} \& {Bahmanyar} \& {Fritz} \&
  {Kallivayalil}}]{Bovy:2016}
{Bovy}, J., {Bahmanyar}, A., {Fritz}, T.~K., \& {Kallivayalil}, N. 2016,
  \href{http://dx.doi.org/10.3847/1538-4357/833/1/31}{\JournalTitle{\apj}, 833,
  31}, \eprint arXiv:{1609.01298}

\bibitem[{{Boylan-Kolchin} {et~al.}(2015){Boylan-Kolchin} \& {Weisz} \&
  {Johnson} \& {Bullock} \& {Conroy} \& {Fitts}}]{Boylan-Kolchin:2015}
{Boylan-Kolchin}, M., {Weisz}, D.~R., {Johnson}, B.~D., {et~al.} 2015,
  \href{http://dx.doi.org/10.1093/mnras/stv1736}{\JournalTitle{\mnras}, 453,
  1503}, \eprint arXiv:{1504.06621}

\bibitem[{{Brandt}(2016)}]{Brandt:2016}
{Brandt}, T.~D. 2016,
  \href{http://dx.doi.org/10.3847/2041-8205/824/2/L31}{\JournalTitle{\apjl},
  824, L31}, \eprint arXiv:{1605.03665}

\bibitem[{{Bullock} \& {Boylan-Kolchin}(2017)}]{Bullock:2017}
{Bullock}, J.~S. \& {Boylan-Kolchin}, M. 2017,
  \href{http://dx.doi.org/10.1146/annurev-astro-091916-055313}{\JournalTitle{\araa},
  55, 343}, \eprint arXiv:{1707.04256}

\bibitem[{{Bullock} \& {Johnston}(2005)}]{Bullock:2005}
{Bullock}, J.~S. \& {Johnston}, K.~V. 2005,
  \href{http://dx.doi.org/10.1086/497422}{\JournalTitle{\apj}, 635, 931},
  \eprint arXiv:{astro-ph/0506467}

\bibitem[{{Bullock} {et~al.}(2000){Bullock} \& {Kravtsov} \&
  {Weinberg}}]{Bullock:2000}
{Bullock}, J.~S., {Kravtsov}, A.~V., \& {Weinberg}, D.~H. 2000,
  \href{http://dx.doi.org/10.1086/309279}{\JournalTitle{\apj}, 539, 517},
  \eprint{astro-ph/0002214}

\bibitem[{{Burke} {et~al.}(2018){Burke} \& {Rykoff} {et~al.}}]{Burke:2018}
{Burke}, D.~L., {Rykoff}, E.~S., {Allam}, S., {et~al.} 2018,
  \href{http://dx.doi.org/10.3847/1538-3881/aa9f22}{\JournalTitle{\aj}, 155,
  41}, \eprint arXiv:{1706.01542}

\bibitem[{{Carlberg}(2013)}]{Carlberg:2013}
{Carlberg}, R.~G. 2013,
  \href{http://dx.doi.org/10.1088/0004-637X/775/2/90}{\JournalTitle{\apj}, 775,
  90}, \eprint arXiv:{1307.1929}

\bibitem[{{Carlin} {et~al.}(2016){Carlin} \& {Sand} {et~al.}}]{Carlin:2016}
{Carlin}, J.~L., {Sand}, D.~J., {Price}, P., {et~al.} 2016,
  \href{http://dx.doi.org/10.3847/2041-8205/828/1/L5}{\JournalTitle{\apj}, 828,
  L5}, \eprint arXiv:{1608.02591}

\bibitem[{{Carlin} {et~al.}(2019){Carlin} \& {Garling} {et~al.}}]{Carlin:2019}
{Carlin}, J.~L., {Garling}, C.~T., {Peter}, A. H.~G., {et~al.} 2019,
  \href{http://dx.doi.org/10.3847/1538-4357/ab4c32}{\JournalTitle{\apj}, 886,
  109}, \eprint arXiv:{1906.08260}

\bibitem[{{Carlin} {et~al.}(2021){Carlin} \& {Mutlu-Pakdil}
  {et~al.}}]{Carlin:2020}
{Carlin}, J.~L., {Mutlu-Pakdil}, B., {Crnojevi{\'c}}, D., {et~al.} 2021,
  \href{http://dx.doi.org/10.3847/1538-4357/abe040}{\JournalTitle{\apj}, 909,
  211}, \eprint arXiv:{2012.09174}

\bibitem[{{Carlsten} {et~al.}(2021){Carlsten} \& {Greene} \& {Peter} \&
  {Beaton} \& {Greco}}]{Carlsten:2020a}
{Carlsten}, S.~G., {Greene}, J.~E., {Peter}, A. H.~G., {Beaton}, R.~L., \&
  {Greco}, J.~P. 2021,
  \href{http://dx.doi.org/10.3847/1538-4357/abd039}{\JournalTitle{\apj}, 908,
  109}, \eprint arXiv:{2006.02443}

\bibitem[{{Carlsten} {et~al.}(2020){Carlsten} \& {Greene} \& {Peter} \& {Greco}
  \& {Beaton}}]{Carlsten:2020b}
{Carlsten}, S.~G., {Greene}, J.~E., {Peter}, A. H.~G., {Greco}, J.~P., \&
  {Beaton}, R.~L. 2020,
  \href{http://dx.doi.org/10.3847/1538-4357/abb60b}{\JournalTitle{\apj}, 902,
  124}, \eprint arXiv:{2006.02444}

\bibitem[{Cañameras {et~al.}(2020)Cañameras \& Schuldt
  {et~al.}}]{Canameras:2020}
Cañameras, R., Schuldt, S., Suyu, S.~H., {et~al.} 2020,
  \href{http://dx.doi.org/10.1051/0004-6361/202038219}{\JournalTitle{\aa}, 644,
  A163}

\bibitem[{{Cerny} {et~al.}(2021){Cerny} \& {Pace} {et~al.}}]{Cerny:2020}
{Cerny}, W., {Pace}, A.~B., {Drlica-Wagner}, A., {et~al.} 2021,
  \href{http://dx.doi.org/10.3847/1538-4357/abe1af}{\JournalTitle{\apj}, 910,
  18}, \eprint arXiv:{2009.08550}

\bibitem[{{Chambers} {et~al.}(2016){Chambers} \& {Magnier}
  {et~al.}}]{Chambers:2016}
{Chambers}, K.~C., {Magnier}, E.~A., {Metcalfe}, N., {et~al.} 2016,
  \JournalTitle{ArXiv e-prints}, \eprint arXiv:{1612.05560}

\bibitem[{{Chartab} {et~al.}(2020){Chartab} \& {Mobasher}
  {et~al.}}]{Chartab:2020}
{Chartab}, N., {Mobasher}, B., {Darvish}, B., {et~al.} 2020,
  \href{http://dx.doi.org/10.3847/1538-4357/ab61fd}{\JournalTitle{\apj}, 890,
  7}, \eprint arXiv:{1912.04890}

\bibitem[{{Chiboucas} {et~al.}(2013){Chiboucas} \& {Jacobs} \& {Tully} \&
  {Karachentsev}}]{Chiboucas:2013}
{Chiboucas}, K., {Jacobs}, B.~A., {Tully}, R.~B., \& {Karachentsev}, I.~D.
  2013,
  \href{http://dx.doi.org/10.1088/0004-6256/146/5/126}{\JournalTitle{\aj}, 146,
  126}, \eprint arXiv:{1309.4130}

\bibitem[{{Choi} {et~al.}(2018{\natexlab{a}}){Choi} \& {Nidever}
  {et~al.}}]{Choi:2018a}
{Choi}, Y., {Nidever}, D.~L., {Olsen}, K., {et~al.} 2018{\natexlab{a}},
  \href{http://dx.doi.org/10.3847/1538-4357/aae083}{\JournalTitle{\apj}, 866,
  90}, \eprint arXiv:{1804.07765}

\bibitem[{{Choi} {et~al.}(2018{\natexlab{b}}){Choi} \& {Nidever}
  {et~al.}}]{Choi:2018b}
{Choi}, Y., {Nidever}, D.~L., {Olsen}, K., {et~al.} 2018{\natexlab{b}},
  \href{http://dx.doi.org/10.3847/1538-4357/aaed1f}{\JournalTitle{\apj}, 869,
  125}, \eprint arXiv:{1805.00481}

\bibitem[{{Collett}(2015)}]{Collett:2015}
{Collett}, T.~E. 2015,
  \href{http://dx.doi.org/10.1088/0004-637X/811/1/20}{\JournalTitle{\apj}, 811,
  20}, \eprint arXiv:{1507.02657}

\bibitem[{{Cooper} {et~al.}(2010){Cooper} \& {Cole} {et~al.}}]{Cooper:2010}
{Cooper}, A.~P., {Cole}, S., {Frenk}, C.~S., {et~al.} 2010,
  \href{http://dx.doi.org/10.1111/j.1365-2966.2010.16740.x}{\JournalTitle{\mnras},
  406, 744}, \eprint arXiv:{0910.3211}

\bibitem[{{Crnojevi{\'c}} {et~al.}(2016){Crnojevi{\'c}} \& {Sand}
  {et~al.}}]{Crnojevic:2016}
{Crnojevi{\'c}}, D., {Sand}, D.~J., {Spekkens}, K., {et~al.} 2016,
  \href{http://dx.doi.org/10.3847/0004-637X/823/1/19}{\JournalTitle{\apj}, 823,
  19}, \eprint arXiv:{1512.05366}

\bibitem[{{Crnojevi{\'c}} {et~al.}(2019){Crnojevi{\'c}} \& {Sand}
  {et~al.}}]{Crnojevic:2019}
{Crnojevi{\'c}}, D., {Sand}, D.~J., {Bennet}, P., {et~al.} 2019,
  \href{http://dx.doi.org/10.3847/1538-4357/aafbe7}{\JournalTitle{\apj}, 872,
  80}, \eprint arXiv:{1809.05103}

\bibitem[{{da Costa} {et~al.}(1998){da Costa} \& {Willmer}
  {et~al.}}]{daCosta:1998}
{da Costa}, L.~N., {Willmer}, C.~N.~A., {Pellegrini}, P.~S., {et~al.} 1998,
  \href{http://dx.doi.org/10.1086/300410}{\JournalTitle{\aj}, 116, 1}, \eprint
  arXiv:{astro-ph/9804064}

\bibitem[{{Dalcanton} {et~al.}(2009){Dalcanton} \& {Williams}
  {et~al.}}]{Dalcanton:2009}
{Dalcanton}, J.~J., {Williams}, B.~F., {Seth}, A.~C., {et~al.} 2009,
  \href{http://dx.doi.org/10.1088/0067-0049/183/1/67}{\JournalTitle{\apjs},
  183, 67}, \eprint arXiv:{0905.3737}

\bibitem[{{Danieli} {et~al.}(2017){Danieli} \& {van Dokkum}
  {et~al.}}]{Danieli:2017}
{Danieli}, S., {van Dokkum}, P., {Merritt}, A., {et~al.} 2017,
  \href{http://dx.doi.org/10.3847/1538-4357/aa615b}{\JournalTitle{\apj}, 837,
  136}, \eprint arXiv:{1702.04727}

\bibitem[{{Darvish} {et~al.}(2018){Darvish} \& {Martin} \& {Gon{\c{c}}alves} \&
  {Mobasher} \& {Scoville} \& {Sobral}}]{Darvish:2018}
{Darvish}, B., {Martin}, C., {Gon{\c{c}}alves}, T.~S., {et~al.} 2018,
  \href{http://dx.doi.org/10.3847/1538-4357/aaa5a4}{\JournalTitle{\apj}, 853,
  155}, \eprint arXiv:{1801.02618}

\bibitem[{{Davis} {et~al.}(2021){Davis} \& {Nierenberg} {et~al.}}]{Davis:2021}
{Davis}, A.~B., {Nierenberg}, A.~M., {Peter}, A. H.~G., {et~al.} 2021,
  \href{http://dx.doi.org/10.1093/mnras/staa3246}{\JournalTitle{\mnras}, 500,
  3854}, \eprint arXiv:{2003.08352}

\bibitem[{{de Vaucouleurs} {et~al.}(1991){de Vaucouleurs} \& {de Vaucouleurs}
  \& {Corwin} \& {Buta} \& {Paturel} \& {Fouque}}]{deVaucouleurs:1991}
{de Vaucouleurs}, G., {de Vaucouleurs}, A., {Corwin}, Herold~G., J., {et~al.}
  1991, {Third Reference Catalogue of Bright Galaxies} (Springer)

\bibitem[{{Deason} {et~al.}(2015){Deason} \& {Wetzel} \& {Garrison-Kimmel} \&
  {Belokurov}}]{Deason:2015}
{Deason}, A.~J., {Wetzel}, A.~R., {Garrison-Kimmel}, S., \& {Belokurov}, V.
  2015, \href{http://dx.doi.org/10.1093/mnras/stv1939}{\JournalTitle{\mnras},
  453, 3568}, \eprint arXiv:{1504.04372}

\bibitem[{{DES Collaboration}(2005){DES Collaboration} \& Abbott \& Aldering \&
  Annis {et~al.}}]{DES:2005}
{DES Collaboration}. 2005, \eprint arXiv:{astro-ph/0510346}

\bibitem[{{DES Collaboration}(2018{\natexlab{a}}){DES Collaboration} \&
  {Abbott} \& {Abdalla} \& {Allam} {et~al.}}]{DES-DR1:2018}
{DES Collaboration}. 2018{\natexlab{a}},
  \href{http://dx.doi.org/10.3847/1538-4365/aae9f0}{\JournalTitle{\apjs}, 239,
  18}, \eprint arXiv:{1801.03181}

\bibitem[{{DES Collaboration}(2016){DES Collaboration} \& {Abbott}
  {et~al.}}]{DES:2016}
{DES Collaboration}. 2016,
  \href{http://dx.doi.org/10.1093/mnras/stw641}{\JournalTitle{\mnras}, 460,
  1270}, \eprint arXiv:{1601.00329}

\bibitem[{{DES Collaboration}(2018{\natexlab{b}}){DES Collaboration} \&
  {Abbott} {et~al.}}]{DES:2018}
{DES Collaboration}. 2018{\natexlab{b}},
  \href{http://dx.doi.org/10.1103/PhysRevD.98.043526}{\JournalTitle{\prd}, 98,
  043526}, \eprint arXiv:{1708.01530}

\bibitem[{{DES Collaboration}(2020){DES Collaboration} \& {Abbott}
  {et~al.}}]{DESY1ClusterCosmology}
{DES Collaboration}. 2020,
  \href{http://dx.doi.org/10.1103/PhysRevD.102.023509}{\JournalTitle{\prd},
  102, 023509}, \eprint arXiv:{2002.11124}

\bibitem[{{DES Collaboration}(2021){DES Collaboration} \& {Abbott}
  {et~al.}}]{DES-DR2:2021}
{DES Collaboration}. 2021, \JournalTitle{arXiv e-prints}, arXiv:2101.05765,
  \eprint arXiv:{2101.05765}

\bibitem[{{Desai} {et~al.}(2012){Desai} \& {Armstrong} {et~al.}}]{Desai:2012}
{Desai}, S., {Armstrong}, R., {Mohr}, J.~J., {et~al.} 2012,
  \href{http://dx.doi.org/10.1088/0004-637X/757/1/83}{\JournalTitle{\apj}, 757,
  83}, \eprint arXiv:{1204.1210}

\bibitem[{{Dey} {et~al.}(2019){Dey} \& {Schlegel} {et~al.}}]{Dey:2019}
{Dey}, A., {Schlegel}, D.~J., {Lang}, D., {et~al.} 2019,
  \href{http://dx.doi.org/10.3847/1538-3881/ab089d}{\JournalTitle{\aj}, 157,
  168}, \eprint arXiv:{1804.08657}

\bibitem[{{Diehl} {et~al.}(2017){Diehl} \& {Buckley-Geer}
  {et~al.}}]{Diehl:2017}
{Diehl}, H.~T., {Buckley-Geer}, E.~J., {Lindgren}, K.~A., {et~al.} 2017,
  \href{http://dx.doi.org/10.3847/1538-4365/aa8667}{\JournalTitle{\apjs}, 232,
  15}

\bibitem[{{D'Onghia} \& {Lake}(2008)}]{DOnghia:2008}
{D'Onghia}, E. \& {Lake}, G. 2008,
  \href{http://dx.doi.org/10.1086/592995}{\JournalTitle{\apjl}, 686, L61},
  \eprint arXiv:{0802.0001}

\bibitem[{{Dooley} {et~al.}(2017{\natexlab{a}}){Dooley} \& {Peter} \& {Carlin}
  \& {Frebel} \& {Bechtol} \& {Willman}}]{Dooley:2017a}
{Dooley}, G.~A., {Peter}, A. H.~G., {Carlin}, J.~L., {et~al.}
  2017{\natexlab{a}},
  \href{http://dx.doi.org/10.1093/mnras/stx2001}{\JournalTitle{\mnras}, 472,
  1060}, \eprint arXiv:{1703.05321}

\bibitem[{{Dooley} {et~al.}(2017{\natexlab{b}}){Dooley} \& {Peter} \& {Yang} \&
  {Willman} \& {Griffen} \& {Frebel}}]{Dooley:2017b}
{Dooley}, G.~A., {Peter}, A. H.~G., {Yang}, T., {et~al.} 2017{\natexlab{b}},
  \href{http://dx.doi.org/10.1093/mnras/stx1900}{\JournalTitle{\mnras}, 471,
  4894}, \eprint arXiv:{1610.00708}

\bibitem[{{Dotter} {et~al.}(2008){Dotter} \& {Chaboyer} \& {Jevremovi{\'c}} \&
  {Kostov} \& {Baron} \& {Ferguson}}]{Dotter:2008}
{Dotter}, A., {Chaboyer}, B., {Jevremovi{\'c}}, D., {et~al.} 2008,
  \href{http://dx.doi.org/10.1086/589654}{\JournalTitle{\apjs}, 178, 89},
  \eprint arXiv:{0804.4473}

\bibitem[{{Drlica-Wagner} {et~al.}(2015){Drlica-Wagner} \& {Bechtol}
  {et~al.}}]{Drlica-Wagner:2015}
{Drlica-Wagner}, A., {Bechtol}, K., {Rykoff}, E.~S., {et~al.} 2015,
  \href{http://dx.doi.org/10.1088/0004-637X/813/2/109}{\JournalTitle{\apj},
  813, 109}, \eprint arXiv:{1508.03622}

\bibitem[{{Drlica-Wagner} {et~al.}(2016){Drlica-Wagner} \& {Bechtol}
  {et~al.}}]{Drlica-Wagner:2016}
{Drlica-Wagner}, A., {Bechtol}, K., {Allam}, S., {et~al.} 2016,
  \href{http://dx.doi.org/10.3847/2041-8205/833/1/L5}{\JournalTitle{\apjl},
  833, L5}, \eprint arXiv:{1609.02148}

\bibitem[{{Drlica-Wagner} {et~al.}(2018){Drlica-Wagner} \& {Sevilla-Noarbe}
  {et~al.}}]{Drlica-Wagner:2018}
{Drlica-Wagner}, A., {Sevilla-Noarbe}, I., {Rykoff}, E.~S., {et~al.} 2018,
  \href{http://dx.doi.org/10.3847/1538-4365/aab4f5}{\JournalTitle{\apjs}, 235,
  33}, \eprint arXiv:{1708.01531}

\bibitem[{{Drlica-Wagner} {et~al.}(2020){Drlica-Wagner} \& {Bechtol}
  {et~al.}}]{Drlica-Wagner:2020}
{Drlica-Wagner}, A., {Bechtol}, K., {Mau}, S., {et~al.} 2020,
  \href{http://dx.doi.org/10.3847/1538-4357/ab7eb9}{\JournalTitle{\apj}, 893,
  47}, \eprint arXiv:{1912.03302}

\bibitem[{{Erkal} {et~al.}(2016{\natexlab{a}}){Erkal} \& {Belokurov} \& {Bovy}
  \& {Sanders}}]{Erkal:2016b}
{Erkal}, D., {Belokurov}, V., {Bovy}, J., \& {Sanders}, J.~L.
  2016{\natexlab{a}},
  \href{http://dx.doi.org/10.1093/mnras/stw1957}{\JournalTitle{\mnras}, 463,
  102}, \eprint arXiv:{1606.04946}

\bibitem[{{Erkal} \& {Belokurov}(2020)}]{Erkal:2020}
{Erkal}, D. \& {Belokurov}, V.~A. 2020,
  \href{http://dx.doi.org/10.1093/mnras/staa1238}{\JournalTitle{\mnras}, 495,
  2554}, \eprint arXiv:{1907.09484}

\bibitem[{{Erkal} {et~al.}(2020){Erkal} \& {Belokurov} \&
  {Parkin}}]{Erkal:2020b}
{Erkal}, D., {Belokurov}, V.~A., \& {Parkin}, D.~L. 2020,
  \href{http://dx.doi.org/10.1093/mnras/staa2840}{\JournalTitle{\mnras}, 498,
  5574}, \eprint arXiv:{2001.11030}

\bibitem[{{Erkal} {et~al.}(2016{\natexlab{b}}){Erkal} \& {Sanders} \&
  {Belokurov}}]{Erkal:2016}
{Erkal}, D., {Sanders}, J.~L., \& {Belokurov}, V. 2016{\natexlab{b}},
  \href{http://dx.doi.org/10.1093/mnras/stw1400}{\JournalTitle{\mnras}, 461,
  1590}, \eprint arXiv:{1603.08922}

\bibitem[{{Fingerhut} {et~al.}(2010){Fingerhut} \& {McCall}
  {et~al.}}]{Fingerhut:2010}
{Fingerhut}, R.~L., {McCall}, M.~L., {Argote}, M., {et~al.} 2010,
  \href{http://dx.doi.org/10.1088/0004-637X/716/1/792}{\JournalTitle{\apj},
  716, 792}

\bibitem[{{Fitzpatrick} {et~al.}(2016){Fitzpatrick} \& {Graham}
  {et~al.}}]{Fitzpatrick:2016}
{Fitzpatrick}, M.~J., {Graham}, M.~J., {Mighell}, K.~J., {et~al.} 2016, Society
  of Photo-Optical Instrumentation Engineers (SPIE) Conference Series, Vol.
  9913, {The NOAO data lab: science-driven development}, ed. G.~{Chiozzi} \&
  J.~C. {Guzman}, 99130L

\bibitem[{{Flaugher} {et~al.}(2015){Flaugher} \& {Diehl}
  {et~al.}}]{Flaugher:2015}
{Flaugher}, B., {Diehl}, H.~T., {Honscheid}, K., {et~al.} 2015,
  \href{http://dx.doi.org/10.1088/0004-6256/150/5/150}{\JournalTitle{\aj}, 150,
  150}, \eprint arXiv:{1504.02900}

\bibitem[{{Frebel} \& {Norris}(2015)}]{Frebel:2015}
{Frebel}, A. \& {Norris}, J.~E. 2015,
  \href{http://dx.doi.org/10.1146/annurev-astro-082214-122423}{\JournalTitle{\araa},
  53, 631}, \eprint arXiv:{1501.06921}

\bibitem[{{Freedman} {et~al.}(2001){Freedman} \& {Madore}
  {et~al.}}]{Freedman:2001}
{Freedman}, W.~L., {Madore}, B.~F., {Gibson}, B.~K., {et~al.} 2001,
  \href{http://dx.doi.org/10.1086/320638}{\JournalTitle{\apj}, 553, 47},
  \eprint arXiv:{astro-ph/0012376}

\bibitem[{{Gaia Collaboration}(2018{\natexlab{a}}){Gaia Collaboration} \&
  {Brown} {et~al.}}]{Gaia:2018}
{Gaia Collaboration}. 2018{\natexlab{a}},
  \href{http://dx.doi.org/10.1051/0004-6361/201833051}{\JournalTitle{\aap},
  616, A1}, \eprint arXiv:{1804.09365}

\bibitem[{{Gaia Collaboration}(2018{\natexlab{b}}){Gaia Collaboration} \&
  {Mignard} {et~al.}}]{Gaia:2018b}
{Gaia Collaboration}. 2018{\natexlab{b}},
  \href{http://dx.doi.org/10.1051/0004-6361/201832916}{\JournalTitle{\aap},
  616, A14}, \eprint arXiv:{1804.09377}

\bibitem[{{Gaia Collaboration}(2021{\natexlab{a}}){Gaia Collaboration} \&
  {Luri} {et~al.}}]{Gaia:2020b}
{Gaia Collaboration}. 2021{\natexlab{a}},
  \href{http://dx.doi.org/10.1051/0004-6361/202039588}{\JournalTitle{\aap},
  649, A7}, \eprint arXiv:{2012.01771}

\bibitem[{{Gaia Collaboration}(2021{\natexlab{b}}){Gaia Collaboration} \&
  {Brown} {et~al.}}]{Gaia:2020a}
{Gaia Collaboration}. 2021{\natexlab{b}},
  \href{http://dx.doi.org/10.1051/0004-6361/202039657}{\JournalTitle{\aap},
  649, A1}, \eprint arXiv:{2012.01533}

\bibitem[{{Garavito-Camargo} {et~al.}(2019){Garavito-Camargo} \& {Besla} \&
  {Laporte} \& {Johnston} \& {G{\'o}mez} \& {Watkins}}]{Garavito-Camargo:2019}
{Garavito-Camargo}, N., {Besla}, G., {Laporte}, C. F.~P., {et~al.} 2019,
  \href{http://dx.doi.org/10.3847/1538-4357/ab32eb}{\JournalTitle{\apj}, 884,
  51}, \eprint arXiv:{1902.05089}

\bibitem[{{Garling} {et~al.}(2020){Garling} \& {Peter} \& {Kochanek} \& {Sand}
  \& {Crnojevi{\'c}}}]{Garling:2020}
{Garling}, C.~T., {Peter}, A. H.~G., {Kochanek}, C.~S., {Sand}, D.~J., \&
  {Crnojevi{\'c}}, D. 2020,
  \href{http://dx.doi.org/10.1093/mnras/stz3526}{\JournalTitle{\mnras}, 492,
  1713}, \eprint arXiv:{1908.11367}

\bibitem[{{Garrison-Kimmel} {et~al.}(2014){Garrison-Kimmel} \& {Boylan-Kolchin}
  \& {Bullock} \& {Lee}}]{Garrison-Kimmel:2014}
{Garrison-Kimmel}, S., {Boylan-Kolchin}, M., {Bullock}, J.~S., \& {Lee}, K.
  2014, \href{http://dx.doi.org/10.1093/mnras/stt2377}{\JournalTitle{\mnras},
  438, 2578}, \eprint arXiv:{1310.6746}

\bibitem[{{Garrison-Kimmel} {et~al.}(2017){Garrison-Kimmel} \& {Wetzel}
  {et~al.}}]{Garrison-Kimmel:2017}
{Garrison-Kimmel}, S., {Wetzel}, A., {Bullock}, J.~S., {et~al.} 2017,
  \href{http://dx.doi.org/10.1093/mnras/stx1710}{\JournalTitle{\mnras}, 471,
  1709}, \eprint arXiv:{1701.03792}

\bibitem[{{Geha} {et~al.}(2017){Geha} \& {Wechsler} {et~al.}}]{Geha:2017}
{Geha}, M., {Wechsler}, R.~H., {Mao}, Y.-Y., {et~al.} 2017,
  \href{http://dx.doi.org/10.3847/1538-4357/aa8626}{\JournalTitle{\apj}, 847,
  4}, \eprint arXiv:{1705.06743}

\bibitem[{{Geringer-Sameth} {et~al.}(2015){Geringer-Sameth} \& {Koushiappas} \&
  {Walker}}]{Geringer-Sameth:2015}
{Geringer-Sameth}, A., {Koushiappas}, S.~M., \& {Walker}, M.~G. 2015,
  \href{http://dx.doi.org/10.1103/PhysRevD.91.083535}{\JournalTitle{\prd}, 91,
  083535}, \eprint arXiv:{1410.2242}

\bibitem[{Gilman {et~al.}(2019)Gilman \& Birrer \& Treu \& Nierenberg \&
  Benson}]{Gilman:2019}
Gilman, D., Birrer, S., Treu, T., Nierenberg, A., \& Benson, A. 2019,
  \href{http://dx.doi.org/10.1093/mnras/stz1593}{\JournalTitle{Monthly Notices
  of the Royal Astronomical Society}, 487, 5721–5738}

\bibitem[{{G{\'o}rski} {et~al.}(2005){G{\'o}rski} \& {Hivon}
  {et~al.}}]{Gorski:2005}
{G{\'o}rski}, K.~M., {Hivon}, E., {Banday}, A.~J., {et~al.} 2005,
  \href{http://dx.doi.org/10.1086/427976}{\JournalTitle{\apj}, 622, 759},
  \eprint{astro-ph/0409513}

\bibitem[{Gough(2009)}]{Gough:2009}
Gough, B. 2009, GNU Scientific Library Reference Manual - Third Edition, 3rd
  edn. (Network Theory Ltd.)

\bibitem[{{Grandis} {et~al.}(2021){Grandis} \& {Mohr} {et~al.}}]{Grandis2021}
{Grandis}, S., {Mohr}, J.~J., {Costanzi}, M., {et~al.} 2021,
  \JournalTitle{arXiv e-prints}, arXiv:2101.04984, \eprint arXiv:{2101.04984}

\bibitem[{{Graus} {et~al.}(2019){Graus} \& {Bullock} \& {Kelley} \&
  {Boylan-Kolchin} \& {Garrison-Kimmel} \& {Qi}}]{Graus:2019}
{Graus}, A.~S., {Bullock}, J.~S., {Kelley}, T., {et~al.} 2019,
  \href{http://dx.doi.org/10.1093/mnras/stz1992}{\JournalTitle{\mnras}, 488,
  4585}, \eprint arXiv:{1808.03654}

\bibitem[{{Grillmair}(2006)}]{Grillmair:2006}
{Grillmair}, C.~J. 2006,
  \href{http://dx.doi.org/10.1086/505863}{\JournalTitle{\apjl}, 645, L37},
  \eprint{astro-ph/0605396}

\bibitem[{{Grillmair}(2017)}]{Grillmair:2017}
{Grillmair}, C.~J. 2017,
  \href{http://dx.doi.org/10.3847/1538-4357/aa8872}{\JournalTitle{\apj}, 847,
  119}, \eprint arXiv:{1708.09029}

\bibitem[{{Hargis} {et~al.}(2014){Hargis} \& {Willman} \&
  {Peter}}]{Hargis:2014}
{Hargis}, J.~R., {Willman}, B., \& {Peter}, A. H.~G. 2014,
  \href{http://dx.doi.org/10.1088/2041-8205/795/1/L13}{\JournalTitle{\apjl},
  795, L13}, \eprint arXiv:{1407.4470}

\bibitem[{{Harris} {et~al.}(2020){Harris} \& {Millman} {et~al.}}]{NumPy:2020}
{Harris}, C.~R., {Millman}, K.~J., {van der Walt}, S.~J., {et~al.} 2020,
  \href{http://dx.doi.org/10.1038/s41586-020-2649-2}{\JournalTitle{\nat}, 585,
  357}, \eprint arXiv:{2006.10256}

\bibitem[{{Hartley} {et~al.}(2020){Hartley} \& {Choi}
  {et~al.}}]{HartleyChoi:2020}
{Hartley}, W.~G., {Choi}, A., {et~al.} 2020, \JournalTitle{arXiv e-prints},
  arXiv:2012.12824, \eprint arXiv:{2012.12824}

\bibitem[{Hennig {et~al.}(2017)Hennig \& Mohr {et~al.}}]{Hennig2017}
Hennig, C., Mohr, J.~J., Zenteno, A., {et~al.} 2017,
  \href{http://dx.doi.org/10.1093/mnras/stx175}{\JournalTitle{Monthly Notices
  of the Royal Astronomical Society}, 467, 4015},
  \eprint{https://academic.oup.com/mnras/article-pdf/467/4/4015/10914049/stx175.pdf}

\bibitem[{{Hildebrandt} {et~al.}(2020){Hildebrandt} \& {K{\"o}hlinger}
  {et~al.}}]{kidsviking}
{Hildebrandt}, H., {K{\"o}hlinger}, F., {van den Busch}, J.~L., {et~al.} 2020,
  \href{http://dx.doi.org/10.1051/0004-6361/201834878}{\JournalTitle{\aap},
  633, A69}, \eprint arXiv:{1812.06076}

\bibitem[{{Hillis} {et~al.}(2016){Hillis} \& {Williams} \& {Dolphin} \&
  {Dalcanton} \& {Skillman}}]{Hillis:2016}
{Hillis}, T.~J., {Williams}, B.~F., {Dolphin}, A.~E., {Dalcanton}, J.~J., \&
  {Skillman}, E.~D. 2016,
  \href{http://dx.doi.org/10.3847/0004-637X/831/2/191}{\JournalTitle{\apj},
  831, 191}, \eprint arXiv:{1609.02106}

\bibitem[{{Hilton} {et~al.}(2021){Hilton} \& {Sif{\'o}n}
  {et~al.}}]{Hilton:2020}
{Hilton}, M., {Sif{\'o}n}, C., {Naess}, S., {et~al.} 2021,
  \href{http://dx.doi.org/10.3847/1538-4365/abd023}{\JournalTitle{\apjs}, 253,
  3}, \eprint arXiv:{2009.11043}

\bibitem[{Huang {et~al.}(2020)Huang \& Storfer {et~al.}}]{Huang:2020}
Huang, X., Storfer, C., Ravi, V., {et~al.} 2020,
  \href{http://dx.doi.org/10.3847/1538-4357/ab7ffb}{\JournalTitle{The
  Astrophysical Journal}, 894, 78}

\bibitem[{Huang {et~al.}(2021)Huang \& Storfer {et~al.}}]{Huang:2021}
Huang, X., Storfer, C., Gu, A., {et~al.} 2021, Discovering New Strong
  Gravitational Lenses in the DESI Legacy Imaging Surveys, \eprint
  arXiv:{2005.04730}

\bibitem[{{Huchra} {et~al.}(2012){Huchra} \& {Macri} {et~al.}}]{Huchra:2012}
{Huchra}, J.~P., {Macri}, L.~M., {Masters}, K.~L., {et~al.} 2012,
  \href{http://dx.doi.org/10.1088/0067-0049/199/2/26}{\JournalTitle{\apjs},
  199, 26}, \eprint arXiv:{1108.0669}

\bibitem[{Hunter(2007)}]{Hunter:2007}
Hunter, J.~D. 2007,
  \href{http://dx.doi.org/10.1109/MCSE.2007.55}{\JournalTitle{Computing In
  Science \& Engineering}, 9, 90}

\bibitem[{{Ibata} {et~al.}(2001{\natexlab{a}}){Ibata} \& {Irwin} \& {Lewis} \&
  {Ferguson} \& {Tanvir}}]{Ibata:2001b}
{Ibata}, R., {Irwin}, M., {Lewis}, G., {Ferguson}, A. M.~N., \& {Tanvir}, N.
  2001{\natexlab{a}}, \JournalTitle{\nat}, 412, 49, \eprint
  arXiv:{astro-ph/0107090}

\bibitem[{{Ibata} {et~al.}(2001{\natexlab{b}}){Ibata} \& {Lewis} \& {Irwin} \&
  {Totten} \& {Quinn}}]{Ibata:2001a}
{Ibata}, R., {Lewis}, G.~F., {Irwin}, M., {Totten}, E., \& {Quinn}, T.
  2001{\natexlab{b}},
  \href{http://dx.doi.org/10.1086/320060}{\JournalTitle{\apj}, 551, 294},
  \eprint arXiv:{astro-ph/0004011}

\bibitem[{{Irwin} {et~al.}(1990){Irwin} \& {Bunclark} \& {Bridgeland} \&
  {McMahon}}]{1990MNRAS.244P..16I}
{Irwin}, M.~J., {Bunclark}, P.~S., {Bridgeland}, M.~T., \& {McMahon}, R.~G.
  1990, \JournalTitle{\mnras}, 244, 16P

\bibitem[{{Ishiyama} {et~al.}(2016){Ishiyama} \& {Sudo} \& {Yokoi} \&
  {Hasegawa} \& {Tominaga} \& {Susa}}]{Ishiyama:2016}
{Ishiyama}, T., {Sudo}, K., {Yokoi}, S., {et~al.} 2016,
  \href{http://dx.doi.org/10.3847/0004-637X/826/1/9}{\JournalTitle{\apj}, 826,
  9}, \eprint arXiv:{1602.00465}

\bibitem[{{Jacobs} {et~al.}(2009){Jacobs} \& {Rizzi} \& {Tully} \& {Shaya} \&
  {Makarov} \& {Makarova}}]{Jacobs:2009}
{Jacobs}, B.~A., {Rizzi}, L., {Tully}, R.~B., {et~al.} 2009,
  \href{http://dx.doi.org/10.1088/0004-6256/138/2/332}{\JournalTitle{\aj}, 138,
  332}, \eprint arXiv:{0902.3675}

\bibitem[{Jacobs {et~al.}(2019{\natexlab{a}})Jacobs \& Collett
  {et~al.}}]{Jacobs:2019b}
Jacobs, C., Collett, T., Glazebrook, K., {et~al.} 2019{\natexlab{a}},
  \href{http://dx.doi.org/10.3847/1538-4365/ab26b6}{\JournalTitle{The
  Astrophysical Journal Supplement Series}, 243, 17}

\bibitem[{Jacobs {et~al.}(2019{\natexlab{b}})Jacobs \& Collett
  {et~al.}}]{Jacobs:2019a}
Jacobs, C., Collett, T., Glazebrook, K., {et~al.} 2019{\natexlab{b}},
  \href{http://dx.doi.org/10.1093/mnras/stz272}{\JournalTitle{Monthly Notices
  of the Royal Astronomical Society}, 484, 5330–5349}

\bibitem[{{Jahn} {et~al.}(2019){Jahn} \& {Sales} {et~al.}}]{Jahn:2019}
{Jahn}, E.~D., {Sales}, L.~V., {Wetzel}, A., {et~al.} 2019,
  \href{http://dx.doi.org/10.1093/mnras/stz2457}{\JournalTitle{\mnras}, 489,
  5348}, \eprint arXiv:{1907.02979}

\bibitem[{{Jang} {et~al.}(2020){Jang} \& {de Jong} {et~al.}}]{Jang:2020}
{Jang}, I.~S., {de Jong}, R.~S., {Minchev}, I., {et~al.} 2020,
  \href{http://dx.doi.org/10.1051/0004-6361/202038651}{\JournalTitle{\aap},
  640, L19}, \eprint arXiv:{2007.13749}

\bibitem[{{Jethwa} {et~al.}(2016){Jethwa} \& {Erkal} \&
  {Belokurov}}]{Jethwa:2016}
{Jethwa}, P., {Erkal}, D., \& {Belokurov}, V. 2016,
  \href{http://dx.doi.org/10.1093/mnras/stw1343}{\JournalTitle{\mnras}, 461,
  2212}, \eprint arXiv:{1603.04420}

\bibitem[{{Jethwa} {et~al.}(2018{\natexlab{a}}){Jethwa} \& {Erkal} \&
  {Belokurov}}]{Jethwa:2018}
{Jethwa}, P., {Erkal}, D., \& {Belokurov}, V. 2018{\natexlab{a}},
  \href{http://dx.doi.org/10.1093/mnras/stx2330}{\JournalTitle{\mnras}, 473,
  2060}, \eprint arXiv:{1612.07834}

\bibitem[{{Jethwa} {et~al.}(2018{\natexlab{b}}){Jethwa} \& {Torrealba}
  {et~al.}}]{Jethwa:2018b}
{Jethwa}, P., {Torrealba}, G., {Navarrete}, C., {et~al.} 2018{\natexlab{b}},
  \href{http://dx.doi.org/10.1093/mnras/sty2226}{\JournalTitle{\mnras}, 480,
  5342}, \eprint arXiv:{1711.09103}

\bibitem[{{Ji} {et~al.}(2016){Ji} \& {Frebel} \& {Chiti} \& {Simon}}]{Ji:2016}
{Ji}, A.~P., {Frebel}, A., {Chiti}, A., \& {Simon}, J.~D. 2016,
  \href{http://dx.doi.org/10.1038/nature17425}{\JournalTitle{\nat}, 531, 610},
  \eprint arXiv:{1512.01558}

\bibitem[{{Johnston} {et~al.}(2002){Johnston} \& {Spergel} \&
  {Haydn}}]{Johnston:2002}
{Johnston}, K.~V., {Spergel}, D.~N., \& {Haydn}, C. 2002,
  \href{http://dx.doi.org/10.1086/339791}{\JournalTitle{\apj}, 570, 656},
  \eprint arXiv:{astro-ph/0111196}

\bibitem[{{Johnston} {et~al.}(1999){Johnston} \& {Zhao} \& {Spergel} \&
  {Hernquist}}]{Johnston:1999}
{Johnston}, K.~V., {Zhao}, H., {Spergel}, D.~N., \& {Hernquist}, L. 1999,
  \href{http://dx.doi.org/10.1086/311876}{\JournalTitle{\apjl}, 512, L109},
  \eprint arXiv:{astro-ph/9807243}

\bibitem[{{Kallivayalil} {et~al.}(2018){Kallivayalil} \& {Sales}
  {et~al.}}]{Kallivayalil:2018}
{Kallivayalil}, N., {Sales}, L.~V., {Zivick}, P., {et~al.} 2018,
  \href{http://dx.doi.org/10.3847/1538-4357/aadfee}{\JournalTitle{\apj}, 867,
  19}, \eprint arXiv:{1805.01448}

\bibitem[{{Karachentsev} {et~al.}(2013){Karachentsev} \& {Makarov} \&
  {Kaisina}}]{Karachentsev:2013}
{Karachentsev}, I.~D., {Makarov}, D.~I., \& {Kaisina}, E.~I. 2013,
  \href{http://dx.doi.org/10.1088/0004-6256/145/4/101}{\JournalTitle{\aj}, 145,
  101}, \eprint arXiv:{1303.5328}

\bibitem[{{Karachentsev} {et~al.}(2002){Karachentsev} \& {Sharina}
  {et~al.}}]{Karachentsev:2002}
{Karachentsev}, I.~D., {Sharina}, M.~E., {Makarov}, D.~I., {et~al.} 2002,
  \href{http://dx.doi.org/10.1051/0004-6361:20020649}{\JournalTitle{\aap}, 389,
  812}, \eprint arXiv:{astro-ph/0204507}

\bibitem[{{Karachentsev} {et~al.}(2003){Karachentsev} \& {Grebel}
  {et~al.}}]{Karachentsev:2003}
{Karachentsev}, I.~D., {Grebel}, E.~K., {Sharina}, M.~E., {et~al.} 2003,
  \href{http://dx.doi.org/10.1051/0004-6361:20030170}{\JournalTitle{\aap}, 404,
  93}, \eprint arXiv:{astro-ph/0302045}

\bibitem[{{Karunakaran} {et~al.}(2020){Karunakaran} \& {Spekkens} \& {Zaritsky}
  \& {Donnerstein} \& {Kadowaki} \& {Dey}}]{Karunakaran:2020}
{Karunakaran}, A., {Spekkens}, K., {Zaritsky}, D., {et~al.} 2020,
  \href{http://dx.doi.org/10.3847/1538-4357/abb464}{\JournalTitle{\apj}, 902,
  39}, \eprint arXiv:{2005.14202}

\bibitem[{{Katz} {et~al.}(2020){Katz} \& {Ramsoy} {et~al.}}]{Katz:2019}
{Katz}, H., {Ramsoy}, M., {Rosdahl}, J., {et~al.} 2020,
  \href{http://dx.doi.org/10.1093/mnras/staa639}{\JournalTitle{\mnras}, 494,
  2200}, \eprint arXiv:{1905.11414}

\bibitem[{{Kim} {et~al.}(2016){Kim} \& {Jerjen} \& {Mackey} \& {Da Costa} \&
  {Milone}}]{2016ApJ...820..119K}
{Kim}, D., {Jerjen}, H., {Mackey}, D., {Da Costa}, G.~S., \& {Milone}, A.~P.
  2016,
  \href{http://dx.doi.org/10.3847/0004-637X/820/2/119}{\JournalTitle{\apj},
  820, 119}, \eprint arXiv:{1512.03530}

\bibitem[{{Kim} {et~al.}(2018){Kim} \& {Peter} \& {Hargis}}]{Kim:2018}
{Kim}, S.~Y., {Peter}, A. H.~G., \& {Hargis}, J.~R. 2018,
  \href{http://dx.doi.org/10.1103/PhysRevLett.121.211302}{\JournalTitle{\prl},
  121, 211302}, \eprint arXiv:{1711.06267}

\bibitem[{{Koopmans}(2005)}]{Koopmans:2005}
{Koopmans}, L.~V.~E. 2005,
  \href{http://dx.doi.org/10.1111/j.1365-2966.2005.09523.x}{\JournalTitle{\mnras},
  363, 1136}, \eprint arXiv:{astro-ph/0501324}

\bibitem[{{Koposov} {et~al.}(2015){Koposov} \& {Belokurov} \& {Torrealba} \&
  {Evans}}]{Koposov:2015}
{Koposov}, S.~E., {Belokurov}, V., {Torrealba}, G., \& {Evans}, N.~W. 2015,
  \href{http://dx.doi.org/10.1088/0004-637X/805/2/130}{\JournalTitle{\apj},
  805, 130}, \eprint arXiv:{1503.02079}

\bibitem[{{Koposov} {et~al.}(2014){Koposov} \& {Irwin} {et~al.}}]{Koposov:2014}
{Koposov}, S.~E., {Irwin}, M., {Belokurov}, V., {et~al.} 2014,
  \href{http://dx.doi.org/10.1093/mnrasl/slu060}{\JournalTitle{\mnras}, 442,
  L85}, \eprint arXiv:{1403.3409}

\bibitem[{{Koposov} {et~al.}(2018){Koposov} \& {Walker}
  {et~al.}}]{Koposov:2018}
{Koposov}, S.~E., {Walker}, M.~G., {Belokurov}, V., {et~al.} 2018,
  \href{http://dx.doi.org/10.1093/mnras/sty1772}{\JournalTitle{\mnras}, 479,
  5343}, \eprint arXiv:{1804.06430}

\bibitem[{{Laevens} {et~al.}(2014){Laevens} \& {Martin}
  {et~al.}}]{Laevens:2014a}
{Laevens}, B.~P.~M., {Martin}, N.~F., {Sesar}, B., {et~al.} 2014,
  \href{http://dx.doi.org/10.1088/2041-8205/786/1/L3}{\JournalTitle{\apjl},
  786, L3}, \eprint arXiv:{1403.6593}

\bibitem[{{Lancaster} {et~al.}(2020){Lancaster} \& {Giovanetti} \& {Mocz} \&
  {Kahn} \& {Lisanti} \& {Spergel}}]{Lancaster:2020}
{Lancaster}, L., {Giovanetti}, C., {Mocz}, P., {et~al.} 2020,
  \href{http://dx.doi.org/10.1088/1475-7516/2020/01/001}{\JournalTitle{\jcap},
  2020, 001}, \eprint arXiv:{1909.06381}

\bibitem[{{Li} {et~al.}(2019){Li} \& {Koposov} {et~al.}}]{Li:2019}
{Li}, T.~S., {Koposov}, S.~E., {Zucker}, D.~B., {et~al.} 2019,
  \href{http://dx.doi.org/10.1093/mnras/stz2731}{\JournalTitle{\mnras}, 490,
  3508}, \eprint arXiv:{1907.09481}

\bibitem[{{Li} {et~al.}(2021){Li} \& {Koposov} {et~al.}}]{Li:2020}
{Li}, T.~S., {Koposov}, S.~E., {Erkal}, D., {et~al.} 2021,
  \href{http://dx.doi.org/10.3847/1538-4357/abeb18}{\JournalTitle{\apj}, 911,
  149}, \eprint arXiv:{2006.10763}

\bibitem[{{Limousin} {et~al.}(2007){Limousin} \& {Richard}
  {et~al.}}]{Limousin2007}
{Limousin}, M., {Richard}, J., {Jullo}, E., {et~al.} 2007,
  \href{http://dx.doi.org/10.1086/521293}{\JournalTitle{\apj}, 668, 643},
  \eprint arXiv:{astro-ph/0612165}

\bibitem[{{Lovisari} {et~al.}(2020{\natexlab{a}}){Lovisari} \& {Ettori}
  {et~al.}}]{Lovisari:2020a}
{Lovisari}, L., {Ettori}, S., {Sereno}, M., {et~al.} 2020{\natexlab{a}},
  \href{http://dx.doi.org/10.1051/0004-6361/202038718}{\JournalTitle{\aap},
  644, A78}, \eprint arXiv:{2010.03582}

\bibitem[{{Lovisari} {et~al.}(2020{\natexlab{b}}){Lovisari} \& {Schellenberger}
  {et~al.}}]{Lovisari:2020b}
{Lovisari}, L., {Schellenberger}, G., {Sereno}, M., {et~al.}
  2020{\natexlab{b}},
  \href{http://dx.doi.org/10.3847/1538-4357/ab7997}{\JournalTitle{\apj}, 892,
  102}, \eprint arXiv:{2002.11740}

\bibitem[{{Lucey} {et~al.}(2018){Lucey} \& {Smith} \& {Schechter} \& {Bosh} \&
  {Levine}}]{Lucey:2018}
{Lucey}, J.~R., {Smith}, R.~J., {Schechter}, P.~L., {Bosh}, A.~S., \& {Levine},
  S.~E. 2018,
  \href{http://dx.doi.org/10.3847/2515-5172/aacea8}{\JournalTitle{Research
  Notes of the American Astronomical Society}, 2, 62}, \eprint
  arXiv:{1806.06861}

\bibitem[{{Lynden-Bell}(1976)}]{Lynden-Bell:1976}
{Lynden-Bell}, D. 1976,
  \href{http://dx.doi.org/10.1093/mnras/174.3.695}{\JournalTitle{\mnras}, 174,
  695}

\bibitem[{{Mackey} {et~al.}(2018){Mackey} \& {Koposov} \& {Da Costa} \&
  {Belokurov} \& {Erkal} \& {Kuzma}}]{Mackey:2018}
{Mackey}, D., {Koposov}, S., {Da Costa}, G., {et~al.} 2018,
  \href{http://dx.doi.org/10.3847/2041-8213/aac175}{\JournalTitle{\apjl}, 858,
  L21}, \eprint arXiv:{1804.06431}

\bibitem[{{Malin} \& {Hadley}(1997)}]{Malin:1997}
{Malin}, D. \& {Hadley}, B. 1997,
  \href{http://dx.doi.org/10.1071/AS97052}{\JournalTitle{\pasa}, 14, 52}

\bibitem[{{Malyshev} {et~al.}(2014){Malyshev} \& {Neronov} \&
  {Eckert}}]{Malyshev:2014}
{Malyshev}, D., {Neronov}, A., \& {Eckert}, D. 2014,
  \href{http://dx.doi.org/10.1103/PhysRevD.90.103506}{\JournalTitle{\prd}, 90,
  103506}, \eprint arXiv:{1408.3531}

\bibitem[{{Mao} {et~al.}(2021){Mao} \& {Geha} {et~al.}}]{Mao:2020}
{Mao}, Y.-Y., {Geha}, M., {Wechsler}, R.~H., {et~al.} 2021,
  \href{http://dx.doi.org/10.3847/1538-4357/abce58}{\JournalTitle{\apj}, 907,
  85}, \eprint arXiv:{2008.12783}

\bibitem[{{Martin} {et~al.}(2013){Martin} \& {Ibata} {et~al.}}]{Martin:2013}
{Martin}, N.~F., {Ibata}, R.~A., {McConnachie}, A.~W., {et~al.} 2013,
  \href{http://dx.doi.org/10.1088/0004-637X/776/2/80}{\JournalTitle{\apj}, 776,
  80}, \eprint arXiv:{1307.7626}

\bibitem[{{Martin} {et~al.}(2015){Martin} \& {Nidever} {et~al.}}]{Martin:2015}
{Martin}, N.~F., {Nidever}, D.~L., {Besla}, G., {et~al.} 2015,
  \href{http://dx.doi.org/10.1088/2041-8205/804/1/L5}{\JournalTitle{\apjl},
  804, L5}, \eprint arXiv:{1503.06216}

\bibitem[{{Mart{\'\i}nez-Delgado} {et~al.}(2008){Mart{\'\i}nez-Delgado} \&
  {Pe{\~n}arrubia} \& {Gabany} \& {Trujillo} \& {Majewski} \&
  {Pohlen}}]{Martinez-Delgado:2008}
{Mart{\'\i}nez-Delgado}, D., {Pe{\~n}arrubia}, J., {Gabany}, R.~J., {et~al.}
  2008, \href{http://dx.doi.org/10.1086/592555}{\JournalTitle{\apj}, 689, 184},
  \eprint arXiv:{0805.1137}

\bibitem[{{Mart{\'\i}nez-Delgado} {et~al.}(2010){Mart{\'\i}nez-Delgado} \&
  {Gabany} {et~al.}}]{Martinez-Delgado:2010}
{Mart{\'\i}nez-Delgado}, D., {Gabany}, R.~J., {Crawford}, K., {et~al.} 2010,
  \href{http://dx.doi.org/10.1088/0004-6256/140/4/962}{\JournalTitle{\aj}, 140,
  962}, \eprint arXiv:{1003.4860}

\bibitem[{{Martinez-Delgado} {et~al.}(2021){Martinez-Delgado} \& {Cooper}
  {et~al.}}]{Martinez-Delgado:2021}
{Martinez-Delgado}, D., {Cooper}, A.~P., {Roman}, J., {et~al.} 2021,
  \JournalTitle{arXiv e-prints}, arXiv:2104.06071, \eprint arXiv:{2104.06071}

\bibitem[{{Mashchenko} {et~al.}(2008){Mashchenko} \& {Wadsley} \&
  {Couchman}}]{Maschenko:2007}
{Mashchenko}, S., {Wadsley}, J., \& {Couchman}, H.~M.~P. 2008,
  \href{http://dx.doi.org/10.1126/science.1148666}{\JournalTitle{Science}, 319,
  174}, \eprint arXiv:{0711.4803}

\bibitem[{{Massana} {et~al.}(2020){Massana} \& {No{\"e}l}
  {et~al.}}]{Massana:2020}
{Massana}, P., {No{\"e}l}, N. E.~D., {Nidever}, D.~L., {et~al.} 2020,
  \href{http://dx.doi.org/10.1093/mnras/staa2451}{\JournalTitle{\mnras}, 498,
  1034}, \eprint arXiv:{2008.00012}

\bibitem[{{Mau} {et~al.}(2019){Mau} \& {Drlica-Wagner} {et~al.}}]{Mau:2019}
{Mau}, S., {Drlica-Wagner}, A., {Bechtol}, K., {et~al.} 2019,
  \href{http://dx.doi.org/10.3847/1538-4357/ab0bb8}{\JournalTitle{\apj}, 875,
  154}, \eprint arXiv:{1812.06318}

\bibitem[{{Mau} {et~al.}(2020){Mau} \& {Cerny} {et~al.}}]{Mau:2020}
{Mau}, S., {Cerny}, W., {Pace}, A.~B., {et~al.} 2020,
  \href{http://dx.doi.org/10.3847/1538-4357/ab6c67}{\JournalTitle{\apj}, 890,
  136}, \eprint arXiv:{1912.03301}

\bibitem[{{McClintock} {et~al.}(2019){McClintock} \& {Varga}
  {et~al.}}]{McClintock2019}
{McClintock}, T., {Varga}, T.~N., {Gruen}, D., {et~al.} 2019,
  \href{http://dx.doi.org/10.1093/mnras/sty2711}{\JournalTitle{\mnras}, 482,
  1352}, \eprint arXiv:{1805.00039}

\bibitem[{{McConnachie}(2012)}]{McConnachie:2012}
{McConnachie}, A.~W. 2012,
  \href{http://dx.doi.org/10.1088/0004-6256/144/1/4}{\JournalTitle{\aj}, 144,
  4}, \eprint arXiv:{1204.1562}

\bibitem[{{Merritt} {et~al.}(2014){Merritt} \& {van Dokkum} \&
  {Abraham}}]{Merritt:2014}
{Merritt}, A., {van Dokkum}, P., \& {Abraham}, R. 2014,
  \href{http://dx.doi.org/10.1088/2041-8205/787/2/L37}{\JournalTitle{\apjl},
  787, L37}, \eprint arXiv:{1406.2315}

\bibitem[{{Morales} {et~al.}(2018){Morales} \& {Mart{\'\i}nez-Delgado} \&
  {Grebel} \& {Cooper} \& {Javanmardi} \& {Miskolczi}}]{Morales:2018}
{Morales}, G., {Mart{\'\i}nez-Delgado}, D., {Grebel}, E.~K., {et~al.} 2018,
  \href{http://dx.doi.org/10.1051/0004-6361/201732271}{\JournalTitle{\aap},
  614, A143}, \eprint arXiv:{1804.03330}

\bibitem[{{Morganson} {et~al.}(2018){Morganson} \& {Gruendl}
  {et~al.}}]{Morganson:2018}
{Morganson}, E., {Gruendl}, R.~A., {Menanteau}, F., {et~al.} 2018,
  \href{http://dx.doi.org/10.1088/1538-3873/aab4ef}{\JournalTitle{\pasp}, 130,
  074501}, \eprint arXiv:{1801.03177}

\bibitem[{{Moster} {et~al.}(2013){Moster} \& {Naab} \& {White}}]{Moster:2013}
{Moster}, B.~P., {Naab}, T., \& {White}, S. D.~M. 2013,
  \href{http://dx.doi.org/10.1093/mnras/sts261}{\JournalTitle{\mnras}, 428,
  3121}, \eprint arXiv:{1205.5807}

\bibitem[{{Mouhcine} {et~al.}(2010){Mouhcine} \& {Ibata} \&
  {Rejkuba}}]{Mouhcine:2010}
{Mouhcine}, M., {Ibata}, R., \& {Rejkuba}, M. 2010,
  \href{http://dx.doi.org/10.1088/2041-8205/714/1/L12}{\JournalTitle{\apjl},
  714, L12}, \eprint arXiv:{1002.0461}

\bibitem[{{M{\"u}ller} {et~al.}(2015){M{\"u}ller} \& {Jerjen} \&
  {Binggeli}}]{Muller:2015}
{M{\"u}ller}, O., {Jerjen}, H., \& {Binggeli}, B. 2015,
  \href{http://dx.doi.org/10.1051/0004-6361/201526748}{\JournalTitle{\aap},
  583, A79}, \eprint arXiv:{1509.04931}

\bibitem[{{M{\"u}ller} {et~al.}(2019){M{\"u}ller} \& {Rejkuba}
  {et~al.}}]{Muller:2019}
{M{\"u}ller}, O., {Rejkuba}, M., {Pawlowski}, M.~S., {et~al.} 2019,
  \href{http://dx.doi.org/10.1051/0004-6361/201935807}{\JournalTitle{\aap},
  629, A18}, \eprint arXiv:{1907.02012}

\bibitem[{{Munshi} {et~al.}(2021){Munshi} \& {Brooks} \& {Applebaum} \&
  {Christensen} \& {Sligh} \& {Quinn}}]{Munshi:2021}
{Munshi}, F., {Brooks}, A., {Applebaum}, E., {et~al.} 2021, \JournalTitle{arXiv
  e-prints}, arXiv:2101.05822, \eprint arXiv:{2101.05822}

\bibitem[{{Munshi} {et~al.}(2019){Munshi} \& {Brooks} {et~al.}}]{Munshi:2018}
{Munshi}, F., {Brooks}, A.~M., {Christensen}, C., {et~al.} 2019,
  \href{http://dx.doi.org/10.3847/1538-4357/ab0085}{\JournalTitle{\apj}, 874,
  40}, \eprint arXiv:{1810.12417}

\bibitem[{{Mutlu-Pakdil} {et~al.}(2019){Mutlu-Pakdil} \& {Sand}
  {et~al.}}]{2019ApJ...885...53M}
{Mutlu-Pakdil}, B., {Sand}, D.~J., {Walker}, M.~G., {et~al.} 2019,
  \href{http://dx.doi.org/10.3847/1538-4357/ab45ec}{\JournalTitle{\apj}, 885,
  53}, \eprint arXiv:{1907.07233}

\bibitem[{{Nadler} {et~al.}(2019){Nadler} \& {Gluscevic} \& {Boddy} \&
  {Wechsler}}]{Nadler:2019b}
{Nadler}, E.~O., {Gluscevic}, V., {Boddy}, K.~K., \& {Wechsler}, R.~H. 2019,
  \href{http://dx.doi.org/10.3847/2041-8213/ab1eb2}{\JournalTitle{\apjl}, 878,
  L32}, \eprint arXiv:{1904.10000}

\bibitem[{{Nadler} {et~al.}(2020){Nadler} \& {Wechsler} {et~al.}}]{Nadler:2020}
{Nadler}, E.~O., {Wechsler}, R.~H., {Bechtol}, K., {et~al.} 2020,
  \href{http://dx.doi.org/10.3847/1538-4357/ab846a}{\JournalTitle{\apj}, 893,
  48}, \eprint arXiv:{1912.03303}

\bibitem[{Nadler {et~al.}(2021)Nadler \& Drlica-Wagner {et~al.}}]{Nadler:2020b}
Nadler, E.~O., Drlica-Wagner, A., Bechtol, K., {et~al.} 2021,
  \href{http://dx.doi.org/10.1103/PhysRevLett.126.091101}{\JournalTitle{\prl},
  126, 091101}, \eprint arXiv:{2008.00022}

\bibitem[{Neilsen {et~al.}(2015)Neilsen \& Bernstein \& Gruendl \&
  Kent}]{Neilsen:2015}
Neilsen, E., Bernstein, G., Gruendl, R., \& Kent, S. 2015, ``Limiting
  magnitude, $\tau$, $T_{eff}$, and image quality in DES Year 1'', Tech. Rep.
  FERMILAB-TM-2610-AE-CD, Fermi National Accelerator Laboratory

\bibitem[{{Newberg} {et~al.}(2002){Newberg} \& {Yanny} {et~al.}}]{Newberg:2002}
{Newberg}, H.~J., {Yanny}, B., {Rockosi}, C., {et~al.} 2002,
  \href{http://dx.doi.org/10.1086/338983}{\JournalTitle{\apj}, 569, 245},
  \eprint{astro-ph/0111095}

\bibitem[{Nidever \& Dorta(2020)}]{NideverDorta:2020}
Nidever, D. \& Dorta, A. 2020,
  \href{http://dx.doi.org/10.5281/zenodo.4291682}{\JournalTitle{Zenodo},
  4291682}

\bibitem[{{Nidever} {et~al.}(2020){Nidever} \& {Astro Data Lab Team}
  {et~al.}}]{Nidever:2020b}
{Nidever}, D.~L., {Astro Data Lab Team}, {et~al.} 2020, {NSCG: NOIRLab Source
  Catalog Generator}, \eprint ascl:{2012.002}

\bibitem[{{Nidever} {et~al.}(2017){Nidever} \& {Olsen} {et~al.}}]{Nidever:2017}
{Nidever}, D.~L., {Olsen}, K., {Walker}, A.~R., {et~al.} 2017,
  \href{http://dx.doi.org/10.3847/1538-3881/aa8d1c}{\JournalTitle{\aj}, 154,
  199}, \eprint arXiv:{1701.00502}

\bibitem[{{Nidever} {et~al.}(2018){Nidever} \& {Dey} {et~al.}}]{Nidever:2018}
{Nidever}, D.~L., {Dey}, A., {Olsen}, K., {et~al.} 2018,
  \href{http://dx.doi.org/10.3847/1538-3881/aad68f}{\JournalTitle{\aj}, 156,
  131}, \eprint arXiv:{1801.01885}

\bibitem[{{Nidever} {et~al.}(2019){Nidever} \& {Olsen}
  {et~al.}}]{Nidever:2019a}
{Nidever}, D.~L., {Olsen}, K., {Choi}, Y., {et~al.} 2019,
  \href{http://dx.doi.org/10.3847/1538-4357/aafaf7}{\JournalTitle{\apj}, 874,
  118}, \eprint arXiv:{1805.02671}

\bibitem[{{Nidever} {et~al.}(2021{\natexlab{a}}){Nidever} \& {Dey}
  {et~al.}}]{Nidever:2020a}
{Nidever}, D.~L., {Dey}, A., {Fasbender}, K., {et~al.} 2021{\natexlab{a}},
  \href{http://dx.doi.org/10.3847/1538-3881/abd6e1}{\JournalTitle{\aj}, 161,
  192}, \eprint arXiv:{2011.08868}

\bibitem[{{Nidever} {et~al.}(2021{\natexlab{b}}){Nidever} \& {Olsen}
  {et~al.}}]{Nidever:2021}
{Nidever}, D.~L., {Olsen}, K., {Choi}, Y., {et~al.} 2021{\natexlab{b}},
  \href{http://dx.doi.org/10.3847/1538-3881/abceb7}{\JournalTitle{\aj}, 161,
  74}, \eprint arXiv:{2011.13943}

\bibitem[{Nikutta {et~al.}(2020)Nikutta \& Fitzpatrick \& Scott \&
  Weaver}]{Nikutta:2020}
Nikutta, R., Fitzpatrick, M., Scott, A., \& Weaver, B. 2020,
  \href{http://dx.doi.org/https://doi.org/10.1016/j.ascom.2020.100411}{\JournalTitle{Astronomy
  and Computing}, 33, 100411}

\bibitem[{Nord {et~al.}(2016)Nord \& Buckley-Geer {et~al.}}]{Nord:2016}
Nord, B., Buckley-Geer, E., Lin, H., {et~al.} 2016,
  \href{http://dx.doi.org/10.3847/0004-637x/827/1/51}{\JournalTitle{The
  Astrophysical Journal}, 827, 51}

\bibitem[{{Odenkirchen} {et~al.}(2001){Odenkirchen} \& {Grebel}
  {et~al.}}]{Odenkirchen:2001}
{Odenkirchen}, M., {Grebel}, E.~K., {Rockosi}, C.~M., {et~al.} 2001,
  \href{http://dx.doi.org/10.1086/319095}{\JournalTitle{\apjl}, 548, L165},
  \eprint arXiv:{astro-ph/0012311}

\bibitem[{{Oguri} \& {Marshall}(2010)}]{Oguri:2010}
{Oguri}, M. \& {Marshall}, P.~J. 2010,
  \href{http://dx.doi.org/10.1111/j.1365-2966.2010.16639.x}{\JournalTitle{\mnras},
  405, 2579}, \eprint arXiv:{1001.2037}

\bibitem[{{Okamoto} {et~al.}(2015){Okamoto} \& {Arimoto}
  {et~al.}}]{Okamoto:2015}
{Okamoto}, S., {Arimoto}, N., {Ferguson}, A. M.~N., {et~al.} 2015,
  \href{http://dx.doi.org/10.1088/2041-8205/809/1/L1}{\JournalTitle{\apjl},
  809, L1}, \eprint arXiv:{1507.04889}

\bibitem[{{Okamoto} {et~al.}(2019){Okamoto} \& {Arimoto} \& {Ferguson} \&
  {Irwin} \& {Bernard} \& {Utsumi}}]{Okamoto:2019}
{Okamoto}, S., {Arimoto}, N., {Ferguson}, A. M.~N., {et~al.} 2019,
  \href{http://dx.doi.org/10.3847/1538-4357/ab44a7}{\JournalTitle{\apj}, 884,
  128}, \eprint arXiv:{1909.12997}

\bibitem[{{Oke}(1974)}]{Oke:1974}
{Oke}, J.~B. 1974,
  \href{http://dx.doi.org/10.1086/190287}{\JournalTitle{\apjs}, 27, 21}

\bibitem[{{Padmanabhan} {et~al.}(2008){Padmanabhan} \& {Schlegel}
  {et~al.}}]{Padmanabhan:2008}
{Padmanabhan}, N., {Schlegel}, D.~J., {Finkbeiner}, D.~P., {et~al.} 2008,
  \href{http://dx.doi.org/10.1086/524677}{\JournalTitle{\apj}, 674, 1217},
  \eprint arXiv:{astro-ph/0703454}

\bibitem[{{Pardy} {et~al.}(2020){Pardy} \& {D'Onghia} {et~al.}}]{Pardy:2020}
{Pardy}, S.~A., {D'Onghia}, E., {Navarro}, J.~F., {et~al.} 2020,
  \href{http://dx.doi.org/10.1093/mnras/stz3192}{\JournalTitle{\mnras}, 492,
  1543}, \eprint arXiv:{1904.01028}

\bibitem[{{Patel} {et~al.}(2020){Patel} \& {Kallivayalil}
  {et~al.}}]{Patel:2020}
{Patel}, E., {Kallivayalil}, N., {Garavito-Camargo}, N., {et~al.} 2020,
  \href{http://dx.doi.org/10.3847/1538-4357/ab7b75}{\JournalTitle{\apj}, 893,
  121}, \eprint arXiv:{2001.01746}

\bibitem[{{Pimbblet} {et~al.}(2002){Pimbblet} \& {Smail}
  {et~al.}}]{Pimbblet2002}
{Pimbblet}, K.~A., {Smail}, I., {Kodama}, T., {et~al.} 2002,
  \href{http://dx.doi.org/10.1046/j.1365-8711.2002.05186.x}{\JournalTitle{\mnras},
  331, 333}, \eprint arXiv:{astro-ph/0111461}

\bibitem[{{Planck Collaboration}(2020){Planck Collaboration} \& {Aghanim}
  {et~al.}}]{Planck:2018}
{Planck Collaboration}. 2020,
  \href{http://dx.doi.org/10.1051/0004-6361/201833910}{\JournalTitle{\aap},
  641, A6}, \eprint arXiv:{1807.06209}

\bibitem[{{Pogson}(1856)}]{Pogson:1856}
{Pogson}, N. 1856,
  \href{http://dx.doi.org/10.1093/mnras/17.1.12}{\JournalTitle{\mnras}, 17, 12}

\bibitem[{{Predehl} {et~al.}(2021){Predehl} \& {Andritschke}
  {et~al.}}]{Predehl:2020}
{Predehl}, P., {Andritschke}, R., {Arefiev}, V., {et~al.} 2021,
  \href{http://dx.doi.org/10.1051/0004-6361/202039313}{\JournalTitle{\aap},
  647, A1}, \eprint arXiv:{2010.03477}

\bibitem[{{Pucha} {et~al.}(2019){Pucha} \& {Carlin} {et~al.}}]{Pucha:2019}
{Pucha}, R., {Carlin}, J.~L., {Willman}, B., {et~al.} 2019,
  \href{http://dx.doi.org/10.3847/1538-4357/ab29fb}{\JournalTitle{\apj}, 880,
  104}, \eprint arXiv:{1905.02210}

\bibitem[{{Roderick} {et~al.}(2016){Roderick} \& {Jerjen} \& {Da Costa} \&
  {Mackey}}]{Roderick:2016}
{Roderick}, T.~A., {Jerjen}, H., {Da Costa}, G.~S., \& {Mackey}, A.~D. 2016,
  \href{http://dx.doi.org/10.1093/mnras/stw949}{\JournalTitle{\mnras}, 460,
  30}, \eprint arXiv:{1604.06214}

\bibitem[{{Rodr{\'\i}guez} {et~al.}(2016){Rodr{\'\i}guez} \& {Baume} \&
  {Feinstein}}]{Rodriguez:2016}
{Rodr{\'\i}guez}, M.~J., {Baume}, G., \& {Feinstein}, C. 2016,
  \href{http://dx.doi.org/10.1051/0004-6361/201527876}{\JournalTitle{\aap},
  594, A34}

\bibitem[{{Roederer} {et~al.}(2016){Roederer} \& {Mateo}
  {et~al.}}]{Roederer:2016}
{Roederer}, I.~U., {Mateo}, M., {Bailey}, John~I., I., {et~al.} 2016,
  \href{http://dx.doi.org/10.3847/0004-6256/151/3/82}{\JournalTitle{\aj}, 151,
  82}, \eprint arXiv:{1601.04070}

\bibitem[{{Romanowsky} {et~al.}(2016){Romanowsky} \& {Mart{\'\i}nez-Delgado}
  {et~al.}}]{Romanowsky:2016}
{Romanowsky}, A.~J., {Mart{\'\i}nez-Delgado}, D., {Martin}, N.~F., {et~al.}
  2016, \href{http://dx.doi.org/10.1093/mnrasl/slv207}{\JournalTitle{\mnras},
  457, L103}, \eprint arXiv:{1512.03815}

\bibitem[{{Rykoff} {et~al.}(2015){Rykoff} \& {Rozo} \& {Keisler}}]{Rykoff:2015}
{Rykoff}, E.~S., {Rozo}, E., \& {Keisler}, R. 2015, \JournalTitle{arXiv
  e-prints}, arXiv:1509.00870, \eprint arXiv:{1509.00870}

\bibitem[{{Rykoff} {et~al.}(2014){Rykoff} \& {Rozo}
  {et~al.}}]{2014ApJ...785..104R}
{Rykoff}, E.~S., {Rozo}, E., {Busha}, M.~T., {et~al.} 2014,
  \href{http://dx.doi.org/10.1088/0004-637X/785/2/104}{\JournalTitle{\apj},
  785, 104}, \eprint arXiv:{1303.3562}

\bibitem[{{Sand} {et~al.}(2015){Sand} \& {Spekkens} {et~al.}}]{Sand:2015}
{Sand}, D.~J., {Spekkens}, K., {Crnojevi{\'c}}, D., {et~al.} 2015,
  \href{http://dx.doi.org/10.1088/2041-8205/812/1/L13}{\JournalTitle{\apjl},
  812, L13}, \eprint arXiv:{1508.01800}

\bibitem[{{Sand} {et~al.}(2014){Sand} \& {Crnojevi{\'c}} {et~al.}}]{Sand:2014}
{Sand}, D.~J., {Crnojevi{\'c}}, D., {Strader}, J., {et~al.} 2014,
  \href{http://dx.doi.org/10.1088/2041-8205/793/1/L7}{\JournalTitle{\apjl},
  793, L7}, \eprint arXiv:{1406.6687}

\bibitem[{{Santana-Silva} {et~al.}(2020){Santana-Silva} \& {Gon{\c{c}}alves}
  {et~al.}}]{Santana-Silva:2020}
{Santana-Silva}, L., {Gon{\c{c}}alves}, T.~S., {Basu-Zych}, A., {et~al.} 2020,
  \href{http://dx.doi.org/10.1093/mnras/staa2757}{\JournalTitle{\mnras}, 498,
  5183}, \eprint arXiv:{2002.07828}

\bibitem[{{Schlafly} \& {Finkbeiner}(2011)}]{Schlafly:2011}
{Schlafly}, E.~F. \& {Finkbeiner}, D.~P. 2011,
  \href{http://dx.doi.org/10.1088/0004-637X/737/2/103}{\JournalTitle{\apj},
  737, 103}, \eprint arXiv:{1012.4804}

\bibitem[{{Schlafly} {et~al.}(2012){Schlafly} \& {Finkbeiner}
  {et~al.}}]{Schlafly:2012}
{Schlafly}, E.~F., {Finkbeiner}, D.~P., {Juri{\'c}}, M., {et~al.} 2012,
  \href{http://dx.doi.org/10.1088/0004-637X/756/2/158}{\JournalTitle{\apj},
  756, 158}, \eprint arXiv:{1201.2208}

\bibitem[{{Schlegel} {et~al.}(1998){Schlegel} \& {Finkbeiner} \&
  {Davis}}]{Schlegel:1998}
{Schlegel}, D.~J., {Finkbeiner}, D.~P., \& {Davis}, M. 1998,
  \href{http://dx.doi.org/10.1086/305772}{\JournalTitle{\apj}, 500, 525},
  \eprint{astro-ph/9710327}

\bibitem[{{Sevilla-Noarbe} {et~al.}(2021){Sevilla-Noarbe} \& {Bechtol}
  {et~al.}}]{Sevilla-Noarbe:2020}
{Sevilla-Noarbe}, I., {Bechtol}, K., {Carrasco Kind}, M., {et~al.} 2021,
  \href{http://dx.doi.org/10.3847/1538-4365/abeb66}{\JournalTitle{\apjs}, 254,
  24}, \eprint arXiv:{2011.03407}

\bibitem[{Shapiro {et~al.}(2004)Shapiro \& Iliev \& Raga}]{Shaprio:2004}
Shapiro, P.~R., Iliev, I.~T., \& Raga, A.~C. 2004,
  \href{http://dx.doi.org/10.1111/j.1365-2966.2004.07364.x}{\JournalTitle{Monthly
  Notices of the Royal Astronomical Society}, 348, 753},
  \eprint{http://oup.prod.sis.lan/mnras/article-pdf/348/3/753/4103465/348-3-753.pdf}

\bibitem[{Shipp {et~al.}(2020)Shipp \& Price-Whelan \& Tavangar \& Mateu \&
  Drlica-Wagner}]{Shipp:2020}
Shipp, N., Price-Whelan, A.~M., Tavangar, K., Mateu, C., \& Drlica-Wagner, A.
  2020, \href{http://dx.doi.org/10.3847/1538-3881/abbd3a}{\JournalTitle{The
  Astronomical Journal}, 160, 244}

\bibitem[{{Shipp} {et~al.}(2018){Shipp} \& {Drlica-Wagner}
  {et~al.}}]{Shipp:2018}
{Shipp}, N., {Drlica-Wagner}, A., {Balbinot}, E., {et~al.} 2018,
  \href{http://dx.doi.org/10.3847/1538-4357/aacdab}{\JournalTitle{\apj}, 862,
  114}, \eprint arXiv:{1801.03097}

\bibitem[{Shipp {et~al.}(2019)Shipp \& Li {et~al.}}]{Shipp:2019}
Shipp, N., Li, T.~S., Pace, A.~B., {et~al.} 2019,
  \href{http://dx.doi.org/10.3847/1538-4357/ab44bf}{\JournalTitle{The
  Astrophysical Journal}, 885, 3}

\bibitem[{{Simon}(2019)}]{Simon:2019}
{Simon}, J.~D. 2019,
  \href{http://dx.doi.org/10.1146/annurev-astro-091918-104453}{\JournalTitle{\araa},
  57, 375}, \eprint arXiv:{1901.05465}

\bibitem[{{Sluse} {et~al.}(2003){Sluse} \& {Surdej} {et~al.}}]{Sluse:2003}
{Sluse}, D., {Surdej}, J., {Claeskens}, J.~F., {et~al.} 2003,
  \href{http://dx.doi.org/10.1051/0004-6361:20030904}{\JournalTitle{\aap}, 406,
  L43}, \eprint arXiv:{astro-ph/0307345}

\bibitem[{{Smercina} {et~al.}(2018){Smercina} \& {Bell}
  {et~al.}}]{Smercina:2018}
{Smercina}, A., {Bell}, E.~F., {Price}, P.~A., {et~al.} 2018,
  \href{http://dx.doi.org/10.3847/1538-4357/aad2d6}{\JournalTitle{\apj}, 863,
  152}, \eprint arXiv:{1807.03779}

\bibitem[{{Smercina} {et~al.}(2020){Smercina} \& {Bell}
  {et~al.}}]{Smercina:2020}
{Smercina}, A., {Bell}, E.~F., {Price}, P.~A., {et~al.} 2020,
  \href{http://dx.doi.org/10.3847/1538-4357/abc485}{\JournalTitle{\apj}, 905,
  60}, \eprint arXiv:{1910.14672}

\bibitem[{{Spekkens} {et~al.}(2013){Spekkens} \& {Mason} \& {Aguirre} \&
  {Nhan}}]{Spekkens:2013}
{Spekkens}, K., {Mason}, B.~S., {Aguirre}, J.~E., \& {Nhan}, B. 2013,
  \href{http://dx.doi.org/10.1088/0004-637X/773/1/61}{\JournalTitle{\apj}, 773,
  61}, \eprint arXiv:{1301.5306}

\bibitem[{{Stetson}(1987)}]{Stetson:1987}
{Stetson}, P.~B. 1987,
  \href{http://dx.doi.org/10.1086/131977}{\JournalTitle{\pasp}, 99, 191}

\bibitem[{{Stetson}(1994)}]{Stetson:1994}
{Stetson}, P.~B. 1994,
  \href{http://dx.doi.org/10.1086/133378}{\JournalTitle{\pasp}, 106, 250}

\bibitem[{{Swanson} {et~al.}(2008){Swanson} \& {Tegmark} \& {Hamilton} \&
  {Hill}}]{Swanson:2008}
{Swanson}, M.~E.~C., {Tegmark}, M., {Hamilton}, A.~J.~S., \& {Hill}, J.~C.
  2008,
  \href{http://dx.doi.org/10.1111/j.1365-2966.2008.13296.x}{\JournalTitle{\mnras},
  387, 1391}, \eprint arXiv:{0711.4352}

\bibitem[{{Tanaka} {et~al.}(2011){Tanaka} \& {Chiba} \& {Komiyama} \&
  {Guhathakurta} \& {Kalirai}}]{Tanaka:2011}
{Tanaka}, M., {Chiba}, M., {Komiyama}, Y., {Guhathakurta}, P., \& {Kalirai},
  J.~S. 2011,
  \href{http://dx.doi.org/10.1088/0004-637X/738/2/150}{\JournalTitle{\apj},
  738, 150}, \eprint arXiv:{1107.0911}

\bibitem[{{Tollerud} {et~al.}(2008){Tollerud} \& {Bullock} \& {Strigari} \&
  {Willman}}]{Tollerud:2008}
{Tollerud}, E.~J., {Bullock}, J.~S., {Strigari}, L.~E., \& {Willman}, B. 2008,
  \href{http://dx.doi.org/10.1086/592102}{\JournalTitle{\apj}, 688, 277},
  \eprint arXiv:{0806.4381}

\bibitem[{Tollerud \& Peek(2018)}]{Tollerud:2018}
Tollerud, E.~J. \& Peek, J. E.~G. 2018,
  \href{http://dx.doi.org/10.3847/1538-4357/aab3e4}{\JournalTitle{The
  Astrophysical Journal}, 857, 45}

\bibitem[{{Toloba} {et~al.}(2016){Toloba} \& {Sand} {et~al.}}]{Toloba:2016}
{Toloba}, E., {Sand}, D.~J., {Spekkens}, K., {et~al.} 2016,
  \href{http://dx.doi.org/10.3847/2041-8205/816/1/L5}{\JournalTitle{\apjl},
  816, L5}, \eprint arXiv:{1512.03816}

\bibitem[{{Tonry} {et~al.}(2018){Tonry} \& {Denneau} {et~al.}}]{Tonry:2018}
{Tonry}, J.~L., {Denneau}, L., {Flewelling}, H., {et~al.} 2018,
  \href{http://dx.doi.org/10.3847/1538-4357/aae386}{\JournalTitle{\apj}, 867,
  105}, \eprint arXiv:{1809.09157}

\bibitem[{{Torrealba} {et~al.}(2018){Torrealba} \& {Belokurov}
  {et~al.}}]{Torrealba:2018}
{Torrealba}, G., {Belokurov}, V., {Koposov}, S.~E., {et~al.} 2018,
  \href{http://dx.doi.org/10.1093/mnras/sty170}{\JournalTitle{\mnras}, 475,
  5085}, \eprint arXiv:{1801.07279}

\bibitem[{{Tosi} {et~al.}(1991){Tosi} \& {Greggio} \& {Marconi} \&
  {Focardi}}]{Tosi:1991}
{Tosi}, M., {Greggio}, L., {Marconi}, G., \& {Focardi}, P. 1991,
  \href{http://dx.doi.org/10.1086/115925}{\JournalTitle{\aj}, 102, 951}

\bibitem[{{Treu}(2010)}]{Treu:2010}
{Treu}, T. 2010,
  \href{http://dx.doi.org/10.1146/annurev-astro-081309-130924}{\JournalTitle{\araa},
  48, 87}, \eprint arXiv:{1003.5567}

\bibitem[{Treu {et~al.}(2018)Treu \& Agnello {et~al.}}]{Treu:2018}
Treu, T., Agnello, A., Baumer, M.~A., {et~al.} 2018,
  \href{http://dx.doi.org/10.1093/mnras/sty2329}{\JournalTitle{Monthly Notices
  of the Royal Astronomical Society}, 481, 1041–1054}

\bibitem[{{Tully} {et~al.}(2006){Tully} \& {Rizzi} {et~al.}}]{Tully:2006}
{Tully}, R.~B., {Rizzi}, L., {Dolphin}, A.~E., {et~al.} 2006,
  \href{http://dx.doi.org/10.1086/505466}{\JournalTitle{\aj}, 132, 729},
  \eprint arXiv:{astro-ph/0603380}

\bibitem[{{Vegetti} {et~al.}(2012){Vegetti} \& {Lagattuta} \& {McKean} \&
  {Auger} \& {Fassnacht} \& {Koopmans}}]{Vegetti:2012}
{Vegetti}, S., {Lagattuta}, D.~J., {McKean}, J.~P., {et~al.} 2012,
  \href{http://dx.doi.org/10.1038/nature10669}{\JournalTitle{\nat}, 481, 341},
  \eprint arXiv:{1201.3643}

\bibitem[{{Virtanen} {et~al.}(2020){Virtanen} \& {Gommers}
  {et~al.}}]{Scipy:2020}
{Virtanen}, P., {Gommers}, R., {Oliphant}, T.~E., {et~al.} 2020,
  \href{http://dx.doi.org/10.1038/s41592-019-0686-2}{\JournalTitle{Nature
  Methods}, 17, 261}, \eprint arXiv:{1907.10121}

\bibitem[{{Vivas} {et~al.}(2019){Vivas} \& {Alonso-Garc{\'\i}a} \& {Mateo} \&
  {Walker} \& {Howard}}]{Vivas:2019}
{Vivas}, A.~K., {Alonso-Garc{\'\i}a}, J., {Mateo}, M., {Walker}, A., \&
  {Howard}, B. 2019,
  \href{http://dx.doi.org/10.3847/1538-3881/aaf4f3}{\JournalTitle{\aj}, 157,
  35}, \eprint arXiv:{1811.12207}

\bibitem[{{Watkins} {et~al.}(2009){Watkins} \& {Evans} {et~al.}}]{Watkins:2009}
{Watkins}, L.~L., {Evans}, N.~W., {Belokurov}, V., {et~al.} 2009,
  \href{http://dx.doi.org/10.1111/j.1365-2966.2009.15242.x}{\JournalTitle{\mnras},
  398, 1757}, \eprint arXiv:{0906.0498}

\bibitem[{{Weisz} \& {Boylan-Kolchin}(2017)}]{Weisz:2017}
{Weisz}, D.~R. \& {Boylan-Kolchin}, M. 2017,
  \href{http://dx.doi.org/10.1093/mnrasl/slx043}{\JournalTitle{\mnras}, 469,
  L83}, \eprint arXiv:{1702.06129}

\bibitem[{{Weisz} {et~al.}(2014{\natexlab{a}}){Weisz} \& {Dolphin}
  {et~al.}}]{Weisz:2014a}
{Weisz}, D.~R., {Dolphin}, A.~E., {Skillman}, E.~D., {et~al.}
  2014{\natexlab{a}},
  \href{http://dx.doi.org/10.1088/0004-637X/789/2/147}{\JournalTitle{\apj},
  789, 147}, \eprint arXiv:{1404.7144}

\bibitem[{{Weisz} {et~al.}(2014{\natexlab{b}}){Weisz} \& {Dolphin}
  {et~al.}}]{Weisz:2014b}
{Weisz}, D.~R., {Dolphin}, A.~E., {Skillman}, E.~D., {et~al.}
  2014{\natexlab{b}},
  \href{http://dx.doi.org/10.1088/0004-637X/789/2/148}{\JournalTitle{\apj},
  789, 148}, \eprint arXiv:{1405.3281}

\bibitem[{{Wetzel} {et~al.}(2015){Wetzel} \& {Deason} \&
  {Garrison-Kimmel}}]{Wetzel:2015}
{Wetzel}, A.~R., {Deason}, A.~J., \& {Garrison-Kimmel}, S. 2015,
  \href{http://dx.doi.org/10.1088/0004-637X/807/1/49}{\JournalTitle{\apj}, 807,
  49}, \eprint arXiv:{1501.01972}

\bibitem[{{Wheeler} {et~al.}(2015){Wheeler} \& {O{\~n}orbe}
  {et~al.}}]{Wheeler:2015}
{Wheeler}, C., {O{\~n}orbe}, J., {Bullock}, J.~S., {et~al.} 2015,
  \href{http://dx.doi.org/10.1093/mnras/stv1691}{\JournalTitle{\mnras}, 453,
  1305}, \eprint arXiv:{1504.02466}

\bibitem[{{Wheeler} {et~al.}(2019){Wheeler} \& {Hopkins}
  {et~al.}}]{Wheeler:2019}
{Wheeler}, C., {Hopkins}, P.~F., {Pace}, A.~B., {et~al.} 2019,
  \href{http://dx.doi.org/10.1093/mnras/stz2887}{\JournalTitle{\mnras}, 490,
  4447}, \eprint arXiv:{1812.02749}

\bibitem[{{Wolf} {et~al.}(2018){Wolf} \& {Onken} {et~al.}}]{Wolf:2018}
{Wolf}, C., {Onken}, C.~A., {Luvaul}, L.~C., {et~al.} 2018,
  \href{http://dx.doi.org/10.1017/pasa.2018.5}{\JournalTitle{\pasa}, 35, e010},
  \eprint arXiv:{1801.07834}

\bibitem[{{Wong} {et~al.}(2020){Wong} \& {Suyu} {et~al.}}]{Wong:2020}
{Wong}, K.~C., {Suyu}, S.~H., {Chen}, G. C.~F., {et~al.} 2020,
  \href{http://dx.doi.org/10.1093/mnras/stz3094}{\JournalTitle{\mnras}, 498,
  1420}, \eprint arXiv:{1907.04869}

\bibitem[{{York} {et~al.}(2000){York} \& {Adelman} {et~al.}}]{York:2000}
{York}, D.~G., {Adelman}, J., {Anderson}, Jr., J.~E., {et~al.} 2000,
  \href{http://dx.doi.org/10.1086/301513}{\JournalTitle{\aj}, 120, 1579},
  \eprint{astro-ph/0006396}

\bibitem[{{Zenteno} {et~al.}(2016){Zenteno} \& {Mohr} {et~al.}}]{Zenteno16}
{Zenteno}, A., {Mohr}, J.~J., {Desai}, S., {et~al.} 2016,
  \href{http://dx.doi.org/10.1093/mnras/stw1649}{\JournalTitle{\mnras}, 462,
  830}, \eprint arXiv:{1603.05981}

\bibitem[{Zenteno {et~al.}(2020)Zenteno \& Hernández-Lang
  {et~al.}}]{Zenteno20}
Zenteno, A., Hernández-Lang, D., Klein, M., {et~al.} 2020,
  \href{http://dx.doi.org/10.1093/mnras/staa1157}{\JournalTitle{Monthly Notices
  of the Royal Astronomical Society}, 495, 705},
  \eprint{https://academic.oup.com/mnras/article-pdf/495/1/705/33241001/staa1157.pdf}

\bibitem[{{Zijlstra} \& {Minniti}(1999)}]{Zijlstra:1999}
{Zijlstra}, A.~A. \& {Minniti}, D. 1999,
  \href{http://dx.doi.org/10.1086/300802}{\JournalTitle{\aj}, 117, 1743},
  \eprint arXiv:{astro-ph/9812330}

\bibitem[{Zonca {et~al.}(2019)Zonca \& Singer {et~al.}}]{Zonca:2019}
Zonca, A., Singer, L., Lenz, D., {et~al.} 2019,
  \href{http://dx.doi.org/10.21105/joss.01298}{\JournalTitle{Journal of Open
  Source Software}, 4, 1298}

\end{thebibliography}
